\documentclass[11pt,letterpaper,onecolumn]{article}
\usepackage[letterpaper,margin=1in]{geometry}
\usepackage[T1]{fontenc}

\usepackage[UKenglish]{babel}
\usepackage{etoolbox}
\patchcmd{\abstract}{\small}{\normalsize}{}{}

\usepackage{amsmath,amssymb,amsthm}
\usepackage{xcolor}
\usepackage[colorlinks=true, allcolors=blue, hypertexnames=false, hyperfootnotes=false]{hyperref}
\hypersetup{pdftitle={Inapproximability of Euclidean Steiner Tree}, pdfauthor={Simon Bartlmae}}
\usepackage[capitalise,noabbrev]{cleveref}
\usepackage{thmtools}
\usepackage{thm-restate}
\usepackage{enumerate}
\usepackage{tikz}
\usepackage{placeins}
\usepackage{float}

\usetikzlibrary{calc,shapes.geometric,arrows.meta}

\definecolor{cableU}{HTML}{7654A3}
\definecolor{cableV}{HTML}{0072B2}
\definecolor{cableW}{HTML}{009E73}
\definecolor{planeCableW}{HTML}{00866A}
\definecolor{cableVp}{HTML}{C74C16}
\definecolor{cableWp}{HTML}{A47B12}
\definecolor{connectionGreen}{HTML}{FF1F1F}

\makeatletter
\ifdefined\thmt@autorefsetup\renewcommand\thmt@autorefsetup{}\fi
\makeatother
\theoremstyle{plain}
\newtheorem{theorem}{Theorem}
\newtheorem{lemma}[theorem]{Lemma}

\newtheorem{corollary}[theorem]{Corollary}
\newtheorem{definition}[theorem]{Definition}
\newtheorem{hypothesis}{Hypothesis}

\newcommand{\BR}{\mathbb R}
\DeclareMathOperator*{\poly}{poly}
\DeclareMathOperator{\OPT}{OPT}

\title{Inapproximability of Euclidean Steiner Tree}
\author{Simon Bartlmae}
\date{}

\begin{document}

\maketitle
\thispagestyle{empty}
\pagenumbering{gobble}
{\renewcommand{\thefootnote}{}\footnotetext{University of Copenhagen. \texttt{simon.bartlmae7@gmail.com}. Partially supported by the Independent Research Fund Denmark, grant 1054-00032B, and by the Carlsberg Foundation, grant CF24-1929.}}

\begin{abstract}
We consider the Euclidean Steiner Tree problem of connecting $n$ given points in $\mathbb{R}^d$ by a shortest network allowing the addition of Steiner points anywhere in the space. Arora [FOCS'96/'97, J.\ ACM'98] gave a polynomial-time approximation scheme (PTAS) for Euclidean Steiner Tree and TSP in every fixed dimension, and Trevisan [STOC'97, SICOMP'00] proved that Euclidean TSP is APX-hard in $O(\log n)$ dimensions. They left open whether Euclidean Steiner Tree admits a PTAS in high dimension. Our first contribution settles this thirty-year-old question by proving APX-hardness in $O(\log n)$ dimensions. Assuming that not all problems in NP can be solved in subexponential time ($\mathrm{NP}\not\subseteq\mathrm{SUBEXP}$), we also rule out a PTAS for Euclidean Steiner Tree in $O(\log\log n)$ dimensions. This dimension bound is tight up to a factor of $O(\log\log\log n)$, since Arora's scheme is a PTAS in up to $\log\log n/\log\log\log n$ dimensions.

In fixed dimension, Kisfaludi-Bak, Nederlof, and W\k{e}grzycki [FOCS'21, J.\ ACM'25] obtained approximation schemes running in time $2^{O_d(\varepsilon^{-(d-1)})}n\log n$ and proved matching Gap-ETH lower bounds for Euclidean TSP and Rectilinear Steiner Tree, but left a corresponding lower bound for Euclidean Steiner Tree as an open question. Our second contribution answers this question with a nearly tight lower bound on the dependence on accuracy for Euclidean Steiner Tree. However, given the recent claimed refutations of ETH and hence Gap-ETH by OpenAI and Anthropic, we use the weaker assumption that 3SAT has no truly subexponential-time algorithm, that is, no $2^{n^\gamma}$-time algorithm on formulas of size $n$ for any constant $\gamma<1$ (a hypothesis one might call \emph{weak ETH}). Under this assumption, we prove that no $(1+\varepsilon)$-approximation algorithm runs in time $2^{O(\varepsilon^{-\gamma(d-1)})}\poly(n)$ for any such $\gamma$.

An elegant aspect of our approach is that the same framework of gap-preserving reductions from three-dimensional matching applies to all three cases: APX-hardness in high dimensions, and lower bounds for approximation schemes in fixed $d\ge3$ and in the plane.
\end{abstract}

\clearpage
\tableofcontents
\clearpage
\pagenumbering{arabic}

\section{Introduction}
\label{sec:introduction}

We consider the Euclidean Steiner Tree problem of connecting $n$ given points in $\mathbb{R}^d$ by a shortest network. These points are called \emph{terminals} and additional \emph{Steiner points} may be placed anywhere in the space. The length of a network is the sum of its Euclidean edge lengths.

The problem has its roots in the Fermat--Torricelli problem and nineteenth-century work on shortest networks. It was later studied by Jarn\'{i}k and K\"ossler~\cite{JarnikKossler34} and by Gilbert and Pollak~\cite{GilbertPollak68}; see Brazil, Graham, Thomas, and Zachariasen~\cite{BrazilGTZ14} for a historical account. Garey, Graham, and Johnson~\cite{DBLP:conf/stoc/GareyGJ76,GareyGJ77} proved NP-hardness even in the plane.

In the 1990s, Arora~\cite{DBLP:conf/focs/Arora96} obtained a polynomial-time approximation scheme (PTAS) for the plane and a quasi\-polynomial-time approximation scheme for higher fixed dimensions. Independently, Mitchell~\cite{DBLP:journals/siamcomp/Mitchell99} developed a planar PTAS based on guillotine subdivisions. Arora later improved his scheme~\cite{DBLP:conf/focs/Arora97,DBLP:journals/jacm/Arora98}, obtaining a randomised PTAS that runs in time $n(\log n)^{O_d(\varepsilon^{-(d-1)})}$ for every fixed dimension~$d$. His method recursively partitions the space and restricts where and how often a near-optimal network crosses region boundaries. Dynamic programming then combines solutions for the smaller regions.

Rao and Smith~\cite{DBLP:conf/stoc/RaoS98} reduced the running time to $O(n\log n)$ for fixed dimension and accuracy. For Euclidean Steiner Tree, their bound is $O_d(2^{O_d(\varepsilon^{-(d-1)}\log(1/\varepsilon))}n+n\log n)$. Their method combines recursive dissections with \emph{banyans}, sparse geometric graphs containing near-optimal Steiner trees for every subset of the terminals. Kisfaludi-Bak, Nederlof, and W\k{e}grzycki~\cite{DBLP:journals/jacm/KisfaludiBakNW25} later removed the logarithmic factor in the exponent using sparsity-sensitive patching, obtaining an approximation scheme running in time $O_d(2^{O_d(\varepsilon^{-(d-1)})}n\log n)$ for Euclidean TSP, Euclidean Steiner Tree, and Rectilinear Steiner Tree. They also proved matching Gap-ETH lower bounds for Euclidean TSP and Rectilinear Steiner Tree, leaving the corresponding bound for Euclidean Steiner Tree as an open question.

The exponential dependence of these schemes on the dimension led Arora~\cite{DBLP:conf/focs/Arora96} to ask whether his scheme in $\mathbb{R}^d$ could run in time $n^{O(d/\varepsilon)}$, with an exponent linear in $d$. For Euclidean TSP, Trevisan~\cite{DBLP:conf/stoc/Trevisan97,DBLP:journals/siamcomp/Trevisan00} proved APX-hardness in $O(\log n)$ dimensions, so such a scheme is unlikely: in these dimensions, it would run in time $n^{O(\log n)}$ for fixed $\varepsilon$, and would thus solve NP-hard problems in quasipolynomial time~\cite{DBLP:journals/jacm/Arora98}. Trevisan also proved APX-hardness for Rectilinear Steiner Tree in $O(n)$ dimensions.

Ruling out a PTAS for Euclidean Steiner Tree in high dimension has since been described as ``a notoriously long standing open problem''~\cite{DBLP:journals/jocg/FleischmannQ0MP25}. The main difficulty is that Steiner points can be placed anywhere. Fleischmann, Gavva, and Karthik~\cite{DBLP:conf/soda/FleischmannGS24,DBLP:journals/theoretics/FleischmannG025} studied the distinction between the \emph{discrete} problem, where Steiner points must come from a given set of candidates, and the \emph{continuous} problem, where they may lie anywhere. They established APX-hardness for the discrete problem in every $\ell_p$ metric and its continuous counterpart in $\ell_\infty$, and reduced the continuous to the discrete problem in $\ell_p$ metrics. For the continuous Euclidean problem, Fleischmann, Gamboa Quintero, Karthik, Mat\v{e}jka, and Petr~\cite{DBLP:journals/jocg/FleischmannQ0MP25} connected approximation hardness with the structure of optimal Steiner trees for regular simplices and simplicial complexes. They formulated structural conjectures that would imply APX-hardness and revisited a conjecture of Smith~\cite{DBLP:journals/algorithmica/Smith92} about the optimal topology for the regular simplex. We instead prove APX-hardness through a gap-preserving reduction from three-dimensional matching.

\paragraph{Remark.}
This paper was almost finished when the OpenAI and Anthropic preprints~\cite{OpenAI-ETH26,Claude-ETH26} claiming to refute ETH appeared. The proofs of \cref{thm:loglog-hardness,thm:fixed-dimension-hardness} originally relied on ETH and Gap-ETH, respectively, but weaker assumptions on the running time of 3SAT, and even $\mathrm{NP}\not\subseteq\mathrm{SUBEXP}$, now suffice for some of the bounds. We believe these reductions remain valuable, as discussed throughout the paper and in \cref{sec:open-problems}.

\paragraph{Our contributions.}
We prove that Euclidean Steiner Tree admits no PTAS in $O(\log n)$ dimensions unless $\mathrm{P}=\mathrm{NP}$, and give conditional lower bounds showing that the exponent $\varepsilon^{-(d-1)}$ is nearly optimal in every fixed dimension. First, we establish APX-hardness in logarithmic dimension.

\begin{restatable}{theorem}{GrowingDimensionHardness}
\label{thm:growing-dimension-hardness}
Euclidean Steiner Tree is APX-hard in $O(\log n)$ dimensions.
\end{restatable}

This rules out a PTAS in arbitrary dimension unless $\mathrm{P}=\mathrm{NP}$, answering Arora's question. The dimension is close to optimal for APX-hardness: for fixed accuracy, Arora's scheme~\cite{DBLP:journals/jacm/Arora98} runs in subexponential time $2^{n^{o(1)}}$ in $o(\log n/\log\log n)$ dimensions, so APX-hardness there would imply $\mathrm{NP}\subseteq\mathrm{SUBEXP}$. Padding the construction also rules out a PTAS in much lower dimension under $\mathrm{NP}\not\subseteq\mathrm{SUBEXP}$.

\begin{restatable}{theorem}{LogLogHardness}
\label{thm:loglog-hardness}
Unless $\mathrm{NP}\subseteq\mathrm{SUBEXP}$, Euclidean Steiner Tree admits no PTAS in $O(\log\log n)$ dimensions.
\end{restatable}

\Cref{thm:loglog-hardness} is tight up to a factor of $O(\log\log\log n)$ in the dimension. Arora's scheme~\cite{DBLP:journals/jacm/Arora98} runs in time $O\bigl(n(\log n)^{(O(\sqrt d/\varepsilon))^{d-1}}\bigr)$, which remains polynomial for every fixed $\varepsilon$ whenever $d\le\log\log n/\log\log\log n$. In this range, $d\log d\le\log\log n$, so for sufficiently large $d$ the exponent of $\log n$ is at most $2^{\frac34d\log d}\le(\log n)^{3/4}$, giving $(\log n)^{(\log n)^{3/4}}=2^{(\log n)^{3/4}\log\log n}=n^{o(1)}$. Thus Euclidean Steiner Tree admits a PTAS in up to $\log\log n/\log\log\log n$ dimensions, but, unless $\mathrm{NP}\subseteq\mathrm{SUBEXP}$, not in $O(\log\log n)$ dimensions. Under the same assumption, a doubly exponential dependence of the running time on the dimension is therefore unavoidable; \cref{sec:open-problems} discusses the remaining gap.

Second, we establish lower bounds on the dependence on accuracy in every fixed dimension. We obtain nearly tight bounds under the following weaker form of ETH, which has not been refuted by the announced SAT algorithms.

\begin{hypothesis}[Weak ETH]
\label{hyp:weak-eth}
3SAT has no truly subexponential-time algorithm, that is, no algorithm running in time $2^{n^\gamma}$ on formulas of size $n$ for any constant $\gamma<1$.
\end{hypothesis}

\begin{restatable}{theorem}{FixedDimensionHardness}
\label{thm:fixed-dimension-hardness}
Under weak ETH, for every fixed $d\ge2$ and every constant $\gamma<1$, Euclidean Steiner Tree in $\mathbb{R}^d$ admits no $(1+\varepsilon)$-approximation algorithm running in time
\[
    2^{O(\varepsilon^{-\gamma(d-1)})}\poly(n).
\]
\end{restatable}

Under Gap-ETH, the same reduction would have ruled out running time $2^{o(\varepsilon^{-(d-1)})}\poly(n)$ for every fixed $d\ge2$. This would have matched the known scheme's dependence on accuracy in the exponent, answering the question of Kisfaludi-Bak, Nederlof, and W\k{e}grzycki~\cite{DBLP:journals/jacm/KisfaludiBakNW25}.

Under weak ETH, the power of $1/\varepsilon$ in the exponent can be made arbitrarily close to $d-1$, the power achieved by the known scheme. This leaves room for subpolynomial savings in the exponent, but excludes any fixed reduction in this power. For every fixed $d\ge3$, it rules out running times such as $2^{2^d/\varepsilon}\poly(n)$, which APX-hardness in $O(\log n)$ dimensions does not exclude, as Arora~\cite{DBLP:journals/jacm/Arora98} observed. The restriction $d\ge3$ in this example is necessary: in the plane, the known scheme already runs in time $2^{O(1/\varepsilon)}\poly(n)$.

Even under $\mathrm{NP}\not\subseteq\mathrm{SUBEXP}$, the same reduction rules out running time $2^{o(\varepsilon^{-c(d-1)})}\poly(n)$ for some constant $c>0$ independent of $d$. Thus the power of $1/\varepsilon$ in the exponent cannot be bounded independently of the dimension.

Even if Gap-ETH fails, improving the PTAS appears highly nontrivial: such an improvement would yield a faster gap-3SAT algorithm, while the claimed ETH refutations rely on sophisticated algebraic methods.

As a by-product, our reduction proves APX-hardness for a completion version of Euclidean Steiner Tree, where parts of the network are already built and must be connected by additional segments of minimum total length. The result holds already in $\mathbb{R}^3$ (\cref{cor:cable-apx}), answering a question of Bartlmae, J\"unger, and Langetepe~\cite{DBLP:journals/corr/abs-2412-07046} negatively for $d\ge3$.

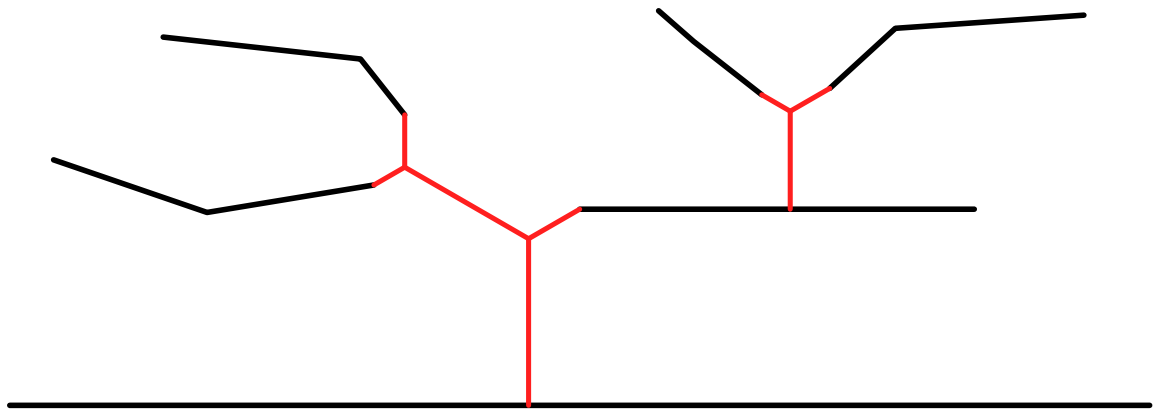
\begin{figure}[!tb]
\centering
\definecolor{introConnectionGreen}{HTML}{FF1F1F}
\begin{tikzpicture}[x=.29cm,y=.29cm,
  cable/.style={draw=black,line width=1.05pt,line cap=round,line join=round},
  connection/.style={draw=introConnectionGreen,line width=.95pt,line cap=round,line join=round}]
\draw[cable] (-13.000000,0.000000)--(13.000000,0.000000);
\draw[cable] (0.000000,4.473944)--(9.000000,4.473944);
\draw[cable] (-3.991561,6.626231)--(-5.000000,7.900000)--(-9.500000,8.400000);
\draw[cable] (-4.694128,5.026599)--(-8.500000,4.400000)--(-12.000000,5.600000);
\draw[cable] (5.702765,7.229449)--(7.200000,8.600000)--(11.500000,8.900000);
\draw[cable] (4.152658,7.084189)--(2.600000,8.300000)--(1.800000,9.000000);
\draw[connection] (4.801913,4.473944)--(4.801913,6.709342);
\draw[connection] (4.801913,6.709342)--(5.702765,7.229449);
\draw[connection] (4.801913,6.709342)--(4.152658,7.084189);
\draw[connection] (-1.165884,-0.000000)--(-1.165884,3.800820);
\draw[connection] (-1.165884,3.800820)--(0.000000,4.473944);
\draw[connection] (-1.165884,3.800820)--(-3.991561,5.432226);
\draw[connection] (-3.991561,5.432226)--(-3.991561,6.626231);
\draw[connection] (-3.991561,5.432226)--(-4.694128,5.026599);
\end{tikzpicture}
\caption{Cables (black) and a shortest network connecting them (red). Connections may attach in cable interiors, and only the red segments are paid for.}
\label{fig:cable-problem}
\end{figure}

\paragraph{Techniques.}
All our results follow the same reduction strategy. We first reduce three-dimensional matching to \emph{Euclidean Cable Steiner Tree}, a generalization of Euclidean Steiner Tree in which each terminal is a connected union of line segments rather than a point. We call these unions \emph{cables}; they are pairwise disjoint, and only the length added outside them is counted (\cref{fig:cable-problem}). A separation condition rules out unintended connections between the cables, ensuring that the minimum cost of connecting them encodes the size of a maximum matching.

We then replace each cable by densely spaced terminals and show that a short tree spanning them can be made to follow the cables at small extra cost. Up to this error, the optimum equals the total cable length plus the minimum connection cost. Each result therefore reduces to a layout problem: drawing separated cables of small total length so that the approximation gap survives.

In linear dimension, we construct such a layout by giving each gadget and each cable its own coordinate direction. We reduce the dimension to $O(\log n)$ by applying the Johnson--Lindenstrauss lemma~\cite{JohnsonL84} to the terminals and Kirszbraun's extension theorem~\cite{Kirszbraun34} to the Steiner points. This preserves the Steiner optimum up to small distortion, following the argument of Karthik C.~S.~\cite[Section~4.2]{DBLP:journals/corr/abs-2602-03837}. Padding the resulting instance gives \cref{thm:loglog-hardness} under $\mathrm{NP}\not\subseteq\mathrm{SUBEXP}$.

For fixed $d\ge3$, the cables must be longer, and we use the cube-wiring theorem of de Berg, Bodlaender, Kisfaludi-Bak, Marx, and van der Zanden~\cite{deBergBBMZ20} to route them with total length small enough to match the Gap-ETH bound. In the plane, we use the crossovers of Garey, Graham, and Johnson~\cite{GareyGJ77} to split crossing cables into pieces that are again separated. An elegant aspect of our approach is that the same reduction framework handles high dimensions, fixed $d\ge3$, and the plane.

\paragraph{Further related work.}
{In general metrics, Steiner Tree is APX-hard~\cite{DBLP:journals/ipl/BernP89}, NP-hard to approximate within $96/95$~\cite{DBLP:journals/tcs/ChlebikC08}, and approximable within $\ln4+\varepsilon$~\cite{DBLP:journals/jacm/ByrkaGRS13}. In the plane, Euclidean Steiner Tree is NP-hard even when the terminals lie on two parallel lines~\cite{DBLP:journals/siamdm/RubinsteinTW97}.} Approximation schemes are also known for TSP~\cite{DBLP:journals/siamcomp/BartalGK16} and Steiner Forest~\cite{DBLP:journals/siamcomp/ChanHJ18} in metrics of bounded doubling dimension. For Euclidean TSP, de Berg, Bodlaender, Kisfaludi-Bak, and Kolay~\cite{DBLP:conf/focs/BergBKK18,DBLP:journals/siamcomp/BergBKK23} gave an ETH-tight exact algorithm in every fixed dimension. Our routing uses the cube-wiring theorem from the related framework of de Berg, Bodlaender, Kisfaludi-Bak, Marx, and van der Zanden~\cite{deBergBBMZ20} for ETH-tight bounds in geometric intersection graphs. Related questions arise in clustering, where centres may lie anywhere in the space. Euclidean $k$-means is APX-hard~\cite{DBLP:conf/compgeom/AwasthiCKS15}, and Cohen-Addad and Karthik~\cite{DBLP:conf/focs/Cohen-AddadS19}, followed by Cohen-Addad, Karthik, and Lee~\cite{DBLP:conf/soda/Cohen-AddadSL22}, established stronger inapproximability bounds for $k$-means and $k$-median in $\ell_p$ metrics.

\section{Preliminaries}
\label{sec:preliminaries}

In a three-dimensional matching instance $H=(U,E)$, the element set $U$ is partitioned into three parts of size $r$, and each triple in $E$ contains one element from each part. A matching consists of pairwise disjoint triples and is perfect if it has size $r$. We consider \emph{3DM-3}, where each element occurs in at most three triples, implying $|E|\le3r$. Write $m=|U|+|E|$ for the instance size and $\nu(H)$ for the maximum size of a matching. All our reductions use the following gap version, in which YES instances have a perfect matching.

The proof is given in \cref{app:gap-source}.

\begin{lemma}
\label{lem:gap-source}
For some constant $\delta>0$, distinguishing $\nu(H)=r$ from $\nu(H)\le(1-\delta)r$ is NP-hard for instances $H$ of 3DM-3. Moreover:
\begin{enumerate}[(i)]
\item Under Gap-ETH, no algorithm distinguishes these cases in time $2^{o(r)}$.
\item Under weak ETH, no algorithm distinguishes these cases in time $2^{O(r^\gamma)}$ for any constant $\gamma<1$.
\item Under $\mathrm{NP}\not\subseteq\mathrm{SUBEXP}$, there is a constant $c>0$ such that no algorithm distinguishes these cases in time $2^{r^c}$.
\end{enumerate}
\end{lemma}

We may assume that every element occurs in at least one triple, since otherwise $H$ has no perfect matching. This gives $r\le|E|\le3r$ and hence $4r\le m\le6r$.

\paragraph{Coordinates.}
We use irrational coordinates for convenience; rounding them to polynomial bit length preserves our bounds, since moving each of $n$ terminals by at most some very small $\varepsilon>0$ changes the optimum by at most $n\varepsilon$.

\section{Technical Overview}
\label{sec:technical-overview}

All our reductions use the gap version of three-dimensional matching from \cref{lem:gap-source}, which asks whether an instance with parts of size $r$ admits a perfect matching or has no matching larger than $(1-\delta)r$.

\subsection{APX-hardness in linear dimension}
\label{sec:overview-linear-dimension}

\paragraph{Warm-up: connecting cables.}
We first consider a variant of the problem in which the terminals are trees of line segments, which we call \emph{cables}.

\begin{restatable}[Euclidean Cable Steiner Tree]{definition}{CableSteinerTree}
\label{def:cable-steiner-tree}
\leavevmode\\
\textbf{Input:} a finite collection of pairwise disjoint \emph{cables} in a Euclidean space, each a connected union of finitely many line segments.\\
\textbf{Feasible solutions:} finite sets of line segments, possibly meeting at Steiner points, that together with the cables form a connected network.\\
\textbf{Objective:} minimize the total length of the added segments.
\end{restatable}

We first explain how cables encode a matching, then replace them by densely spaced point terminals while preserving the approximation gap.

Given $H$, our goal is to construct a cable instance with minimum connection cost $\frac43|U|-\Delta\nu(H)$, where $\Delta=4-\sqrt2-\sqrt6>0$. Equivalently, $H$ has a matching of size at least $k$ if and only if the cables can be connected at cost at most $\frac43|U|-\Delta k$.

This identity yields a constant hardness gap for Euclidean Cable Steiner Tree, since each missing matching triple increases the connection cost by $\Delta$. With $|U|=3r$, the minimum connection cost is $4r-\Delta\nu(H)$, so
\[
\begin{aligned}
    \nu(H)=r
        &\quad\Longrightarrow\quad \text{cost}=4r-\Delta r=(4-\Delta)r,\\
    \nu(H)\le(1-\delta)r
        &\quad\Longrightarrow\quad \text{cost}\ge4r-\Delta(1-\delta)r=(4-\Delta)r+\Delta\delta r.
\end{aligned}
\]
The ratio between these bounds is the constant $1+\Delta\delta/(4-\Delta)$.

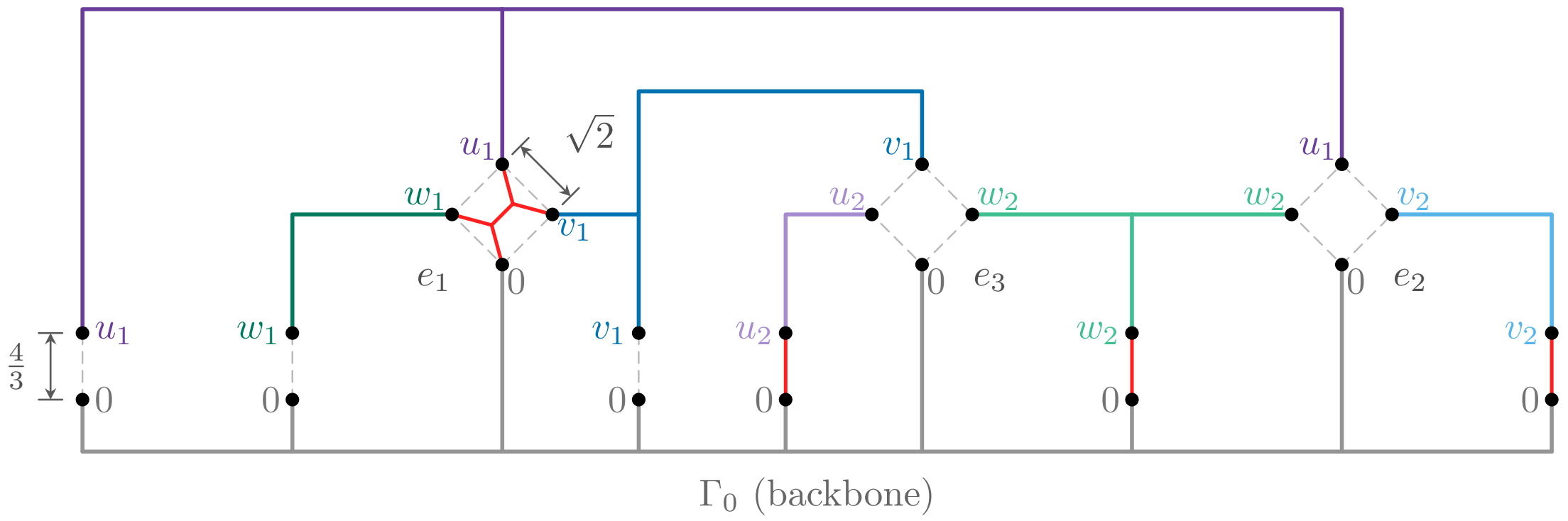
\begin{figure}[!h]
\centering
\definecolor{instUa}{HTML}{6A3D9A}
\definecolor{instUb}{HTML}{A68BD0}
\definecolor{instVa}{HTML}{0072B2}
\definecolor{instVb}{HTML}{56B4E9}
\definecolor{instWa}{HTML}{007A5E}
\definecolor{instWb}{HTML}{3FBF8F}
\begin{tikzpicture}[
  x=.85cm,y=.85cm,font=\small,line cap=round,line join=round,
  cable/.style={line width=1.05pt},
  backbone/.style={draw=black!42,line width=1.05pt},
  connection/.style={draw=connectionGreen,line width=.95pt},
  guide/.style={draw=black!28,densely dashed,line width=.45pt},
  dimension/.style={draw=black!65,line width=.5pt,{Stealth[length=3pt,width=3pt]}-{Stealth[length=3pt,width=3pt]}},
  dimension tick/.style={draw=black!65,line width=.5pt},
  port/.style={circle,fill=black,inner sep=0pt,minimum size=3.6pt},
  lab/.style={inner sep=1pt}
]
\def\h{.55}
\pgfmathsetmacro{\a}{\h*(1-1/sqrt(3))/2}
\def\yt{2.25}
\def\ys{1.35}
\def\yf{-1.3}
\def\yg{-2.03}
\def\yb{-2.6}

\draw[backbone] (-9.2,\yb)--(6.9,\yb);
\foreach \x in {-4.6,0,4.6} \draw[backbone] (\x,\yb)--(\x,-\h);
\foreach \x in {-9.2,-6.9,-3.105,-1.495,2.3,6.9} \draw[backbone] (\x,\yb)--(\x,\yg);
\node[below=3pt,text=black!60] at (-1.15,\yb) {$\Gamma_0$ (backbone)};

\draw[cable,instUa] (-4.6,\h)--(-4.6,\yt)--(4.6,\yt)--(4.6,\h) (-4.6,\yt)--(-9.2,\yt)--(-9.2,\yf);
\draw[cable,instVa] (-4.6+\h,0)--(-3.105,0)--(-3.105,\ys)--(0,\ys)--(0,\h) (-3.105,0)--(-3.105,\yf);
\draw[cable,instWa] (-4.6-\h,0)--(-6.9,0)--(-6.9,\yf);
\draw[cable,instUb] (-\h,0)--(-1.495,0)--(-1.495,\yf);
\draw[cable,instWb] (\h,0)--(4.6-\h,0) (2.3,0)--(2.3,\yf);
\draw[cable,instVb] (4.6+\h,0)--(6.9,0)--(6.9,\yf);

\foreach \cx in {-4.6,0,4.6}
  \draw[guide] (\cx,\h)--(\cx+\h,0)--(\cx,-\h)--(\cx-\h,0)--cycle;
\foreach \x in {-9.2,-6.9,-3.105,-1.495,2.3,6.9} \draw[guide] (\x,\yf)--(\x,\yg);

\draw[dimension] (-9.55,\yf)--node[left=2pt,text=black!70] {$\frac43$} (-9.55,\yg);
\foreach \y in {\yf,\yg}
  \draw[dimension tick] (-9.68,\y)--(-9.42,\y);
\coordinate (instDimA) at (-4.4,\h+.2);
\coordinate (instDimB) at (-4.6+\h+.2,.2);
\draw[dimension] (instDimA)--node[above right=1pt,text=black!70] {$\sqrt2$} (instDimB);
\foreach \p in {instDimA,instDimB}
  \draw[dimension tick] ($(\p)+(-.09,-.09)$)--($(\p)+(.09,.09)$);

\draw[connection] (-4.6,\h)--(-4.6+\a,\a)--(-4.6+\h,0) (-4.6+\a,\a)--(-4.6-\a,-\a)
  (-4.6-\h,0)--(-4.6-\a,-\a)--(-4.6,-\h);
\draw[connection] (-1.495,\yf)--(-1.495,\yg) (2.3,\yf)--(2.3,\yg) (6.9,\yf)--(6.9,\yg);

\foreach \cx/\tl/\ll/\rl/\tc/\lc/\rc in {
  -4.6/{u_1}/{w_1}/{v_1}/instUa/instWa/instVa,
   0/{v_1}/{u_2}/{w_2}/instVa/instUb/instWb,
   4.6/{u_1}/{w_2}/{v_2}/instUa/instWb/instVb} {
  \foreach \p in {(\cx,\h),(\cx-\h,0),(\cx+\h,0),(\cx,-\h)} \node[port] at \p {};
  \node[lab,above left,text=\tc] at (\cx,\h) {$\tl$};
  \node[lab,above left,text=\lc] at (\cx-\h,0) {$\ll$};
  \ifdim\cx pt=-4.6pt
    \node[lab,below right,text=\rc] at (\cx+\h,0) {$\rl$};
  \else
    \node[lab,above right,text=\rc] at (\cx+\h,0) {$\rl$};
  \fi
  \node[lab,below right,text=black!55] at (\cx,-\h) {$0$};
}
\node[text=black!70] at (-5.35,-.72) {$e_1$};
\node[text=black!70] at (.75,-.72) {$e_3$};
\node[text=black!70] at (5.35,-.72) {$e_2$};
\foreach \x/\name/\colour in {-9.2/{u_1}/instUa,-6.9/{w_1}/instWa,-3.105/{v_1}/instVa,-1.495/{u_2}/instUb,2.3/{w_2}/instWb,6.9/{v_2}/instVb} {
  \node[port] at (\x,\yf) {};
  \node[port] at (\x,\yg) {};
  \ifdim\x pt=-9.2pt
    \node[lab,right=2pt,text=\colour] at (\x,\yf) {$\name$};
    \node[lab,right=2pt,text=black!55] at (\x,\yg) {$0$};
  \else
    \node[lab,left=2pt,text=\colour] at (\x,\yf) {$\name$};
    \node[lab,left=2pt,text=black!55] at (\x,\yg) {$0$};
  \fi
}
\end{tikzpicture}
\caption{A cable instance with $r=2$ and the triples $e_1=\{u_1,v_1,w_1\}$ (left), $e_3=\{u_2,v_1,w_2\}$ (middle), and $e_2=\{u_1,v_2,w_2\}$ (right), so $\nu(H)=1$. Only the red connections contribute to the cost; the element cables and backbone are shown in grey. An optimal network uses the four-port Steiner tree at one square and three fallback segments, with total cost $(4-\Delta)+3\cdot\frac43=8-\Delta$.}
\label{fig:cable-instance}
\end{figure}

The construction, illustrated in \cref{fig:cable-instance}, has one cable for each element and an additional backbone cable. The cables remain disjoint and stay far apart except near gadgets, where short connections can join them.

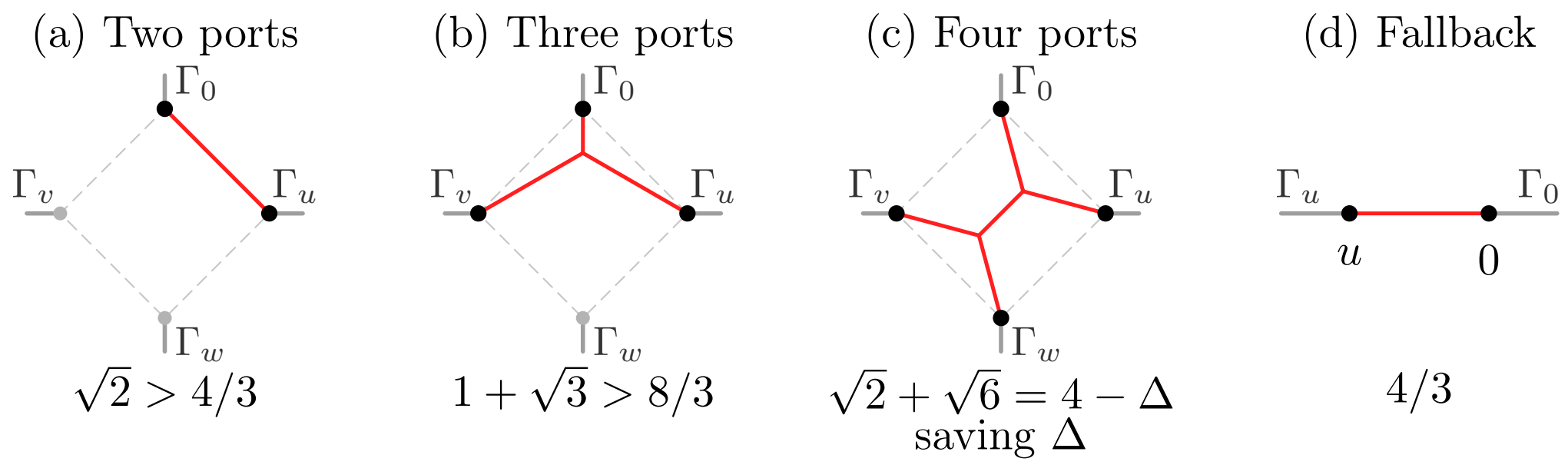
\begin{figure}[!h]
\centering
\begin{tikzpicture}[line cap=round,line join=round,
  x=0.9cm,y=0.9cm,font=\normalsize,
  cable/.style={draw=black!38,line width=1.05pt,line cap=round,line join=round},
  connection/.style={draw=connectionGreen,line width=.95pt,line cap=round,line join=round},
  guide/.style={draw=black!22,densely dashed,line width=0.4pt},
  steiner/.style={draw=none,fill=none,inner sep=0pt,minimum size=0pt},
  cablab/.style={font=\small,text=black!80,inner sep=1pt}
]
\foreach \shift/\heading in {0/{(a) Two ports},4/{(b) Three ports},8/{(c) Four ports}} {
  \begin{scope}[xshift=\shift*0.9cm]
    \node at (0,1.7) {\heading};
    \draw[guide] (1,0)--(0,1)--(-1,0)--(0,-1)--cycle;
    \foreach \x/\y in {1/0,0/1,-1/0,0/-1} {
      \draw[cable] (\x,\y)--(1.32*\x,1.32*\y);
      \fill[black!30] (\x,\y) circle[radius=1.7pt];
    }
    \node[cablab,anchor=west] at (.05,1.25) {$\Gamma_0$};
    \node[cablab,anchor=west] at (.05,-1.25) {$\Gamma_w$};
    \node[cablab,anchor=south] at (-1.25,.05) {$\Gamma_v$};
    \node[cablab,anchor=south] at (1.25,.05) {$\Gamma_u$};
  \end{scope}
}

\draw[connection] (0,1)--(1,0);
\fill (0,1) circle[radius=2pt];
\fill (1,0) circle[radius=2pt];
\node at (0,-1.7) {$\sqrt2>4/3$};

\begin{scope}[xshift=3.6cm]
  \coordinate (S) at (0,{1/sqrt(3)});
  \draw[connection] (-1,0)--(S)--(1,0) (S)--(0,1);
  \foreach \p in {(-1,0),(1,0),(0,1)} \fill \p circle[radius=2pt];
  \node[steiner] at (S) {};
  \node at (0,-1.7) {$1+\sqrt3>8/3$};
\end{scope}

\begin{scope}[xshift=7.2cm]
  \pgfmathsetmacro{\a}{(1-1/sqrt(3))/2}
  \coordinate (S) at (\a,\a);
  \coordinate (T) at (-\a,-\a);
  \draw[connection] (0,1)--(S)--(1,0) (S)--(T) (-1,0)--(T)--(0,-1);
  \foreach \p in {(-1,0),(1,0),(0,1),(0,-1)} \fill \p circle[radius=2pt];
  \node[steiner] at (S) {};
  \node[steiner] at (T) {};
  \node at (0,-1.7) {$\sqrt2+\sqrt6=4-\Delta$};
  \node at (0,-2.15) {saving $\Delta$};
\end{scope}

\begin{scope}[xshift=10.8cm]
  \node at (0,1.7) {(d) Fallback};
  \draw[cable] (-1.32,0)--(-2/3,0) (2/3,0)--(1.32,0);
  \draw[connection] (-2/3,0)--(2/3,0);
  \fill (-2/3,0) circle[radius=2pt];
  \fill (2/3,0) circle[radius=2pt];
  \node[below=4pt] at (-2/3,0) {$u$};
  \node[below=4pt] at (2/3,0) {$0$};
  \node[cablab,anchor=south] at (-1.15,.05) {$\Gamma_u$};
  \node[cablab,anchor=south] at (1.15,.05) {$\Gamma_0$};
  \node at (0,-1.7) {$4/3$};
\end{scope}
\end{tikzpicture}
\caption{Connection costs at the gadgets, with cables in grey and added connections in red. A square for a triple $\{u,v,w\}$ can join $\Gamma_0,\Gamma_u,\Gamma_v,\Gamma_w$, while the fallback for $u$ joins $\Gamma_u$ to the backbone $\Gamma_0$. Only the four-port Steiner tree improves on the baseline of $4/3$ per connection, saving exactly $\Delta$.}
\label{fig:local-gadgets}
\end{figure}

We use two types of gadgets, shown in \cref{fig:local-gadgets}. We first describe the gadgets and specify which cables attach to each one, showing how they encode a matching; the geometric layout will be described later. For each triple $\{u,v,w\}\in E$, draw a \emph{square gadget} of side $\sqrt2$ with corners labelled $0,u,v,w$. For each element $u\in U$, draw a \emph{fallback gadget} consisting of two points at distance $4/3$, labelled $0$ and $u$. We call these labelled points \emph{ports}. For each label $i\in U\cup\{0\}$, a single cable $\Gamma_i$ connects all ports labelled $i$ outside the gadgets. We call $\Gamma_0$ the \emph{backbone}. Thus $\Gamma_u$ contains the corner labelled $u$ of each square gadget for a triple containing $u$ and the point labelled $u$ of its fallback gadget, while the backbone contains the port labelled $0$ of every gadget.

For the forward implication, let $M$ be a matching of size $k$. For each triple in $M$, join its three element cables to the backbone using the four-port Steiner tree in its square (\cref{fig:local-gadgets}(c)), of length $\sqrt2+\sqrt6=4-\Delta$. Since the triples in $M$ are disjoint, this connects $3k$ element cables; join each of the remaining $|U|-3k$ cables to the backbone by its fallback segment, of length $\frac43$. The resulting network is connected and has total cost $(4-\Delta)k+\frac43(|U|-3k)=\frac43|U|-\Delta k$.

For the converse, a network connecting the cables may attach anywhere along them. For the analysis, we therefore require different cables to stay far apart except near shared gadgets, making connections elsewhere expensive. For each gadget $s$, let $z_s$ be its centre and $B_s$ the closed ball of radius $20$ around $z_s$.

\begin{samepage}
\begin{restatable}{definition}{SeparatedCables}
\label{def:separated}
The cables are \emph{separated} if the following conditions hold.
\begin{enumerate}[(S1)]
\item \emph{Between cables.} The balls $B_s$ are pairwise more than $20$ apart. Inside each ball $B_s$, the cables consist exactly of the straight segments extending from the ports of $s$ directly away from $z_s$ to the boundary of $B_s$. Any two points on different cables less than $20$ apart lie in a common ball.
\item \emph{Within a cable.} Every cable segment has length at least $1$. Non-incident segments of the same cable are at least $1$ apart, and incident segments meet at an angle of at least $45^\circ$.
\end{enumerate}
\end{restatable}
\end{samepage}

\begin{figure}[!h]
\centering
\scalebox{.85}{\definecolor{sepCableW}{HTML}{009E73}
\begin{tikzpicture}[line cap=round,line join=round,x=1cm,y=1cm,font=\small,>=latex,
  cable/.style={draw=black,line width=1.05pt,line cap=round,line join=round},
  ball/.style={fill=black!4,draw=black!45,dotted,line width=.8pt},
  gad/.style={fill=black!10,draw=black!70,line width=.5pt,line join=round},
  port/.style={circle,fill=black,inner sep=0pt,minimum size=3.6pt},
  ctr/.style={circle,fill=black!75,inner sep=0pt,minimum size=2.4pt},
  dim/.style={draw=black!60,line width=.45pt,<->},
  note/.style={font=\footnotesize}]
\def\R{1.25}
\def\F{5.2}
\filldraw[ball] (0,0) circle[radius=\R];
\filldraw[ball] (\F,0) circle[radius=\R];
\draw[cable,draw=cableU] (.45,0)--(\F-.38,0);
\draw[cable,draw=black!45] (0,-.45)--(0,-1.6)--(\F+2.1,-1.6)--(\F+2.1,0)--(\F+.38,0);
\draw[cable,draw=cableV] (0,.45)--(0,1.55);
\draw[cable,draw=cableV,line cap=butt,dash pattern=on 2pt off 2.5pt] (0,1.55)--(0,2.05);
\draw[cable,draw=sepCableW] (-.45,0)--(-1.6,0);
\draw[cable,draw=sepCableW,line cap=butt,dash pattern=on 2pt off 2.5pt] (-1.6,0)--(-2.1,0);
\foreach \p in {(.45,0),(0,.45),(-.45,0),(0,-.45)} \node[port] at \p {};
\node[port] at (\F-.38,0) {}; \node[port] at (\F+.38,0) {};
\node[ctr] at (0,0) {};
\node[note,anchor=north west] at (.12,-.5) {$z_s$};
\node[note,anchor=north] at (\F,-.18) {$z_{s'}$};
\node[note,text=black!60,anchor=north east] at (225:\R) {$B_s$};
\node[note,text=black!60,anchor=north west] at ($(\F,0)+(-45:\R)$) {$B_{s'}$};
\node[text=cableU,anchor=south] at (\F/2,.05) {$\Gamma_u$};
\node[text=cableV,anchor=west] at (.1,1.75) {$\Gamma_v$};
\node[text=sepCableW,anchor=south] at (-1.7,.08) {$\Gamma_w$};
\node[text=black!60,anchor=north] at (\F/2+.9,-1.65) {$\Gamma_0$};
\draw[dim] (0,0)--node[note,above left=-2pt] {$20$} (135:\R);
\draw[dim] ({sqrt(\R*\R-.5*.5)},.5)--node[note,above] {$>20$} ({\F-sqrt(\R*\R-.5*.5)},.5);
\draw[dim] (\F/2,-.1)--node[note,right] {$\ge20$} (\F/2,-1.5);
\node[anchor=west] at (-2.1,2.45) {(a) Between cables (S1)};
\begin{scope}[shift={(9.3,0)}]
  \node[anchor=west] at (-.5,2.45) {(b) Within a cable (S2)};
  \coordinate (P0) at (0,-1.4); \coordinate (P1) at (0,1);
  \coordinate (P2) at (1.4,-.4); \coordinate (P3) at (1.4,-1.4);
  \draw[cable,draw=black] (P0)--(P1)--(P2)--(P3);
  \draw[cable,draw=black,line cap=butt,dash pattern=on 2pt off 2.5pt] (P0)--(0,-1.8) (P3)--(1.4,-1.8);
  \foreach \p in {P1,P2} \fill[black] (\p) circle[radius=1.5pt];
  \draw[black!60,line width=.45pt] ($(P1)+(270:.5)$) arc[start angle=270,end angle=315,radius=.5];
  \node[note,anchor=east] at ($(P1)+(-.08,-.45)$) {$\ge45^\circ$};
  \draw[dim] ($(P1)+(.3,.3)$)--node[note,above right=-2pt] {$\ge1$} ($(P2)+(.3,.3)$);
  \draw[dim] (0,-1.1)--node[note,below] {$\ge1$} (1.4,-1.1);
\end{scope}
\end{tikzpicture}}
\small
\caption{Separated cables.}
\label{fig:separation}
\end{figure}
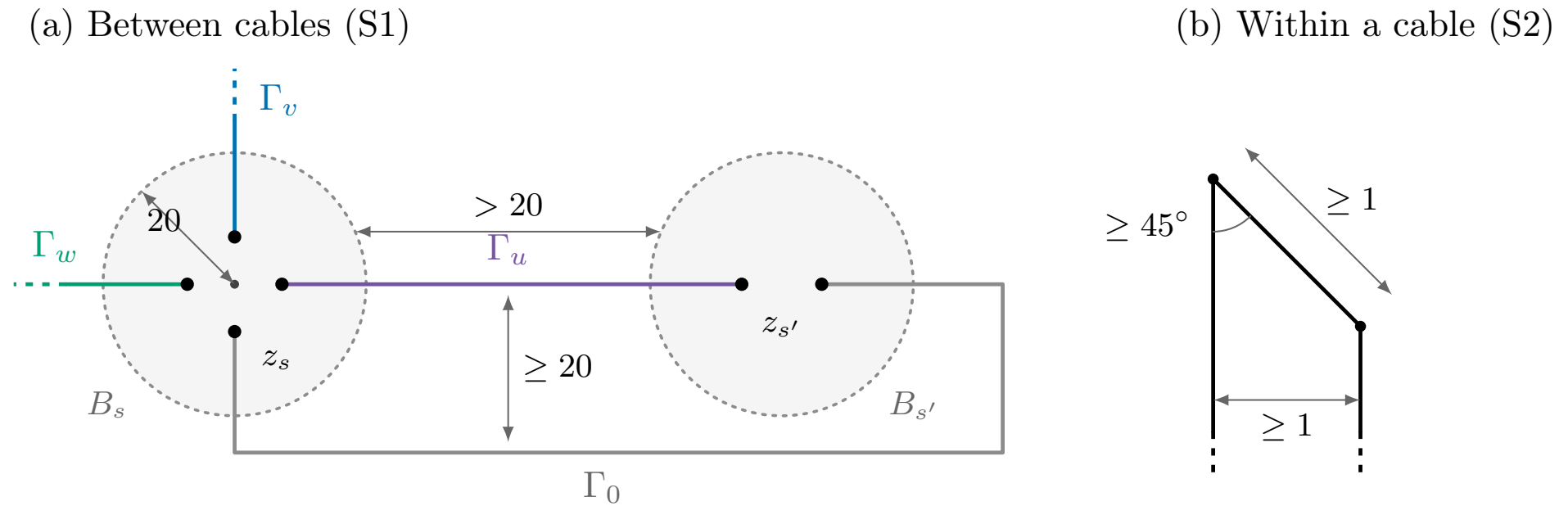

\Cref{fig:separation} illustrates these conditions.

We show that there is an optimal network that is \emph{clean}: it consists precisely of the cables, fallback segments, and Steiner trees joining all four corners of selected squares (\cref{fig:local-gadgets}(c) and~(d)). Removing a part of the network that joins $r$ cables leaves at most $r$ components. One contains the backbone, and each other component contains an element cable that can be joined to the backbone by its fallback segment, so at most $r-1$ fallback segments of total length at most $\frac43(r-1)$ restore connectivity. If the cables share no gadget, separation forces this part of the network to bridge gaps of at least $20$, so replacing it by fallback segments is cheaper.

We may therefore assume that each connection joins cables at points inside the ball $B_s$ of a single gadget $s$; \cref{sec:cable-reduction-proof} gives the formal argument. It is easy to modify them without increasing their length so that they attach only at its ports, as in \cref{fig:local-gadgets}. Of these connections, only the four-port tree at a square beats the fallback cost $\frac43(r-1)$ of joining $r$ cables, saving $\Delta$.

Two such trees cannot use squares sharing an element $u$, since both would join $\Gamma_u$ to the backbone and create a cycle. Their triples therefore form a matching, saving at most $\Delta\nu(H)$ from the fallback cost $\frac43|U|$. This gives a minimum connection cost of at least $\frac43|U|-\Delta\nu(H)$.

With enough dimensions, placing the gadgets and joining equally labelled ports into separated cables is relatively straightforward. We also keep the cables short, which will be useful later when replacing them by point terminals. The following construction achieves $L=O(m)$ in $O(m)$ dimensions, preserving a constant approximation gap.

\paragraph{High-dimensional routing.}
We work in $\BR^{m+|U|+3}$, assigning each gadget $s$ its own unit coordinate vector $e_s$ and each cable $\Gamma_i$ its own unit coordinate vector $f_i$. We draw each gadget in a plane parallel to the $b_1b_2$-plane, where $b_1,b_2$ are the two remaining unit coordinate vectors. For a sufficiently large constant $R$, centre gadget $s$ at $Re_s$, with ports $Re_s\pm b_1$ and $Re_s\pm b_2$ for a square, or $Re_s\pm\frac23b_1$ for a fallback. The centres of any two distinct gadgets are then at distance $R\sqrt2$.

\begin{figure}[!h]
\centering
\begin{tikzpicture}[line cap=round,line join=round,
  x=0.85cm,y=0.85cm,font=\normalsize,>=latex,
  axis/.style={->,densely dashed,draw=black!32,line width=.6pt},
  localaxis/.style={->,draw=black!55,line width=.6pt},
  direction/.style={fill=white,inner sep=1.5pt},
  cable/.style={draw=cableU,line width=1.05pt,line cap=round,line join=round}
]
\coordinate (Origin) at (1.75,-2.55);
\coordinate (H) at (0,2.35);
\coordinate (C1) at (-5.8,0);
\coordinate (C2) at (5.8,0);
\coordinate (C3) at (0,-.75);
\coordinate (U3) at ($(C3)+(-.55,0)$);
\coordinate (B3) at ($(C3)+(.55,0)$);
\coordinate (Q3) at ($(C3)+(-2.65,0)$);

\foreach \s in {1,2} {
  \coordinate (U\s) at ($(C\s)+(0,.6)$);
  \coordinate (Q\s) at ($(C\s)+(0,1.6)$);
  \filldraw[fill=black!3,draw=black!45,line width=.65pt]
    ($(C\s)+(1,0)$)--(U\s)--($(C\s)+(-1,0)$)
    --($(C\s)+(0,-.6)$)--cycle;
}

\foreach \s in {1,2,3} \draw[axis,shorten >=2pt] (Origin)--(C\s);
\draw[axis,shorten >=3pt] (Origin)--(H);
\node[direction,below left=1pt] at ($(Origin)!.72!(C1)$) {$e_1$};
\node[direction,left=2pt] at ($(Origin)!.4!(C2)$) {$e_2$};
\node[direction,right=2pt] at ($(Origin)!.5!(C3)$) {$e_3$};
\node[direction,right=2pt] at ($(Origin)!.66!(H)$) {$f_u$};
\fill[black!65] (Origin) circle[radius=2pt];
\node[below=5pt] at (Origin) {origin $\mathbf{0}$};
\foreach \s in {1,2} {
  \foreach \dx/\dy in {1/0,-1/0,0/-.6}
    \fill[black!55] ($(C\s)+(\dx,\dy)$) circle[radius=1.6pt];
  \draw[localaxis] (C\s)--($(C\s)+(1.4,0)$)
    node[above=2pt] {$b_1$};
  \draw[localaxis] (C\s)--($(C\s)+(0,-1.02)$)
    node[right=2pt] {$b_2$};
  \draw[black!45,line width=.45pt]
    (C\s)--($(C\s)+(-.52,-1.3)$)--($(C\s)+(-.8,-1.3)$)
    node[left=3pt,text=black] {$Re_{\s}$};
  \filldraw[fill=black!3,draw=black!45,line width=.6pt]
    (C\s) circle[radius=1.7pt];
}

\draw[draw=black!40,densely dashed,line width=.55pt] (U3)--(B3);
\fill[black!60] (B3) circle[radius=2pt];
\node[above=3pt] at (B3) {$0$};
\node[above=3pt,text=cableU] at (U3) {$u$};
\node[above=3pt,fill=white,inner sep=1pt] at (C3) {$\frac43$};
\draw[localaxis] (C3)--($(C3)+(1.25,0)$)
  node[right=2pt] {$b_1$};
\fill[black!60] (B3) circle[radius=2pt];
\draw[black!45,line width=.45pt]
  (C3)--($(C3)+(-.52,-1.1)$)--($(C3)+(-.8,-1.1)$)
  node[left=3pt,text=black] {$Re_3$};
\filldraw[fill=black!3,draw=black!45,line width=.6pt]
  (C3) circle[radius=1.7pt];

\foreach \s in {1,2,3} {
  \draw[cable] (U\s)--(Q\s)--(H);
  \fill[cableU] (U\s) circle[radius=2.3pt];
  \fill[cableU] (Q\s) circle[radius=1.6pt];
}
\node[left=4pt,text=cableU] at (U1) {$u$};
\node[left=3pt,text=cableU,font=\footnotesize] at ($(U1)!.5!(Q1)$) {gadget segment};
\node[above=1pt,sloped,text=cableU,font=\footnotesize] at ($(Q1)!.5!(H)$) {hub segment};
\node[right=4pt,text=cableU] at (U2) {$u$};
\node[above right=2pt,text=cableU] at ($(Q2)!.5!(H)$) {cable $\Gamma_u$};
\filldraw[fill=cableU,draw=white,line width=.5pt] (H) circle[radius=2.7pt];
\node[above=5pt,text=cableU] at (H) {hub $Rf_u$};
\end{tikzpicture}
\caption{The cable $\Gamma_u$ joins its ports at two squares and a fallback to the hub $Rf_u$, using a gadget segment followed by a hub segment for each port. Gadgets are centred at $Re_s$ in planes parallel to the $b_1b_2$-plane, and dashed arrows mark the orthogonal directions $e_s$ and $f_u$.}
\label{fig:high-dimensional-routing}
\end{figure}
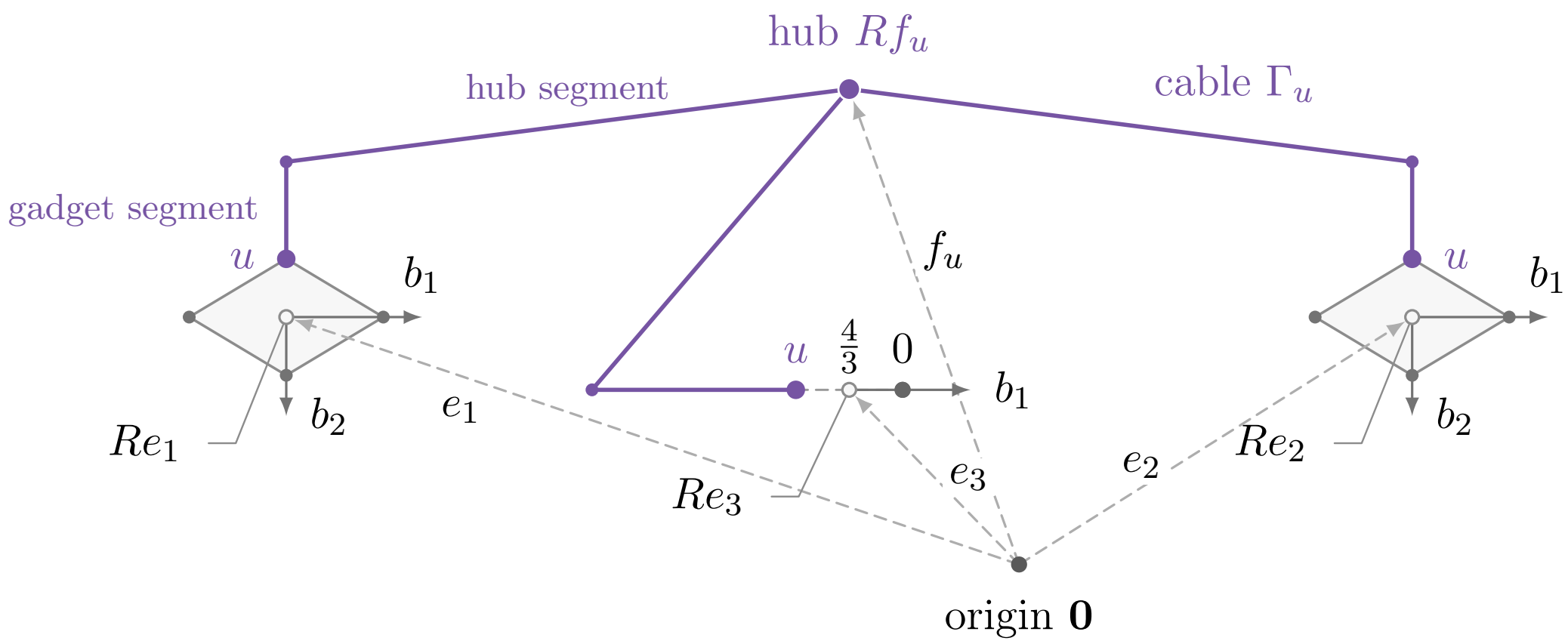

The \emph{hub} of $\Gamma_i$ is $Rf_i$, and each port labelled $i$ is joined to it by two segments. The \emph{gadget segment} extends straight away from the gadget's centre, within its plane, until it reaches distance $R/2$ from the centre; the \emph{hub segment} then joins this point to $Rf_i$. Gadgets and cables in high dimensions are hard to draw, but we try our best in \cref{fig:high-dimensional-routing}.

For sufficiently large $R$, the separate coordinate directions $e_s$ and $f_i$ ensure that different cables come close only near a shared gadget, giving separation. There are $O(m)$ segments, each of constant length, so the total cable length is $L=O(m)$.

\paragraph{From cables to point terminals.}
We now replace the cables by closely spaced point terminals. Connecting the points along each cable adds about $L$ to the cost of joining the cables. Since the routing above has $L=O(m)=O(r)$, the total cost remains $O(r)$, so the additive gap $\Delta\delta r$ still gives a constant approximation gap.

For a small parameter $\rho>0$, sample each cable segment at spacing $\Theta(\rho^2)$, including its endpoints, and let $P_\rho$ be the resulting set of terminals. The edges between consecutive samples have total length $L$.

A short tree spanning $P_\rho$ must essentially follow the cables, paying their total length $L$ in addition to the cost of connecting them. The bounds for the cable instance therefore carry over up to a small error.

\begin{restatable}[Cable reduction]{theorem}{CableReduction}
\label{thm:cable-reduction}
There is an absolute constant $C_2$ such that, for separated cables and every $0<\rho\le1/100$,
\begin{equation}
\label{eq:cable-reduction}
    L+\frac43|U|-\Delta\nu(H)-C_2\rho L
    \;\le\;\OPT(P_\rho)\;\le\;
    L+\frac43|U|-\Delta\nu(H).
\end{equation}
\end{restatable}

For the upper bound, take the union of the four-port trees at the squares of a maximum matching, the fallback segments for uncovered elements, and all edges between consecutive samples. This gives a connected network spanning $P_\rho$ of length $L+\frac43|U|-\Delta\nu(H)$.

For the lower bound, let $T$ be an optimal Steiner tree on $P_\rho$. If $T$ contains all edges between consecutive samples, they contribute length $L$ and the remaining edges connect the cables, so the cable bound applies without an error term. However, $T$ need not contain these edges, but a short tree must follow the cables closely. The following lemma recovers a tree along each cable at an additional cost of $O(\rho L)$, allowing us to apply the cable bound up to this error.

\begin{restatable}[Cable recovery]{lemma}{CableRecovery}
\label{lem:cable-recovery}
Let $T$ be a Steiner tree on $P_\rho$ of length $O(L)$. By adding segments of total length $O(\rho L)$ to $T$, we obtain a network containing pairwise disjoint trees $T_i$, one for each cable $\Gamma_i$. Each $T_i$ spans the sampled terminals on $\Gamma_i$ and lies within distance $\rho$ of that cable; together, these trees have total length at least $L$.
\end{restatable}

\paragraph{The approximation gap.}
Taking all edges between consecutive samples and all fallback segments shows that $T$ has length $O(L)$, since every gadget contributes constant cable length and hence $|U|=O(L)$; so \cref{lem:cable-recovery} applies.

Set $\tau=L+\frac43|U|-\Delta r$. Since both $L$ and $\tau$ are $O(m)=O(r)$, choose constants $0<\rho\le\frac1{100}$ and $\beta>0$ such that $C_2\rho L\le\frac{\Delta\delta}{2}r$ and $\beta\tau\le\frac{\Delta\delta}{2}r$. If $\nu(H)=r$, the upper bound in \cref{thm:cable-reduction} gives $\OPT(P_\rho)\le\tau$. If $\nu(H)\le(1-\delta)r$, then
\begin{align*}
    \OPT(P_\rho)
    &\ge L+\frac43|U|-\Delta\nu(H)-C_2\rho L
        &&\text{by the lower bound in \cref{thm:cable-reduction},}\\
    &\ge\tau+\Delta\delta r-C_2\rho L
        &&\text{since }\nu(H)\le(1-\delta)r,\\
    &\ge\tau+\frac{\Delta\delta}{2}r
        &&\text{by the choice of }\rho,\\
    &\ge(1+\beta)\tau
        &&\text{by the choice of }\beta.
\end{align*}
A $(1+\frac\beta2)$-approximation would therefore distinguish the source cases. Since $\rho$ is constant, there are $n=\Theta(m)$ terminals in $O(m)=O(n)$ dimensions, which proves APX-hardness in linear dimension.

\FloatBarrier

\subsection{Reducing the dimension}
\label{sec:overview-dimension-reduction}

{We reduce the dimension using the Johnson--Lindenstrauss lemma~\cite{JohnsonL84}. For any fixed $\eta>0$, the deterministic construction of Engebretsen, Indyk, and O'Donnell~\cite{DBLP:conf/soda/EngebretsenIO02} maps the terminals $P=\{p_1,\ldots,p_n\}$ in $O(n)$ dimensions to $Q=\{q_1,\ldots,q_n\}$ in $O(\log n)$ dimensions. After rescaling, they satisfy
\[
    \|p_i-p_j\|\le\|q_i-q_j\|\le(1+\eta)\|p_i-p_j\|
    \qquad\text{for all }i,j.
\]

This maps the terminals, but to compare the Steiner tree optima on $P$ and $Q$, we must also map the Steiner points between the two spaces. Following Karthik C.~S.~\cite[Section~4.2]{DBLP:journals/corr/abs-2602-03837}, we use Kirszbraun's extension theorem~\cite{Kirszbraun34} to extend the terminal maps to all points of the two spaces, including the Steiner points. A tree on $Q$ then gives a network connecting $P$ of no greater length, while a tree on $P$ gives a network connecting $Q$ of length at most $1+\eta$ times the original length. Applying this to optimal trees gives
\[
    \OPT(P)\le\OPT(Q)\le(1+\eta)\OPT(P).
\]
Choosing $\eta>0$ to be a sufficiently small constant therefore preserves the constant approximation gap in $O(\log n)$ dimensions.}

\paragraph{No PTAS in dimension $O(\log\log n)$.}
Our reduction maps a 3SAT formula with $s$ variables to $n=s^{O(1)}$ terminals in $O(\log n)$ dimensions with a constant approximation gap. Following Trevisan~\cite{DBLP:journals/siamcomp/Trevisan00} for TSP, we increase the number of terminals by adding copies of an existing terminal, leaving the optimum unchanged. With $N=2^{n^\gamma}$ terminals for a small constant $\gamma>0$, the dimension is $O(\log n)=O(\log\log N)$, while a PTAS would decide the formula in time $N^{O(1)}=2^{s^{O(\gamma)}}$. Since $\gamma$ can be chosen arbitrarily small, a PTAS would imply $\mathrm{NP}\subseteq\mathrm{SUBEXP}$, so the lower bound follows from $\mathrm{NP}\not\subseteq\mathrm{SUBEXP}$, a weaker assumption than ETH.

\subsection{Lower bounds in fixed dimension}
\label{sec:overview-fixed-dimension}

For the fixed-dimensional lower bounds, we retain separation but allow longer cables. For fixed $d\ge3$, we must allow the total cable length $L$ to grow faster than the matching instance size $m$. Only the cable routing changes, so we can use the same gadgets and matching argument: YES and NO instances differ by $\Delta\delta r$ in connection cost, and by at least $\Delta\delta r/2$ after dense sampling. Every tree on the samples also pays for the cables, so both optima are of order $L$ and their relative gap is $\Theta(m/L)$, which is subconstant.

This smaller gap suffices if we can route the cables with total length
\[
    L=O_d(m^{d/(d-1)}).
\]
The relative gap is then $\Omega_d(m^{-1/(d-1)})$, so choosing $\varepsilon=c_d m^{-1/(d-1)}$ for a sufficiently small constant $c_d>0$ distinguishes the YES and NO instances. This gives $\varepsilon^{-(d-1)}=\Theta_d(m)$. For any fixed $0<\gamma<1$, an approximation scheme running in time $2^{O(\varepsilon^{-\gamma(d-1)})}\poly(n)$ would therefore solve the source matching gap in time $2^{O(m^\gamma)}$, provided that sampling produces polynomially many terminals. By \cref{lem:gap-source}, this would contradict weak ETH. It remains to construct a separated routing of this total length.

\paragraph{Cube wiring.}
To route the cables without crossings, we use a consequence of the Cube Wiring Theorem of de Berg, Bodlaender, Kisfaludi-Bak, Marx, and van der Zanden~\cite{deBergBBMZ20} (see~\cite[Theorem~31]{deBergBBMZ18arxiv}). For fixed $d\ge3$, every graph with $n$ vertices and maximum degree at most $2d$, the degree of the integer grid in $\BR^d$, can be drawn in polynomial time in the cube $\{0,\ldots,S\}^d\subseteq\BR^d$ with $S=O(n^{1/(d-1)})$. Vertices map to distinct grid points, and edges follow grid paths that meet only at common endpoints. These paths are edge-disjoint, so their total length is bounded by the number of grid edges in the cube, which is $O_d(n^{d/(d-1)})$.

\paragraph{From the grid drawing to cables.}
For each label $i$, construct a tree $T_i$ whose leaves are exactly the gadgets with a port labelled $i$ and whose internal vertices have degree three. The union of these trees has $O(m)$ vertices and maximum degree four, so the cube-wiring theorem gives a grid drawing in $\BR^d$. We turn the drawing of each $T_i$ into a cable $\Gamma_i$ by modifying the paths near each gadget to attach to its ports. The resulting cables are separated and have total length $L=O_d(m^{d/(d-1)})$.

\paragraph{From cables to point terminals.}
By \cref{thm:cable-reduction}, the optimum on the sampled terminals equals $L+\frac43|U|-\Delta\nu(H)$ up to an error of $O(\rho L)$. We choose $\rho=\Theta(m/L)$ so that this error is at most half the matching gap $\Delta\delta r$, and set $\tau=L+(4-\Delta)r=O_d(m^{d/(d-1)})$. This gives
\[
\begin{aligned}
    \text{YES: }\nu(H)=r
        &\quad\Longrightarrow\quad \OPT\le\tau,\\
    \text{NO: }\nu(H)\le(1-\delta)r
        &\quad\Longrightarrow\quad \OPT\ge\tau+\frac{\Delta\delta}{2}r=\Bigl(1+\frac{\Delta\delta r}{2\tau}\Bigr)\tau.
\end{aligned}
\]
Since $r=\Theta(m)$, the relative gap $\Delta\delta r/(2\tau)$ is $\Omega_d(m^{1/(1-d)})$, so a $(1+\varepsilon)$-approximation with $\varepsilon=c_d\,m^{1/(1-d)}$ distinguishes the two cases for a sufficiently small constant $c_d>0$. Sampling produces $n=O(L^3/m^2)$ terminals, which is polynomial in $m$. Together with $\varepsilon^{-(d-1)}=\Theta_d(m)$, this gives the weak-ETH lower bound above.

Under Gap-ETH, the same reduction would have given a tight lower bound, excluding approximation schemes running in time $2^{o(\varepsilon^{-(d-1)})}\poly(n)$. The weaker assumption $\mathrm{NP}\not\subseteq\mathrm{SUBEXP}$ still rules out time $2^{o(\varepsilon^{-c(d-1)})}\poly(n)$ for some constant $c>0$, by \cref{lem:gap-source}.

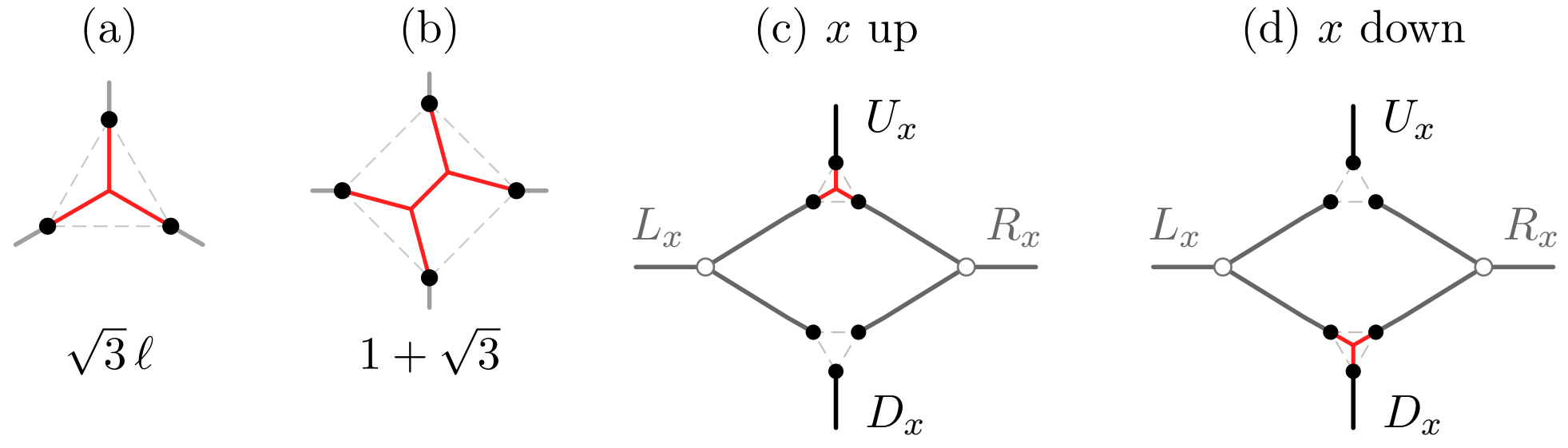
\begin{figure}[!h]
\centering
\begin{tikzpicture}[line cap=round,line join=round,x=1.02cm,y=1.02cm,font=\normalsize,
  row/.style={draw=black!38,line width=1.05pt,line cap=round,line join=round},
  connection/.style={draw=connectionGreen,line width=.95pt,line cap=round,line join=round},
  guide/.style={draw=black!22,densely dashed,line width=0.4pt},
  active/.style={circle,fill=black!30,inner sep=0pt,minimum size=3.4pt},
  steiner/.style={draw=none,fill=none,inner sep=0pt,minimum size=0pt},
  caption/.style={align=center}]
\node at (0,1.3) {(a)};
\draw[guide] (0,0.5774)--(-0.5,-0.2887)--(0.5,-0.2887)--cycle;
\draw[row] (0,0.5774)--++(0,0.3);
\draw[row] (-0.5,-0.2887)--++(-0.26,-0.15);
\draw[row] (0.5,-0.2887)--++(0.26,-0.15);
\draw[connection] (0,0)--(0,0.5774) (0,0)--(-0.5,-0.2887) (0,0)--(0.5,-0.2887);
\foreach \p in {(0,0.5774),(-0.5,-0.2887),(0.5,-0.2887)} \fill \p circle[radius=2pt];
\node[caption] at (0,-1.3) {$\sqrt3\,\ell$};
\begin{scope}[xshift=2.6*1.02cm]
  \node at (0,1.3) {(b)};
  \draw[guide] (0,0.7071)--(0.7071,0)--(0,-0.7071)--(-0.7071,0)--cycle;
  \foreach \x/\y in {0/1,1/0,0/-1,-1/0}
    \draw[row] (0.7071*\x,0.7071*\y)--(0.95*\x,0.95*\y);
  \draw[connection] (0,0.7071)--(0.1494,0.1494)--(0.7071,0)
    (0.1494,0.1494)--(-0.1494,-0.1494)
    (-0.7071,0)--(-0.1494,-0.1494)--(0,-0.7071);
  \foreach \p in {(-0.7071,0),(0,0.7071),(0.7071,0),(0,-0.7071)} \fill \p circle[radius=2pt];
  \node[caption] at (0,-1.3) {$1+\sqrt3$};
\end{scope}
\foreach \sx/\heading/\up in {5.9/{(c) $x$ up}/1,10.1/{(d) $x$ down}/0} {
  \node at (\sx,1.3) {\heading};
  \begin{scope}[shift={(\sx,-0.62)},x=0.72cm,y=0.72cm,
    xrow/.style={draw=black,line width=1.05pt,line cap=round,line join=round},
    xguide/.style={draw=black!25,densely dashed,line width=0.4pt},
    xactive/.style={circle,fill=black,inner sep=0pt,minimum size=3.6pt},
    xjunction/.style={circle,draw=black!55,fill=white,inner sep=0pt,minimum size=4pt,line width=0.5pt}]
    \draw[xrow] (0,1.2)--(0,1.85);
    \draw[xrow] (0,-1.2)--(0,-1.85);
    \draw[xrow,draw=black!60] (-0.26,0.75)--(-0.56,0.58)--(-1.5,0)--(-0.56,-0.58)--(-0.26,-0.75)
      (-1.5,0)--(-2.3,0);
    \draw[xrow,draw=black!60] (0.26,0.75)--(0.56,0.58)--(1.5,0)--(0.56,-0.58)--(0.26,-0.75)
      (1.5,0)--(2.3,0);
    \node[xjunction] at (-1.5,0) {};
    \node[xjunction] at (1.5,0) {};
    \draw[xguide] (0,1.2)--(-0.26,0.75)--(0.26,0.75)--cycle;
    \draw[xguide] (0,-1.2)--(-0.26,-0.75)--(0.26,-0.75)--cycle;
    \ifnum\up=1
      \draw[connection] (0,0.9)--(0,1.2) (0,0.9)--(-0.26,0.75) (0,0.9)--(0.26,0.75);
    \else
      \draw[connection] (0,-0.9)--(0,-1.2) (0,-0.9)--(-0.26,-0.75) (0,-0.9)--(0.26,-0.75);
    \fi
    \foreach \px/\py in {0/1.2,-0.26/0.75,0.26/0.75,0/-1.2,-0.26/-0.75,0.26/-0.75}
      \node[xactive] at (\px,\py) {};
    \node[anchor=west] at (0.15,1.7) {$U_x$};
    \node[anchor=west] at (0.15,-1.7) {$D_x$};
    \node[text=black!60,anchor=south] at (-2.05,0.05) {$L_x$};
    \node[text=black!60,anchor=south] at (2.05,0.05) {$R_x$};
  \end{scope}
}
\end{tikzpicture}
\caption{{Gadgets in the plane, with cables in grey and connections in red. (a, b) Shortest networks joining all corners of an equilateral triangle of side $\ell$ and a unit square. (c, d) At an up crossover $x$, the upper triangle joins $U_x$, $L_x$, and $R_x$; at a down crossover, the lower triangle joins $D_x$, $L_x$, and $R_x$.}}
\label{fig:overview-plane-costs}
\label{fig:overview-plane-crossover}
\end{figure}

{
\subsection{Lower bounds in the plane}
\label{sec:overview-plane}

To handle crossings in the plane, we build on the crossover layout of Garey, Graham, and Johnson~\cite{GareyGJ77}. Our construction is essentially theirs, but the analysis uses our cable framework, with terminal sampling and cable recovery as before. We show that the optima of YES and NO instances differ by $\Omega(r)$; since the total cable length is $O(r^2)$, this yields the relative gap $\Omega(1/r)$ required for $d=2$.

\paragraph{Gadgets and layout.}
Fix $\omega=1/100$. We use two types of gadgets: equilateral triangles of side length $1$ or $1-\omega$, and unit squares (\cref{fig:overview-plane-costs}(a,~b)). A \emph{crossover} $x$ consists of an upper and a lower triangle where a horizontal and a vertical cable meet, with the cables $U_x$ above, $D_x$ below, and $L_x,R_x$ to the left and right at its corners (\cref{fig:overview-plane-crossover}(c,~d)).
}

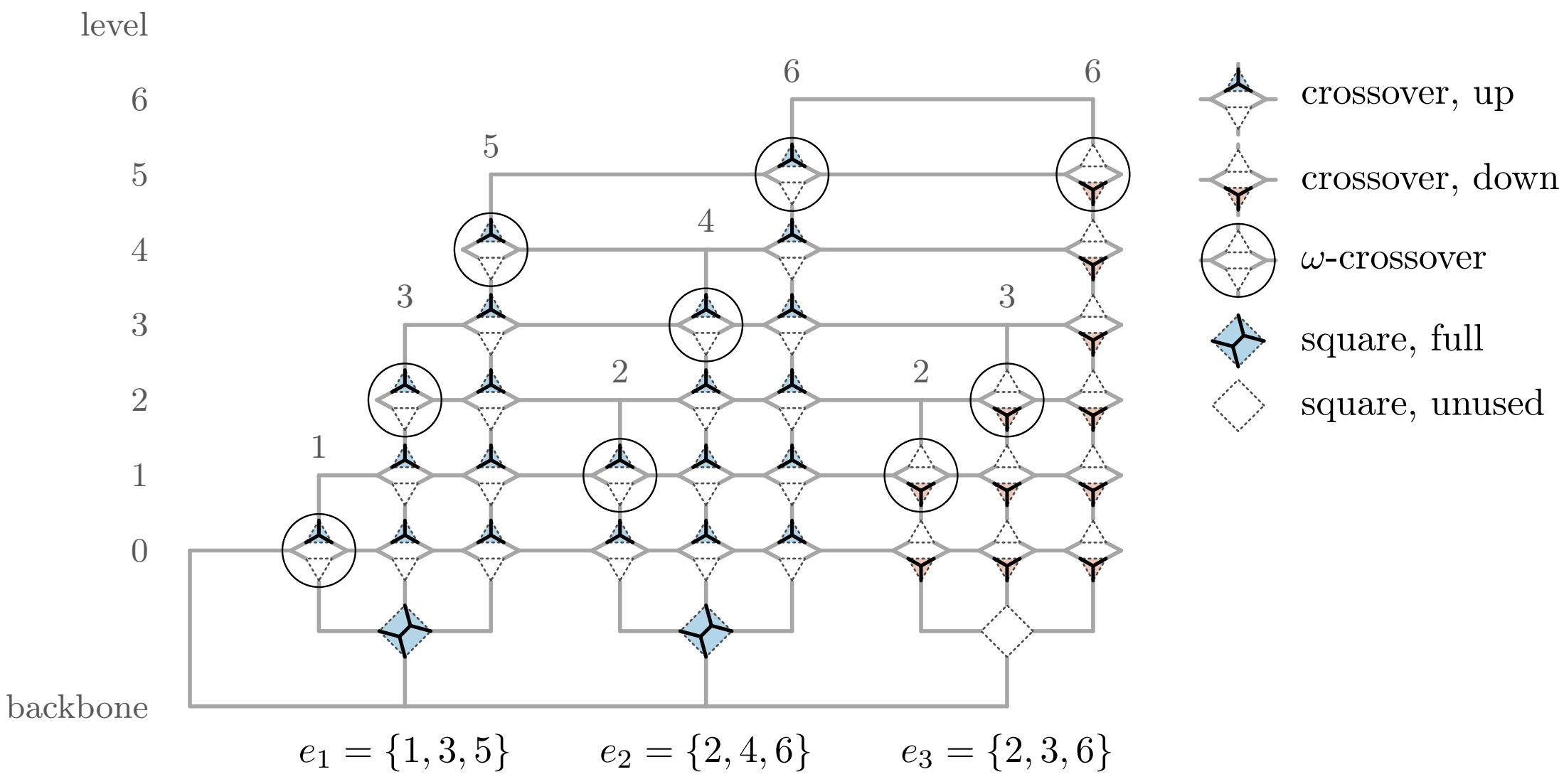
\begin{figure}[!h]
\centering
\begin{tikzpicture}[line cap=round,line join=round,x=1cm,y=1cm,font=\small,
  row/.style={draw=black!35,line width=1.05pt,line cap=round,line join=round},
  stub/.style={draw=black!35,line width=1.05pt,line cap=round},
  tri/.style={draw=black!70,densely dotted,line width=.5pt,line join=round},
  conn/.style={draw=black,line width=.95pt,line cap=round,line join=round},
  warp/.style={draw=black,line width=.45pt},
  sq/.style={diamond,draw=black,inner sep=0pt,minimum size=9pt,line width=.5pt},
  lab/.style={text=black!65,font=\footnotesize}]
\newcommand{\xover}[4]{%
  \begin{scope}[shift={(#1,#2)},x=1mm,y=1mm]
    \fill[white] (-2.6,-2.8) rectangle (2.6,2.8);
    \draw[stub] (-2.6,0)--(-1.2,.75) (-2.6,0)--(-1.2,-.75) (2.6,0)--(1.2,.75) (2.6,0)--(1.2,-.75);
    \filldraw[tri,fill=#3] (0,2.8)--(-1.2,.75)--(1.2,.75)--cycle;
    \filldraw[tri,fill=#4] (0,-2.8)--(-1.2,-.75)--(1.2,-.75)--cycle;
  \end{scope}}
\newcommand{\upx}[2]{\xover{#1}{#2}{cableV!30}{white}%
  \begin{scope}[shift={(#1,#2)},x=1mm,y=1mm]
    \draw[conn] (0,1.43)--(0,2.8) (0,1.43)--(-1.2,.75) (0,1.43)--(1.2,.75);
  \end{scope}}
\newcommand{\downx}[2]{\xover{#1}{#2}{white}{cableVp!30}%
  \begin{scope}[shift={(#1,#2)},x=1mm,y=1mm]
    \draw[conn] (0,-1.43)--(0,-2.8) (0,-1.43)--(-1.2,-.75) (0,-1.43)--(1.2,-.75);
  \end{scope}}
\newcommand{\sqbox}[3]{%
  \begin{scope}[shift={(#1,#2)},x=1mm,y=1mm]
    \filldraw[fill=#3,draw=black!70,densely dotted,line width=.5pt,line join=round] (2.4,0)--(0,2.4)--(-2.4,0)--(0,-2.4)--cycle;
  \end{scope}}
\newcommand{\fullsq}[2]{\sqbox{#1}{#2}{cableV!30}%
  \begin{scope}[shift={(#1,#2)},x=1mm,y=1mm]
    \draw[conn] (0,2.4)--(.507,.507)--(2.4,0) (.507,.507)--(-.507,-.507) (-2.4,0)--(-.507,-.507)--(0,-2.4);
  \end{scope}}
\newcommand{\emptysq}[2]{\sqbox{#1}{#2}{white}}
\draw[row] (0.00,0.00)--(8.66,0.00);
\node[lab,anchor=east] at (-0.25,0.00) {0};
\draw[row] (1.20,0.70)--(8.66,0.70);
\node[lab,anchor=east] at (-0.25,0.70) {1};
\draw[row] (1.74,1.40)--(8.66,1.40);
\node[lab,anchor=east] at (-0.25,1.40) {2};
\draw[row] (2.00,2.10)--(8.66,2.10);
\node[lab,anchor=east] at (-0.25,2.10) {3};
\draw[row] (2.54,2.80)--(8.66,2.80);
\node[lab,anchor=east] at (-0.25,2.80) {4};
\draw[row] (2.80,3.50)--(8.66,3.50);
\node[lab,anchor=east] at (-0.25,3.50) {5};
\draw[row] (5.60,4.20)--(8.40,4.20);
\node[lab,anchor=east] at (-0.25,4.20) {6};
\node[lab,anchor=east] at (-0.25,4.90) {level};
\draw[row] (1.20,-0.75)--(1.20,0.70);
\draw[row] (2.00,-0.75)--(2.00,2.10);
\draw[row] (2.80,-0.75)--(2.80,3.50);
\draw[row] (4.00,-0.75)--(4.00,1.40);
\draw[row] (4.80,-0.75)--(4.80,2.80);
\draw[row] (5.60,-0.75)--(5.60,4.20);
\draw[row] (6.80,-0.75)--(6.80,1.40);
\draw[row] (7.60,-0.75)--(7.60,2.10);
\draw[row] (8.40,-0.75)--(8.40,4.20);
\draw[row] (1.20,-0.75)--(2.80,-0.75) (2.00,-0.75)--(2.00,-1.45);
\draw[row] (4.00,-0.75)--(5.60,-0.75) (4.80,-0.75)--(4.80,-1.45);
\draw[row] (6.80,-0.75)--(8.40,-0.75) (7.60,-0.75)--(7.60,-1.45);
\draw[row] (0,0)--(0,-1.45)--(7.60,-1.45);
\node[lab,anchor=east] at (-0.25,-1.45) {backbone};
\upx{1.20}{0.00}
\draw[warp] (1.20,0.00) circle[radius=3.4mm];
\node[lab,anchor=south] at (1.20,0.73) {1};
\upx{2.00}{0.00}
\upx{2.00}{0.70}
\upx{2.00}{1.40}
\draw[warp] (2.00,1.40) circle[radius=3.4mm];
\node[lab,anchor=south] at (2.00,2.13) {3};
\upx{2.80}{0.00}
\upx{2.80}{0.70}
\upx{2.80}{1.40}
\upx{2.80}{2.10}
\upx{2.80}{2.80}
\draw[warp] (2.80,2.80) circle[radius=3.4mm];
\node[lab,anchor=south] at (2.80,3.53) {5};
\upx{4.00}{0.00}
\upx{4.00}{0.70}
\draw[warp] (4.00,0.70) circle[radius=3.4mm];
\node[lab,anchor=south] at (4.00,1.43) {2};
\upx{4.80}{0.00}
\upx{4.80}{0.70}
\upx{4.80}{1.40}
\upx{4.80}{2.10}
\draw[warp] (4.80,2.10) circle[radius=3.4mm];
\node[lab,anchor=south] at (4.80,2.83) {4};
\upx{5.60}{0.00}
\upx{5.60}{0.70}
\upx{5.60}{1.40}
\upx{5.60}{2.10}
\upx{5.60}{2.80}
\upx{5.60}{3.50}
\draw[warp] (5.60,3.50) circle[radius=3.4mm];
\node[lab,anchor=south] at (5.60,4.23) {6};
\downx{6.80}{0.00}
\downx{6.80}{0.70}
\draw[warp] (6.80,0.70) circle[radius=3.4mm];
\node[lab,anchor=south] at (6.80,1.43) {2};
\downx{7.60}{0.00}
\downx{7.60}{0.70}
\downx{7.60}{1.40}
\draw[warp] (7.60,1.40) circle[radius=3.4mm];
\node[lab,anchor=south] at (7.60,2.13) {3};
\downx{8.40}{0.00}
\downx{8.40}{0.70}
\downx{8.40}{1.40}
\downx{8.40}{2.10}
\downx{8.40}{2.80}
\downx{8.40}{3.50}
\draw[warp] (8.40,3.50) circle[radius=3.4mm];
\node[lab,anchor=south] at (8.40,4.23) {6};
\fullsq{2.00}{-0.75}
\node[anchor=north] at (2.00,-1.57) {$e_1=\{1,3,5\}$};
\fullsq{4.80}{-0.75}
\node[anchor=north] at (4.80,-1.57) {$e_2=\{2,4,6\}$};
\emptysq{7.60}{-0.75}
\node[anchor=north] at (7.60,-1.57) {$e_3=\{2,3,6\}$};
\draw[stub] (9.40,4.20)--(10.10,4.20) (9.75,3.87)--(9.75,4.53);
\upx{9.75}{4.20}
\node[anchor=west] at (10.20,4.20) {crossover, up};
\draw[stub] (9.40,3.45)--(10.10,3.45) (9.75,3.12)--(9.75,3.78);
\downx{9.75}{3.45}
\node[anchor=west] at (10.20,3.45) {crossover, down};
\draw[stub] (9.40,2.70)--(10.10,2.70) (9.75,2.37)--(9.75,3.03);
\xover{9.75}{2.70}{white}{white}
\draw[warp] (9.75,2.70) circle[radius=3.4mm];
\node[anchor=west] at (10.20,2.70) {$\omega$-crossover};
\fullsq{9.75}{1.95}
\node[anchor=west] at (10.20,1.95) {square, full};
\emptysq{9.75}{1.35}
\node[anchor=west] at (10.20,1.35) {square, unused};
\end{tikzpicture}
\caption{{Layout for elements $1,\ldots,6$ and three triples, with optimal connections shown in black. The perfect matching $\{e_1,e_2\}$ corresponds to the columns of $e_1$ and $e_2$ being up and those of $e_3$ being down.}}
\label{fig:overview-plane-layout}
\end{figure}

{
The layout (\cref{fig:overview-plane-layout}) has horizontal \emph{levels} $0,\ldots,3r$, where level~$0$ contains the \emph{backbone}, and a square below level~$0$ for each triple. For each element $j$ of a triple, a vertical \emph{column} runs from a corner of its square up to level~$j$, crossing the lower levels at crossovers, and the remaining corner of the square is joined to the backbone. The top crossover of each column, on level $j-1$, is an \emph{$\omega$-crossover}, with an upper triangle of side $1-\omega$ instead of $1$.
}

{
\paragraph{Clean trees and perfect matchings.}
We first explain how a short tree connecting the cables encodes a matching: up to small errors, such a tree consists of the cables and Steiner trees, each joining all ports of one gadget.

\begin{restatable}{definition}{PlaneCleanTree}
\label{def:overview-plane-clean}
A tree is \emph{clean} if it contains the cables and connects them only at gadgets, using Steiner trees on their corners (\cref{fig:overview-plane-costs}(a,~b)): whenever it uses a square, it joins all four cables, and whenever it uses a triangle, it joins all three.
\end{restatable}

At a crossover $x$, the upper triangle joins $L_x$ and $R_x$ to $U_x$, while the lower joins them to $D_x$ (\cref{fig:overview-plane-crossover}(c,~d)). Using both would join $L_x$ to $R_x$ twice and create a cycle. We call $x$ \emph{up} if the upper triangle is chosen, \emph{down} if the lower one is chosen, and \emph{broken} if neither is chosen. First consider a clean tree with no broken crossovers, so each crossover uses exactly one triangle. A unit triangle costs $\sqrt3$ and the smaller upper triangle of an $\omega$-crossover costs $\sqrt3\,(1-\omega)$, a saving of $\sqrt3\,\omega$. Since the cost at a square is $1+\sqrt3$, such a tree using $q$ squares and having $C_\omega$ up $\omega$-crossovers has length $L+\sqrt3\,C+(1+\sqrt3)\,q-\sqrt3\,\omega C_\omega$, where $C$ is the number of crossovers. We show that every such tree has length at least
\[
    \tau=L+\sqrt3\,C+(1+\sqrt3)\,r-3\sqrt3\,\omega r,
\]
with equality only when the used squares form a perfect matching.
}

\begin{figure}[!h]
\centering
\begin{tikzpicture}[line cap=round,line join=round,
  x=0.9cm,y=0.9cm,font=\small,
  level/.style={draw=black!32,line width=1.05pt,line cap=round,line join=round},
  segment/.style={draw=planeCableW,line width=1.05pt,line cap=round,line join=round},
  loose/.style={draw=planeCableW,line width=1.05pt,line cap=round,dash pattern=on 2pt off 1.5pt},
  pointer/.style={draw=black,line width=.5pt,->,>=latex,shorten >=2pt},
  guide/.style={draw=black!45,densely dotted,line width=0.5pt},
  connection/.style={draw=black,line width=.95pt,line cap=round,line join=round},
  steiner/.style={draw=none,fill=none,inner sep=0pt,minimum size=0pt},
  note/.style={font=\footnotesize,align=left},
  lab/.style={text=black!65,font=\footnotesize}
]
\def\La{0}\def\Lb{0.95}\def\Lc{1.9}\def\Ld{2.85}
\def\yq{-0.85}\def\yb{-1.3}
\newcommand{\tri}[1]{%
  \draw[guide] (0,#1+0.376)--(-0.12,#1+0.168)--(0.12,#1+0.168)--cycle;
  \draw[guide] (0,#1-0.376)--(-0.12,#1-0.168)--(0.12,#1-0.168)--cycle;
  \draw[draw=black!32,line width=1.05pt] (-0.12,#1+0.168)--(-0.336,#1)--(-0.12,#1-0.168)
    (0.12,#1+0.168)--(0.336,#1)--(0.12,#1-0.168);}
\newcommand{\fullup}[1]{%
  \draw[connection] (0,#1+0.24)--(0,#1+0.376) (0,#1+0.24)--(-0.12,#1+0.168) (0,#1+0.24)--(0.12,#1+0.168);
  \node[steiner] at (0,#1+0.24) {};}
\newcommand{\fulldown}[1]{%
  \draw[connection] (0,#1-0.24)--(0,#1-0.376) (0,#1-0.24)--(-0.12,#1-0.168) (0,#1-0.24)--(0.12,#1-0.168);
  \node[steiner] at (0,#1-0.24) {};}
\newcommand{\sqgad}{%
  \draw[guide] (0,\yq+0.16)--(0.16,\yq)--(0,\yq-0.16)--(-0.16,\yq)--cycle;}
\newcommand{\sqfull}{%
  \draw[connection] (0,\yq+0.16)--(0.0336,\yq+0.0336)--(0.16,\yq) (0.0336,\yq+0.0336)--(-0.0336,\yq-0.0336)
    (-0.16,\yq)--(-0.0336,\yq-0.0336)--(0,\yq-0.16);}
\foreach \y/\t in {\Ld/{level 3},\Lc/{level 2},\Lb/{level 1},\La/{level 0},\yb/{backbone}}
  \node[lab,anchor=east] at (-1.0,\y) {\t};
\foreach \sx/\heading in {0/{(a) All down},4.0/{(b) A switch},8.0/{(c) All up},12.0/{(d) Up above down}} {
  \begin{scope}[xshift=\sx*0.9cm]
    \node at (0,\Ld+0.75) {\heading};
    \draw[level] (-0.85,\Ld)--(0.85,\Ld);
    \foreach \y in {\La,\Lb,\Lc} \draw[level] (-0.85,\y)--(-0.336,\y) (0.336,\y)--(0.85,\y);
    \draw[level] (0,\Lc+0.376)--(0,\Ld);
    \draw[segment] (0,\Lc-0.376)--(0,\Lb+0.376) (0,\Lb-0.376)--(0,\La+0.376) (0,\La-0.376)--(0,\yq+0.16);
    \draw[level] (-0.16,\yq)--(-0.5,\yq) (0.16,\yq)--(0.5,\yq) (0,\yq-0.16)--(0,\yb) (-0.85,\yb)--(0.85,\yb);
    \foreach \y in {\La,\Lb,\Lc} {\tri{\y}}
    \sqgad
    \draw[draw=black,line width=.45pt] (0,\Lc) circle[radius=0.5];
  \end{scope}
}
\begin{scope}
  \fulldown{\Lc} \fulldown{\Lb} \fulldown{\La}
  \node[note,anchor=west] at (0.6,\yq) {unused};
\end{scope}
\begin{scope}[xshift=4.0*0.9cm]
  \fulldown{\Lc} \fullup{\Lb} \fullup{\La}
  \sqfull
  \node[note,anchor=west] (nb) at (0.95,1.43) {joins levels\\1 and 2};
  \draw[pointer] (nb.west)--(0,1.43);
\end{scope}
\begin{scope}[xshift=8.0*0.9cm]
  \fullup{\Lc} \fullup{\Lb} \fullup{\La}
  \sqfull
  \node[note,anchor=west] (nc) at (0.95,2.56) {joins levels\\2 and 3};
  \draw[pointer] (nc.west)--(0,2.56);
\end{scope}
\begin{scope}[xshift=12.0*0.9cm]
  \draw[draw=white,line width=2.4pt] (0,\Lc-0.376)--(0,\Lb+0.376);
  \draw[loose] (0,\Lc-0.376)--(0,\Lb+0.376);
  \fullup{\Lc} \fulldown{\Lb} \fulldown{\La}
  \node[note,anchor=west] (nd) at (0.95,1.43) {attached\\by neither};
  \draw[pointer] (nd.west)--(0,1.43);
\end{scope}
\end{tikzpicture}
\caption{{Columns for element $3$, with the top $\omega$-crossover circled. (a) The square is unused and all crossovers are down. (b) At a switch, both crossovers attach to the cable between them, joining levels $1$ and $2$. (c) All crossovers are up, so the top one joins levels $2$ and $3$. (d) An up crossover directly above a down one leaves the cable between them unattached.}}
\label{fig:overview-plane-switch}
\end{figure}
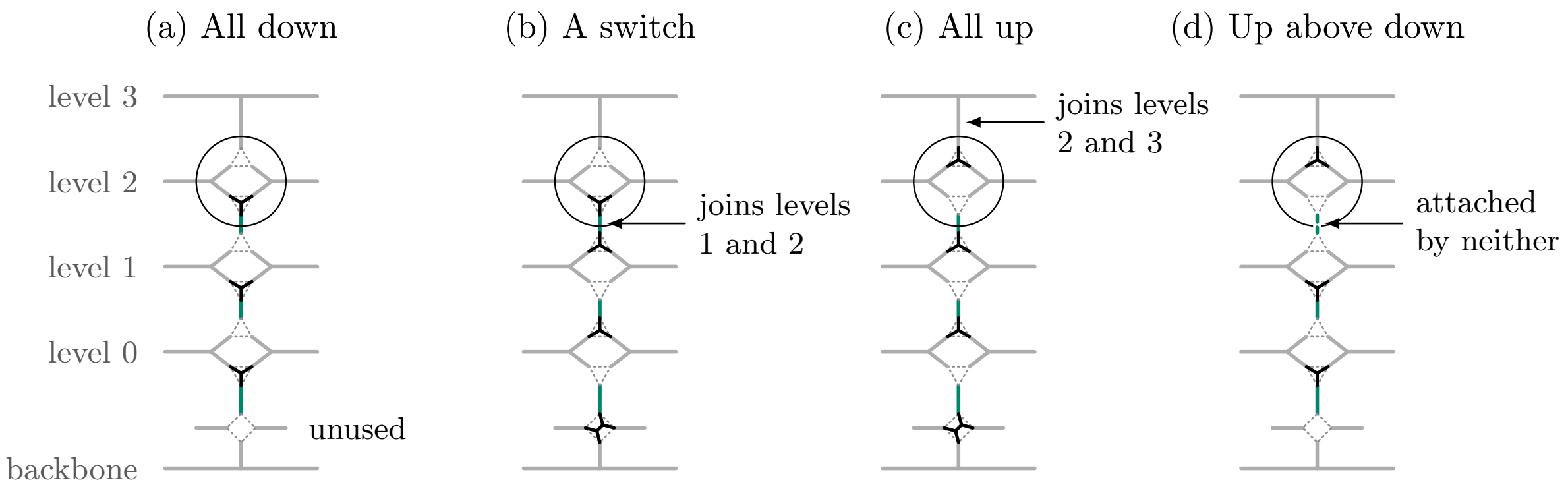

{
Every crossover of such a tree joins $L_x$ to $R_x$, so each level is connected and level $0$ contains the backbone. The argument uses the following properties.
\begin{enumerate}[(P1)]
\item \emph{Columns.} Within a column, all crossovers below an up crossover are up, and all those above a down crossover are down. Otherwise, an up crossover would lie directly above a down one, leaving the cable between them attached by neither (\cref{fig:overview-plane-switch}(d)).
\item \emph{Squares.} If a square is used, the bottom crossover of each of its three columns must be up: a down one would already connect the bottom cable to the backbone through level $0$, so using the square would create a cycle. If the square is unused, all three bottom crossovers must be down, since otherwise the bottom cables would be unattached.
\item \emph{$\omega$-crossovers.} For each element $j$, at most one $\omega$-crossover on level $j-1$ is up. An up crossover joins levels $j-1$ and $j$ by attaching the cable above it, which continues along level $j$; two such crossovers would join the same levels twice and create a cycle (\cref{fig:overview-plane-switch}(c)). Hence $C_\omega\le3r$.
\end{enumerate}
These properties imply that all crossovers in a column of an unused square are down (\cref{fig:overview-plane-switch}(a)). In a column of a used square, the bottom crossover is up, and the crossovers either remain up throughout (\cref{fig:overview-plane-switch}(c)) or change once from up to down at a pair of consecutive crossovers. We call this pair a \emph{switch} (\cref{fig:overview-plane-switch}(b)); both crossovers attach to the cable between them, joining their levels.

\paragraph{Lower bound.}
Columns of unused squares are down throughout and join no two levels, while each column of a used square joins exactly one pair of consecutive levels: either at its switch or, if it is up throughout, at its $\omega$-crossover. All $3r$ pairs must be joined, and each used square has three columns, so $q\ge r$. Together with $C_\omega\le3r$ from (P3), this gives a length of at least $\tau$, with equality requiring $q=r$ and $C_\omega=3r$.

\paragraph{Optimal clean trees.}
Suppose such a tree has length $\tau$, so $q=r$ and $C_\omega=3r$. Each element $j$ then has exactly one up $\omega$-crossover on level $j-1$, and these crossovers already connect every pair of consecutive levels. A switch would therefore create a cycle, so no column has one. By (P1), every column is all up or all down, and is all up if and only if its $\omega$-crossover is up. By (P2), a square is used exactly when all three of its columns are up. If two used squares shared an element $j$, their columns for $j$ would give two up $\omega$-crossovers on level $j-1$, contrary to (P3); hence the corresponding triples form a (perfect) matching.

Conversely, any perfect matching gives a clean tree of length $\tau$ by choosing the corresponding squares and making all crossovers in their columns up and all remaining crossovers down (\cref{fig:overview-plane-layout}). Intuitively, the $\omega$-crossovers make it cheaper to join levels at column tops, discouraging switches that allow overlapping triples.

To prove an approximation gap, we will argue that every tree whose length is close to $\tau$ yields a nearly perfect matching. Since there are no fallbacks, we cannot generally make a tree clean as in higher dimensions. We therefore count broken crossovers, missing discounts, and other ways of connecting the cables. We show that each such deviation costs at least $\sqrt3\,\omega$ and that a tree with few deviations still yields a large matching.
}

\FloatBarrier

\section{The cable reduction}
\label{sec:cable}

This section gives a reduction from three-dimensional matching to Euclidean Steiner Tree that applies to any \emph{separated} cable routing. \Cref{sec:growing,sec:fixed} construct such routings in growing and fixed dimension.

Recall that the matching instance $H=(U,E)$ has three parts of size $r$, total size $m=|U|+|E|$, and maximum matching size $\nu(H)$. The construction takes place in $\BR^D$, with $D$ determined by the cable routing.

\subsection{Cable instances}
\label{sec:cable-routing}

\begin{samepage}
\begingroup
\expandafter\patchcmd\csname thmt@original@definition\endcsname{{theorem}}{{definition}}{}{}
\CableSteinerTree*
\endgroup
\end{samepage}

Euclidean Cable Steiner Tree contains the usual Euclidean Steiner Tree problem as the special case where every cable is a single point, and we believe this generalization is of independent interest. It can also be viewed as a completion version of Euclidean Steiner Tree: parts of the network are already built and must be connected at minimum additional length, as when extending road, pipeline, or cable networks. Althaus, Rauterberg, and Ziegler~\cite{AlthausRZ20} studied exact algorithms for the planar case.

We now give the construction sketched in \cref{sec:overview-linear-dimension} in full; \cref{fig:cable-instance} shows a small example. Our instance has one square gadget for each triple and one fallback gadget for each element; these $m$ gadgets are indexed by $s=1,\ldots,m$ and have centres $z_s\in\BR^D$. We use the same two orthogonal unit coordinate vectors $b_1,b_2$ for every gadget and define the \emph{ports}, where the cables attach, by the following coordinates and labels.
\begin{enumerate}[(a)]
\item For the square of a triple $\{u,v,w\}$, the ports $p_{s,0},p_{s,u},p_{s,v},p_{s,w}$ are the four points $z_s\pm b_1$ and $z_s\pm b_2$. They form a square of side length $\sqrt2$.
\item For the fallback of an element $u$, the ports $p_{s,0},p_{s,u}$ are the points $z_s\pm\frac23b_1$. They are at distance $4/3$.
\end{enumerate}
We use the elements of $U$ and an additional symbol $0$ as labels. The shortest networks joining two, three, or four ports of a square have lengths $\sqrt2$, $1+\sqrt3$, and $\sqrt2+\sqrt6=4-\Delta$, respectively (\cref{fig:local-gadgets}). Only the four-port network costs less than $\frac43$ per joined pair.

For each label $i$, we join the ports labelled $i$ by a tree of line segments, the \emph{cable} $\Gamma_i$, so that different cables are disjoint; we call $\Gamma_0$ the \emph{backbone}. Let $L$ be the total cable length and, for each gadget $s$, let $B_s$ be the closed ball of radius $20$ centred at $z_s$. We require the cables to satisfy the following condition, illustrated in \cref{fig:separation}.

\begin{samepage}
\begingroup
\expandafter\patchcmd\csname thmt@original@definition\endcsname{{theorem}}{{definition}}{}{}
\SeparatedCables*
\endgroup
\end{samepage}

Explicitly, the segments in (S1) are $(z_s+b,\,z_s+20b)$ for $b\in\{\pm b_1,\pm b_2\}$ when $s$ is a square, and $(z_s+\frac23b,\,z_s+20b)$ for $b\in\{\pm b_1\}$ when $s$ is a fallback.
The segments leaving distinct ports of a gadget extend in orthogonal or opposite directions and are therefore at least $\frac43$ apart. Together with (S1), this shows that points on different cables are at least $\frac43$ apart. By (S2), there are at most $L$ cable segments, and by (S1), $|U|<L$.

For $0<\rho\le1/100$, we form $P_\rho$ by placing $\lceil t/\rho^2\rceil+1$ equally spaced terminals along each cable segment of length $t$, including both endpoints. Consecutive terminals along a segment are at distance in $[\rho^2/2,\rho^2]$, and the edges between them have total length $L$, with $|P_\rho|=\Theta(L/\rho^2)$.

The main result of this section shows that the optimum on $P_\rho$ equals the cable length $L$ plus the cost of connecting the cables, up to an error of $O(\rho L)$.

\CableReduction*

\begin{definition}
\label{def:clean}
A network is \emph{clean} if it consists of the cables, the four-port Steiner trees at squares whose triples form a matching, and the fallback segments for all elements outside these triples.
\end{definition}

A clean network joins every element cable to the backbone and is therefore connected; if it uses $q$ squares, its length is
\[
    L+\tfrac43(|U|-3q)+(4-\Delta)\,q=L+\tfrac43|U|-\Delta q.
\]
The clean network corresponding to a maximum matching contains $P_\rho$, proving the upper bound in \cref{thm:cable-reduction}. For the lower bound, we will show that every short tree on $P_\rho$ has length at least that of the clean network of some matching, up to an error of $O(\rho L)$.

\subsection{Recovering the cables}
\label{sec:cable-recovery-proof}

Suppose for a moment that an optimal tree on $P_\rho$ contains all edges between consecutive terminals, and hence all cables. Removing the cables leaves subtrees joining distinct cables. We will show that replacing each subtree by fallback segments or a four-port Steiner tree gives a clean network of no greater length, proving the lower bound of \cref{thm:cable-reduction} even without the error term $C_2\rho L$. In general, a tree on $P_\rho$ need not contain the cables, but it cannot stray far from them cheaply. We show that adding a little length suffices to obtain, for each cable, a tree that stays within distance $\rho$ of it.

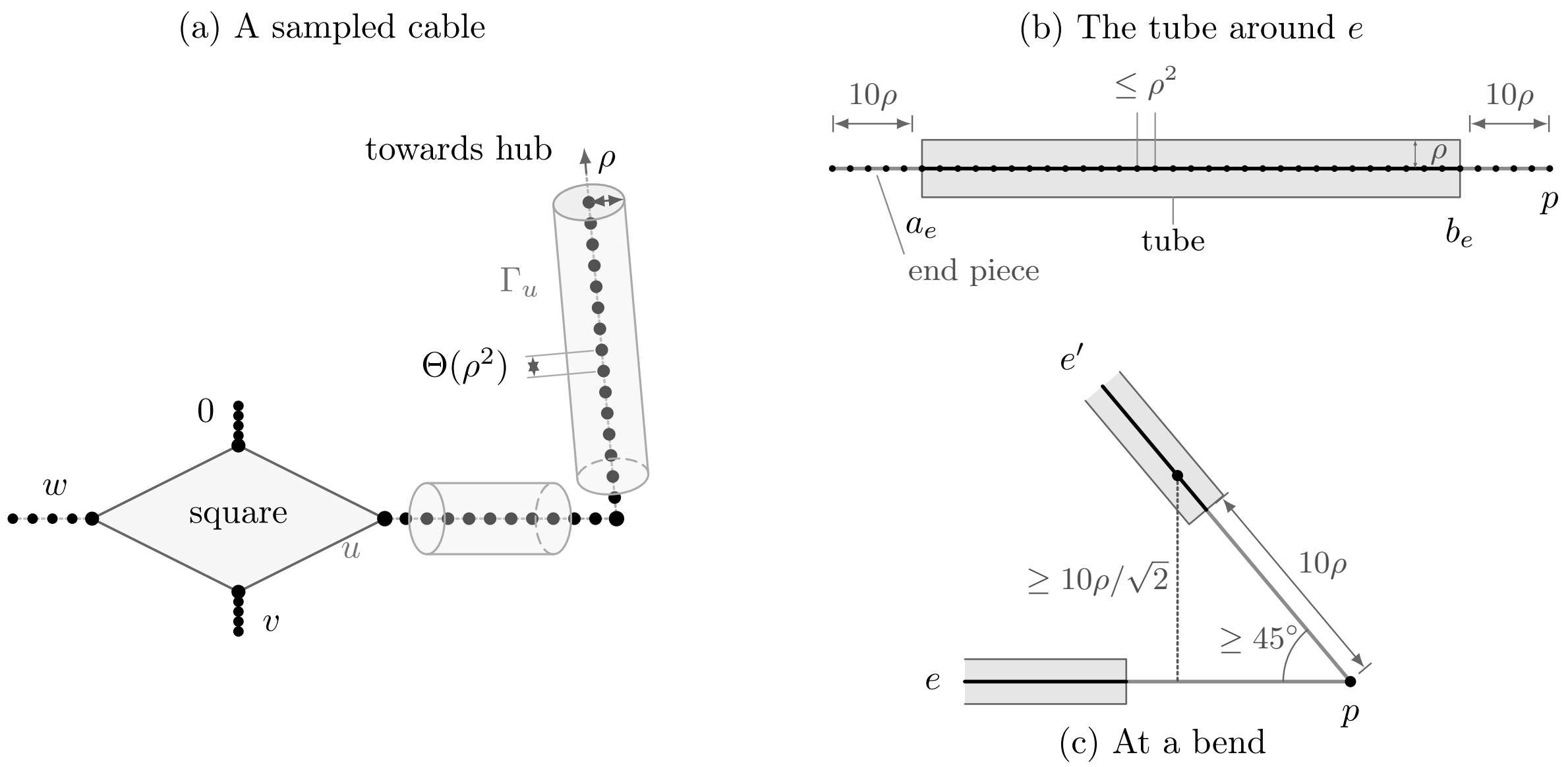
\begin{figure}[!tb]
\centering
\colorlet{recGrey}{black!55}
\begin{tikzpicture}[line cap=round,line join=round,
  x=1cm,y=1cm,font=\small,>=latex,
  endpiece/.style={draw=black!45,line width=1.05pt,line cap=round},
  middle/.style={draw=black,line width=1.05pt,line cap=round,line join=round},
  tube/.style={fill=black!10,draw=black!60,line width=.5pt},
  term/.style={circle,fill=black,inner sep=0pt,minimum size=1.9pt},
  dim/.style={draw=black!60,line width=.45pt,|<->|,>=latex},
  tick/.style={draw=black!45,line width=.4pt},
  note/.style={text=black!70,font=\footnotesize}
]
\begin{scope}[shift={(0,1)},line cap=round,line join=round,
  x=0.85cm,y=0.85cm,font=\small,>=latex,
  original/.style={draw=black!80,line width=1.05pt,line cap=round,line join=round},
  added/.style={draw=connectionGreen,line width=.95pt,line cap=round,line join=round},
  recovered/.style={draw=recGrey,line width=1.05pt,line cap=round,line join=round},
  guide/.style={draw=black!35,densely dashed,line width=.45pt},
  tubetree/.style={draw=black!80,line width=1.05pt,line cap=round,line join=round}]
\node at (3.5,7.12) {(a) A sampled cable};
\begin{scope}[shift={(2.4,2.2)},x={(0.612cm,-0.306cm)},y={(-0.612cm,-0.306cm)},z={(0cm,0.85cm)}]
  \coordinate (O) at (0,0,0);
  \coordinate (U) at (2.4,0,0);
  \coordinate (V) at (2.4,2.4,0);
  \coordinate (W) at (0,2.4,0);
  \coordinate (Q) at (4.3,-1.9,0);
  \coordinate (EO) at (-.65,-.65,0);
  \coordinate (EV) at (3.05,3.05,0);
  \coordinate (EW) at (-.65,3.05,0);
  \foreach \p/\q in {O/EO,V/EV,W/EW} {
    \draw[draw=black!25,densely dotted,line width=.7pt] (\p)--(\q);
    \foreach \j in {1,...,4}
      \fill[black] ($(\p)!{\j/4}!(\q)$) circle[radius=1.5pt];
  }
  \filldraw[fill=black!3,draw=black!60,line width=.7pt]
    (O)--(U)--(V)--(W)--cycle;
  \node at (1.2,1.2,0) {square};
  \draw[draw=recGrey!60,densely dotted,line width=.7pt] (U)--(Q);
  \foreach \j in {0,...,11}
    \fill[black] ($(U)!{\j/11}!(Q)$) circle[radius=1.8pt];
  \begin{scope}[shift={(U)},x=0.85cm,y=0.85cm]
    \pgfmathsetmacro{\tubeleft}{2.736*2/11}
    \pgfmathsetmacro{\tuberight}{2.736*8/11}
    \path[fill=recGrey!14,fill opacity=.35]
      (\tubeleft,-.42) arc[start angle=270,end angle=90,x radius=0.1785cm,y radius=0.357cm]
      --(\tuberight,.42) arc[start angle=90,end angle=-90,x radius=0.1785cm,y radius=0.357cm] --cycle;
    \draw[draw=recGrey!60,line width=.6pt]
      (\tubeleft,-.42)--(\tuberight,-.42)
      (\tubeleft,.42)--(\tuberight,.42);
    \draw[draw=recGrey!60,line width=.6pt]
      (\tubeleft,0) ellipse[x radius=0.1785cm,y radius=0.357cm];
    \draw[draw=recGrey!60,line width=.6pt,densely dashed]
      (\tuberight,-.42) arc[start angle=270,end angle=90,x radius=0.1785cm,y radius=0.357cm];
    \draw[draw=recGrey!60,line width=.6pt]
      (\tuberight,.42) arc[start angle=90,end angle=-90,x radius=0.1785cm,y radius=0.357cm];
  \end{scope}

  \begin{scope}[shift={(Q)},x=0.85cm,y=0.85cm,rotate=5]
    \draw[draw=recGrey!60,densely dotted,line width=.7pt] (0,0)--(0,4.1);
    \foreach \j in {0,...,15}
      \fill[black] (0,{.25*\j}) circle[radius=1.8pt];
    \path[fill=recGrey!14,fill opacity=.35]
      (-.42,.5) arc[start angle=180,end angle=360,x radius=0.357cm,y radius=0.1785cm]
      --(.42,3.75) arc[start angle=0,end angle=180,x radius=0.357cm,y radius=0.1785cm] --cycle;
    \draw[draw=recGrey!60,line width=.6pt,densely dashed]
      (.42,.5) arc[start angle=0,end angle=180,x radius=0.357cm,y radius=0.1785cm];
    \draw[draw=recGrey!60,line width=.6pt]
      (-.42,.5) arc[start angle=180,end angle=360,x radius=0.357cm,y radius=0.1785cm];
    \draw[draw=recGrey!60,line width=.6pt]
      (-.42,.5)--(-.42,3.75) (.42,.5)--(.42,3.75);
    \filldraw[fill=recGrey,fill opacity=.06,draw=recGrey!65,draw opacity=1,line width=.6pt]
      (0,3.75) ellipse[x radius=0.357cm,y radius=0.1785cm];
    \draw[->,draw=black!60,line width=.6pt]
      (0,4.1)--(0,4.4) node[left=5pt] {towards hub};
    \draw[<->,draw=recGrey!80!black,line width=.55pt]
      (0,3.75)--node[above=4pt] {$\rho$} (.42,3.75);
    \draw[draw=recGrey!60,line width=.45pt]
      (-.1,1.75)--(-.9,1.75) (-.1,2)--(-.9,2);
    \draw[<->,draw=recGrey!80!black,line width=.5pt]
      (-.82,1.75)--node[left=3pt] {$\Theta(\rho^2)$} (-.82,2);
    \node[anchor=east,text=recGrey] at (-.5,2.85) {$\Gamma_u$};
  \end{scope}
  \foreach \p in {O,V,W} \fill[black] (\p) circle[radius=2pt];
  \fill[black] (U) circle[radius=2.2pt];
  \fill[black] (Q) circle[radius=2.2pt];
  \node[above left=3pt] at (O) {$0$};
  \node[below left=3pt,text=recGrey] at (U) {$u$};
  \node[below right=3pt] at (V) {$v$};
  \node[above left=3pt] at (W) {$w$};
\end{scope}

\end{scope}
\begin{scope}[shift={(8.0,5.65)},x=.9cm,y=.9cm]
\def\r{0.32}
\coordinate (P) at (8,0);
\filldraw[tube] (1.0,-\r) rectangle (7.0,\r);
\draw[endpiece] (0,0)--(1.0,0) (7.0,0)--(8,0);
\draw[middle] (1.0,0)--(7.0,0);
\foreach \x in {0,0.2,...,8.01} \node[term] at (\x,0) {};
\node[below=3pt] at (P) {$p$};
\node[below=2pt] at (1.0,-\r) {$a_e$};
\node[below=2pt] at (7.0,-\r) {$b_e$};
\draw[dim] (0,.5)--node[above,note] {$10\rho$} (0.9,.5);
\draw[dim] (7.1,.5)--node[above,note] {$10\rho$} (8,.5);
\draw[draw=black!60,line width=.3pt,
  {Latex[length=1.6pt,width=1.2pt]}-{Latex[length=1.6pt,width=1.2pt]}]
  (6.5,0)--(6.5,\r);
\node[note,anchor=west] at (6.52,.16) {$\rho$};
\draw[tick] (3.4,.04)--(3.4,.62) (3.6,.04)--(3.6,.62);
\node[note,anchor=south] at (3.5,.62) {$\le\rho^2$};
\draw[draw=black!45,line width=.45pt] (0.5,-.08)--(0.8,-.95)
  node[anchor=north west,inner sep=1pt,note] {end piece};
\draw[draw=black!45,line width=.45pt] (3.8,-\r)--(3.8,-.62)
  node[anchor=north,inner sep=1pt,note,text=black] {tube};
\node at (4,1.55) {(b) The tube around $e$};

\end{scope}
\begin{scope}[shift={(13.2,.5)},x=.9cm,y=.9cm]
  \def\s{0.25}
  \coordinate (Q) at (0,0);
  \coordinate (E) at (-1,0);                    %
  \coordinate (F) at ({cos(130)},{sin(130)});   %
  \path[fill=black!10] (-2.5,-\s) rectangle (-4.3,\s);
  \draw[draw=black!60,line width=.5pt] (-4.3,-\s)--(-2.5,-\s)--(-2.5,\s)--(-4.3,\s);
  \begin{scope}[rotate=130]
    \path[fill=black!10] (2.5,-\s) rectangle (4.3,\s);
    \draw[draw=black!60,line width=.5pt] (4.3,-\s)--(2.5,-\s)--(2.5,\s)--(4.3,\s);
  \end{scope}
  \draw[endpiece] (Q)--($2.5*(E)$) (Q)--($2.5*(F)$);
  \draw[middle] ($2.5*(E)$)--($4.3*(E)$) ($2.5*(F)$)--($4.3*(F)$);
  \fill (Q) circle[radius=1.6pt];
  \node[below=3pt] at (Q) {$p$};
  \node[left=3pt] at ($4.3*(E)$) {$e$};
  \node[above left=1pt] at ($4.3*(F)$) {$e'$};
  \draw[black!60,line width=.45pt] (180:.75) arc[start angle=180,end angle=130,radius=.75];
  \node[note] at (155:1.12) {$\ge45^\circ$};
  \coordinate (Z) at ($3.0*(F)$);
  \coordinate (W) at ($3.0*cos(50)*(E)$);
  \draw[black!65,densely dotted,line width=.7pt] (Z)--(W);
  \fill (Z) circle[radius=1.6pt];
  \node[note,anchor=east,fill=white,inner sep=1pt] at ($(Z)!.5!(W)+(-.05,0)$) {$\ge10\rho/\sqrt2$};
  \draw[dim] ($2.5*(F)+(40:.22)$)--node[anchor=south west,inner sep=1pt,note] {$10\rho$} ($(Q)+(40:.22)$);
  \node at (-2.1,-.7) {(c) At a bend};
\end{scope}
\end{tikzpicture}
\caption{Sampling and tubes. (a) A sampled cable with tubes around its middle segments. (b) The tube around a sampled segment $e$. (c) Two segments meeting at $p$ at an angle of at least $45^\circ$. Every point on the middle segment of $e'$ lies at least $10\rho$ from $p$ and therefore at least $10\rho/\sqrt2$ from $e$, so the tubes are disjoint.}
\label{fig:sampling}
\end{figure}

For each cable segment $e$, choose terminals $a_e$ and $b_e$ at distances between $10\rho$ and $10\rho+\rho^2$ from its two endpoints, respectively. We call $(a_e,b_e)$ the \emph{middle segment} of $e$ and its two remaining parts the \emph{end pieces}, each of length at most $10\rho+\rho^2$. Each end piece consists of edges between consecutive terminals. We surround each middle segment by a closed cylinder of radius $\rho$ with flat caps, called its \emph{tube} (\cref{fig:sampling}). These tubes are pairwise disjoint: for segments on different cables or non-incident segments on the same cable, this follows because their distance exceeds $2\rho$. If segments $e,e'$ share an endpoint $p$, their angle is at least $45^\circ$ by (S2), so every point on the middle segment of $e'$ is at least $10\rho$ from $p$ and hence at least $10\rho\sin45^\circ>7\rho$ from $e$. The tube of $e'$ therefore has distance greater than $6\rho$ from $e$, so the two tubes are disjoint.

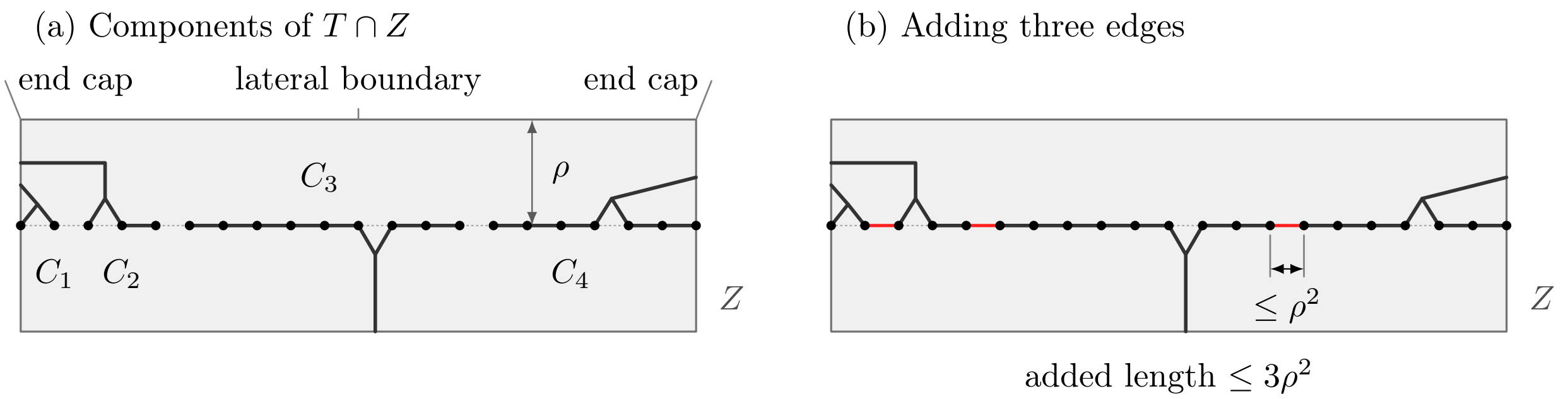
\begin{figure}[!tb]
\centering
\resizebox{\textwidth}{!}{\begin{tikzpicture}[line cap=round,line join=round,
  x=1cm,y=1cm,font=\small,>=latex,
  tree/.style={draw=black!80,line width=1.05pt,line cap=round,line join=round},
  patch/.style={draw=connectionGreen,line width=.95pt,line cap=round,line join=round},
  guide/.style={draw=black!35,densely dotted},
  callout/.style={draw=black!50,line width=.5pt}
]
\foreach \stage/\offset in {1/0,2/8.4} {
  \begin{scope}[shift={(\offset,0)}]
    \fill[black,fill opacity=.06] (0,-1.1) rectangle (7,1.1);
    \draw[draw=black!55,line width=.6pt] (0,-1.1) rectangle (7,1.1);
    \draw[guide] (0,0)--(7,0);
    \draw[tree] (0,0)--(.175,.22)--(.35,0);
    \draw[tree] (.175,.22)--(0,.42);
    \draw[tree] (.7,0)--(.875,.28)--(1.05,0)--(1.4,0);
    \draw[tree] (.875,.28)--(.875,.65)--(0,.65);
    \draw[tree] (1.75,0)--(3.5,0)--(3.675,-.3)--(3.85,0)--(4.55,0);
    \draw[tree] (3.675,-.3)--(3.675,-1.1);
    \draw[tree] (4.9,0)--(5.95,0)--(6.125,.28)--(6.3,0)--(7,0);
    \draw[tree] (6.125,.28)--(7,.5);
    \ifnum\stage=2
      \draw[patch] (.35,0)--(.7,0) (1.4,0)--(1.75,0) (4.55,0)--(4.9,0);
    \fi
    \foreach \j in {0,...,20}
      \fill[black] ({.35*\j},0) circle[radius=1.4pt];
    \node[anchor=west,text=black!70] at (7.1,-.75) {$Z$};
  \end{scope}
}
\node[anchor=west] at (0,2.05) {(a) Components of $T\cap Z$};
\node[anchor=east] (leftcap) at (1.3,1.5) {end cap};
\draw[callout] (leftcap.west)--(0,1.1);
\node (side) at (3.5,1.5) {lateral boundary};
\draw[callout] (side.south)--(3.5,1.1);
\node[anchor=west] (rightcap) at (5.7,1.5) {end cap};
\draw[callout] (rightcap.east)--(7,1.1);
\draw[<->,draw=black!65] (5.3,0)--node[right=2pt] {$\rho$} (5.3,1.1);
\node at (.35,-.5) {$C_1$};
\node at (1.05,-.5) {$C_2$};
\node at (3.1,.5) {$C_3$};
\node at (5.7,-.5) {$C_4$};
\begin{scope}[shift={(8.4,0)}]
  \node[anchor=west] at (0,2.05) {(b) Adding three edges};
  \draw[callout] (4.55,-.1)--(4.55,-.53) (4.9,-.1)--(4.9,-.53);
  \draw[<->,draw=black] (4.55,-.45)--node[below=2pt,text=black] {$\le\rho^2$} (4.9,-.45);
  \node[anchor=north,text=black] at (3.5,-1.25) {added length $\le3\rho^2$};
\end{scope}
\end{tikzpicture}}
\caption{Connecting components inside a tube. Black segments show $T\cap Z$. The components $C_1,C_2$ reach the left cap, $C_3$ the lateral boundary, and $C_4$ the right cap. Three red edges between consecutive samples connect them.}
\label{fig:tube-patching}
\end{figure}

Fix a tree $T$ of length $O(L)$ spanning $P_\rho$. Its intersection with a tube may be disconnected, but adding edges between consecutive samples along the cable segment lets us join its components at small cost (\cref{fig:tube-patching}).

\begin{lemma}
\label{lem:one-tube-patching}
Let $Z$ be the tube around a middle segment and $w$ the length of $T$ inside $Z$. We can connect all sampled terminals on that segment within $Z$ by adding edges between consecutive terminals, with total added length at most $\rho w+2\rho^2+4\rho\sqrt w$.
\end{lemma}

\begin{proof}
Let $k$ be the number of components of $T\cap Z$ containing sampled terminals; see \cref{fig:tube-patching}. We can connect these $k$ components using $k-1$ edges between consecutive samples, each of length at most $\rho^2$, for a total added length of at most $\rho^2(k-1)$.

If $k>1$, each such component meets the tube boundary, since $T$ is connected. A component meeting the lateral boundary contains a path of length at least $\rho$ from a sample on the cable segment to the side of the tube, so there are at most $w/\rho$ such components. Assign every other component to an end cap that it meets. For the $k_l$ components assigned to the left cap, choose one sampled terminal from each and order these samples from left to right. The $j$th sample is at distance at least $(j-1)\rho^2/2$ from the cap, so its component contains a path to the cap of at least this length. These paths lie in distinct components and are therefore disjoint. Their lengths sum to at most $w$, giving
\[
    w\ge\sum_{j=1}^{k_l}\frac{(j-1)\rho^2}{2}
    =\frac{\rho^2 k_l(k_l-1)}{4}.
\]
It follows that $k_l\le1+2\sqrt w/\rho$. The same bound holds at the right cap. Consequently $k\le w/\rho+2+4\sqrt w/\rho$, which proves the claimed bound on the added length. If $k=1$, nothing needs to be added.
\end{proof}

We now apply this bound to all tubes to prove the cable-recovery lemma.

\begingroup
\expandafter\patchcmd\csname thmt@original@lemma\endcsname{{theorem}}{{lemma}}{}{}
\CableRecovery*
\endgroup
\begin{proof}
Let $n_{\mathrm{seg}}\le L$ be the number of cable segments. For each cable segment $e$, let $w_e$ be the length of $T$ inside the tube of $e$. The tubes are disjoint, so $\sum_ew_e\le\operatorname{len}(T)$. By \cref{lem:one-tube-patching} and Cauchy--Schwarz, the total length added inside the tubes is at most $\rho\operatorname{len}(T)+2n_{\mathrm{seg}}\rho^2+4\rho\sqrt{n_{\mathrm{seg}}\operatorname{len}(T)}=O(\rho L)$, since $\operatorname{len}(T)=O(L)$. We also add all edges between consecutive terminals along the end pieces, with total length at most $(20\rho+2\rho^2)n_{\mathrm{seg}}\le21\rho L$.

After adding these edges, the samples on the middle segment of each cable segment $e$ are connected inside its tube. Choose a tree $J_e$ in this network spanning these samples; because it contains $a_e$ and $b_e$, it is at least as long as the middle segment.

For each cable $\Gamma_i$, let $T_i$ be the union of the trees $J_e$ for its segments and all their end pieces. The tubes are disjoint and meet the end pieces only at $a_e$ and $b_e$, so each $T_i$ is a tree spanning the sampled terminals on $\Gamma_i$, and the trees $T_i$ are pairwise disjoint. Moreover, $T_i$ lies within distance $\rho$ of $\Gamma_i$ and has length at least $\operatorname{len}(\Gamma_i)$, so $\sum_i\operatorname{len}(T_i)\ge L$.
\end{proof}

\subsection{Connectors and cleaning}
\label{sec:cable-reduction-proof}

By cable recovery, we can add segments of total length $O(\rho L)$ to $T$ to obtain a connected network containing the forest $F=\bigcup_iT_i$, whose length is at least $L$. We choose a spanning tree $\widehat T$ of this network containing $F$ and repeatedly remove leaves outside $F$, which ensures that every leaf of $\widehat T$ lies in $F$.

\begin{figure}[!t]
\centering
\begin{tikzpicture}[x=.6cm,y=.6cm,font=\small,line cap=round,line join=round,
  tube/.style={draw=black!13,line width=5pt},
  tree/.style={draw=black,line width=.8pt},
  cont/.style={draw=black!45,line width=.8pt,densely dotted},
  conn/.style={draw=connectionGreen,line width=1.2pt},
  att/.style={circle,fill=black,inner sep=0pt,minimum size=4pt},
  port/.style={circle,draw=black,fill=white,inner sep=0pt,minimum size=3.6pt,line width=.6pt},
  gadget/.style={draw=black!40,densely dashed,line width=.45pt},
  tnode/.style={circle,draw=black,fill=white,inner sep=0pt,minimum size=16pt,line width=.7pt,font=\footnotesize},
  snode/.style={rectangle,draw=connectionGreen,fill=white,inner sep=0pt,minimum size=12pt,line width=.8pt},
  tedge/.style={draw=black!50,line width=.7pt},
  lab/.style={font=\footnotesize}]
\node[anchor=west] at (-.7,5) {(a) Connectors};
\foreach \pp in {
    {(1.6,.8)--(1.6,0)},                     %
    {(-.2,0)--(5.6,0)},                      %
    {(4.4,0)--(4.4,.35)},                    %
    {(.7,1.7)--(.2,1.7)--(.2,3.3)},          %
    {(1.6,2.6)--(1.6,3.6)--(3.4,3.6)},       %
    {(2.5,1.7)--(3.2,1.7)},                  %
    {(4.4,1.35)--(4.4,3.6)}} {               %
  \draw[tube] \pp; \draw[tree] \pp;}
\draw[cont] (-.2,0)--(-.7,0) (5.6,0)--(6.1,0) (.2,3.3)--(.2,3.8) (3.4,3.6)--(3.8,3.6)
  (3.2,1.7)--(3.6,1.7) (4.4,3.6)--(4.4,4);
\draw[gadget] (1.6,2.6)--(2.5,1.7)--(1.6,.8)--(.7,1.7)--cycle;
\pgfmathsetmacro{\a}{.9*(1-1/sqrt(3))/2}
\draw[conn] (1.6,2.6)--(1.6+\a,1.7+\a)--(2.5,1.7) (1.6+\a,1.7+\a)--(1.6-\a,1.7-\a)
  (.7,1.7)--(1.6-\a,1.7-\a)--(1.6,.8);
\draw[gadget] (4.4,.35)--(4.4,1.35);
\foreach \p in {(4.4,.35),(4.4,1.35)} \node[port] at \p {};
\coordinate (bV) at (3.1,3.6); \coordinate (bX) at (4.4,2.9);
\draw[conn] (bV)--(bX);
\foreach \p in {(1.6,2.6),(2.5,1.7),(1.6,.8),(.7,1.7),(bV),(bX)} \node[att] at \p {};
\node[lab,anchor=north] at (5.4,-.12) {$T_0$};
\node[lab,anchor=south] at (.2,3.8) {$T_u$};
\node[lab,anchor=south] at (2.4,3.68) {$T_v$};
\node[lab,anchor=north] at (3,1.6) {$T_w$};
\node[lab,anchor=west] at (4.5,2) {$T_x$};
\node[lab,text=connectionGreen,anchor=north east] at (1.05,1.3) {$S_1$};
\node[lab,text=connectionGreen,anchor=north east] at (3.75,3.2) {$S_2$};
\begin{scope}[shift={(8,0)}]
  \node[anchor=west] at (-.5,5) {(b) Incidence tree};
  \foreach \n/\y in {0/4,u/3,w/2,v/1,x/0} \coordinate (T\n) at (0,\y);
  \coordinate (S1) at (2.6,2.5); \coordinate (S2) at (2.6,.5);
  \draw[tedge] (S1)--(T0) (S1)--(Tu) (S1)--(Tw) (S1)--(Tv) (S2)--(Tv) (S2)--(Tx);
  \foreach \n in {0,u,w,v,x} \node[tnode] at (T\n) {$T_\n$};
  \foreach \n in {1,2} \node[snode] at (S\n) {\footnotesize$S_\n$};
\end{scope}
\end{tikzpicture}
\caption{Connectors at the square $s$ for $\{u,v,w\}$ and the fallback of $x$. (a) The recovered trees $T_i$ (black) lie in cable tubes (grey), and connectors (red) attach at the dots. Connector $S_1$ is good at $s$; $S_2$ is bad: its attachments on $T_v$ and $T_x$ lie outside the gadget balls. (b) The incidence tree of the trees $T_i$ and connectors.}
\label{fig:connectors}
\end{figure}
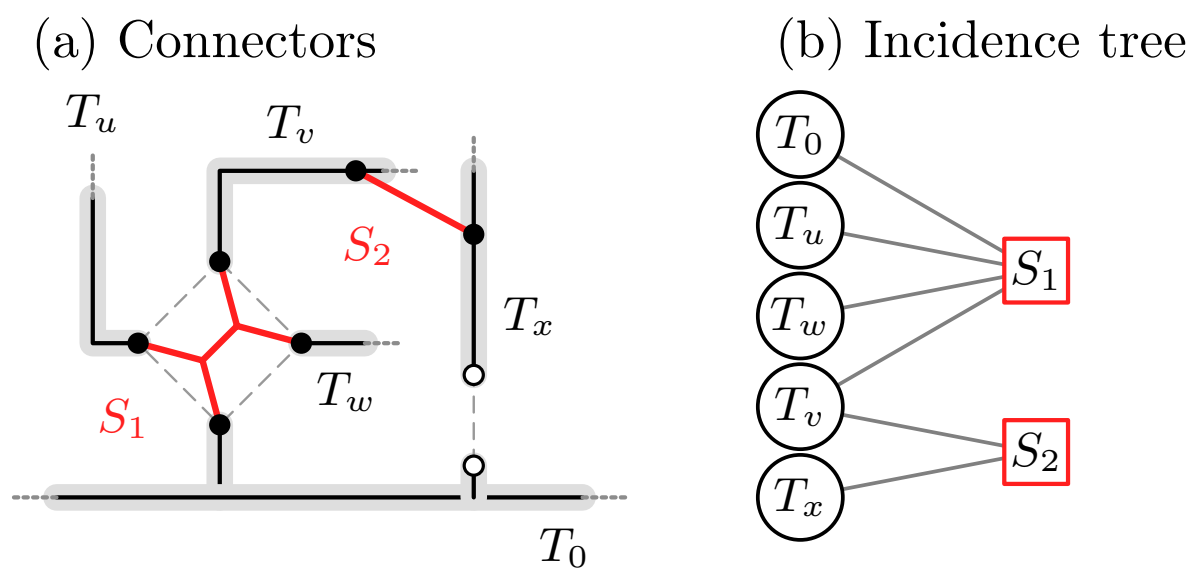

Removing $F$ decomposes the rest of $\widehat T$ into \emph{connectors}: subtrees whose leaves, called \emph{attachments}, lie on $F$ and whose other points lie outside $F$ (\cref{fig:connectors}(a)). Every connector has at least two attachments, and it cannot attach twice to the same $T_i$, since this would create a cycle in $\widehat T$.

The bipartite graph between the trees $T_i$ and the connectors, with one edge per attachment, is also a tree, which we call the \emph{incidence tree} (\cref{fig:connectors}(b)). If $r_S$ denotes the number of attachments of connector $S$, this tree has $\sum_Sr_S$ edges and one vertex for each of the $|U|+1$ trees $T_i$ and each connector. Since a tree has one edge fewer than vertices,
\begin{equation}
\label{eq:connector-count}
    \sum_S(r_S-1)=|U|.
\end{equation}

Connectors attach to the trees $T_i$, so their attachments lie within distance $\rho$ of the cables. An attachment is \emph{near} a port $p$ of a gadget $s$ if it lies within distance $\rho$ of the cable segment from $p$ to the boundary of $B_s$, as specified in (S1). We compare connectors joining all cables at a gadget with the shortest network connecting these cables (at the ports, as in \cref{fig:local-gadgets}), and all other connectors with fallback segments.

\begin{definition}
\label{def:good-connector}
A connector is
\begin{enumerate}[(i)]
\item \emph{good} at the square $s$ of a triple $\{u,v,w\}$ if it has exactly four attachments, one near each of the ports $p_{s,0},p_{s,u},p_{s,v},p_{s,w}$;
\item \emph{good} at the fallback $s$ of an element $u$ if it has exactly two attachments, one near each of the ports $p_{s,0}$ and $p_{s,u}$;
\item \emph{bad} otherwise.
\end{enumerate}
\end{definition}

The following lemma shows that a good connector is at least as long as the shortest network on the ports of its gadget, and a bad connector at least as long as $r_S-1$ fallbacks, up to small errors.

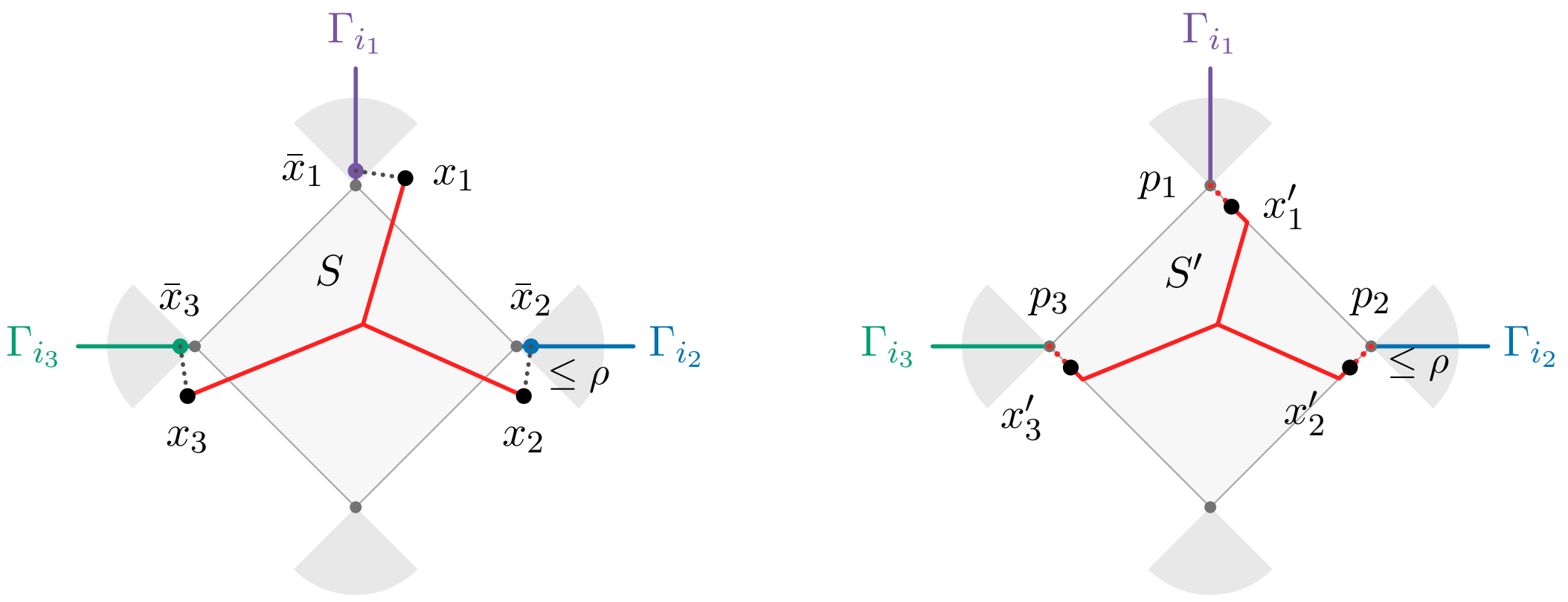
\begin{figure}[!t]
\centering
\begin{tikzpicture}[line cap=round,line join=round,x=1.3cm,y=1.3cm,font=\normalsize,>=latex,
  network/.style={draw=connectionGreen,line width=1.05pt,line cap=round,line join=round},
  added/.style={draw=connectionGreen,line width=1.4pt,dash pattern=on 0pt off 2.6pt,line cap=round,line join=round}]
\foreach \shift in {0,7.6} {
  \begin{scope}[xshift=\shift cm]
    \foreach \cx/\cy/\a in {0/1.1/45,1.1/0/-45,0/-1.1/225,-1.1/0/135}
      \fill[black!9] (\cx,\cy)--++(\a:.6) arc[start angle=\a,end angle=\a+90,radius=.6]--cycle;
    \filldraw[fill=black!3,draw=black!35] (0,1.1)--(1.1,0)--(0,-1.1)--(-1.1,0)--cycle;
    \draw[cableU,line width=1.05pt] (0,1.1)--(0,1.9);
    \draw[cableV,line width=1.05pt] (1.1,0)--(1.9,0);
    \draw[cableW,line width=1.05pt] (-1.1,0)--(-1.9,0);
    \foreach \pt in {(0,1.1),(1.1,0),(0,-1.1),(-1.1,0)}
      \fill[black!55] \pt circle[radius=1.5pt];
  \end{scope}
}
\draw[network] (.34,1.15)--(.05,.15)--(1.15,-.34);
\draw[network] (.05,.15)--(-1.15,-.34);
\node at (-.18,.52) {$S$};
\fill[cableU] (0,1.2) circle[radius=2pt];
\fill[cableV] (1.2,0) circle[radius=2pt];
\fill[cableW] (-1.2,0) circle[radius=2pt];
\draw[black!70,line width=1.1pt,dash pattern=on 0pt off 2.2pt,line cap=round] (0,1.2)--(.34,1.15);
\draw[black!70,line width=1.1pt,dash pattern=on 0pt off 2.2pt,line cap=round] (1.2,0)--(1.15,-.34);
\draw[black!70,line width=1.1pt,dash pattern=on 0pt off 2.2pt,line cap=round] (-1.2,0)--(-1.15,-.34);
\fill (.34,1.15) circle[radius=2pt] node[right=3pt] {$x_1$};
\fill (1.15,-.34) circle[radius=2pt] node[below=4pt] {$x_2$};
\fill (-1.15,-.34) circle[radius=2pt] node[below=4pt] {$x_3$};
\node[left=4pt] at (0,1.2) {$\bar x_1$};
\node[above=4pt] at (1.2,0) {$\bar x_2$};
\node[above=4pt] at (-1.2,0) {$\bar x_3$};
\node[right=1pt,font=\small] at (1.18,-.2) {$\le\rho$};
\node[cableU,above] at (0,1.9) {$\Gamma_{i_1}$};
\node[cableV,right] at (1.9,0) {$\Gamma_{i_2}$};
\node[cableW,left] at (-1.9,0) {$\Gamma_{i_3}$};
\node at (0,-2) {(a) Before projection};
\begin{scope}[xshift=7.6cm]
  \draw[network] (.145,.955)--(.252326,.847674)--(.05,.15)
    --(.880189,-.219811)--(.955,-.145);
  \draw[network] (.05,.15)--(-.873077,-.226923)--(-.955,-.145);
  \node at (-.18,.52) {$S'$};
  \draw[added] (0,1.1)--(.145,.955);
  \draw[added] (1.1,0)--(.955,-.145);
  \draw[added] (-1.1,0)--(-.955,-.145);
  \fill (.145,.955) circle[radius=2pt] node[right=4pt] {$x'_1$};
  \fill (.955,-.145) circle[radius=2pt] node[below left=2pt] {$x'_2$};
  \fill (-.955,-.145) circle[radius=2pt] node[below left=3pt] {$x'_3$};
  \node[left=4pt] at (0,1.1) {$p_1$};
  \node[above=4pt] at (1.1,0) {$p_2$};
  \node[above=4pt] at (-1.1,0) {$p_3$};
  \node[right=2pt,font=\small] at (1.05,-.12) {$\le\rho$};
  \node[cableU,above] at (0,1.9) {$\Gamma_{i_1}$};
  \node[cableV,right] at (1.9,0) {$\Gamma_{i_2}$};
  \node[cableW,left] at (-1.9,0) {$\Gamma_{i_3}$};
  \node at (0,-2) {(b) Project and join to ports};
\end{scope}
\end{tikzpicture}
\caption{Projection onto a square gadget, for three cables. The shaded regions project to the corners, so each cable point $\bar x_j$ projects to its port $p_j$, and $x_j$ to $x'_j$. The projected network $S'$ is no longer than $S$, and the dotted segments, each of length at most $\rho$, join it to the ports.}
\label{fig:gadget-projection}
\end{figure}

\begin{lemma}
\label{lem:displaced-attachments}
Let $S$ be a connector with $r$ attachments.
\begin{enumerate}[(i)]
\item If $S$ is good at a square, then $\operatorname{len}(S)\ge4-\Delta-4\rho$.
\item If $S$ is good at a fallback, then $\operatorname{len}(S)\ge\frac43-2\rho$.
\item If $S$ is bad, then $\operatorname{len}(S)\ge(\frac43-2\rho)(r-1)$.
\end{enumerate}
\end{lemma}

\begin{proof}
Let $x_1,\ldots,x_r$ be the attachments of $S$. They lie within distance $\rho$ of distinct cables, so for each $j$ we can choose a point $\bar x_j$ on the corresponding cable with $\|x_j-\bar x_j\|\le\rho$, and if $S$ is good at a gadget $s$, we choose it inside $B_s$. We group the attachments according to where these points lie: $x_j$ and $x_k$ are in the same group if $\bar x_j$ and $\bar x_k$ lie in a common ball $B_s$, and $x_j$ forms a group on its own if $\bar x_j$ lies in no ball. By (S1), attachments in different groups are at distance at least $20-2\rho>18$, and since a ball meets at most four cables, every group has at most four attachments.

If there are $g\ge2$ groups, then $S$ is bad, since a good connector has all its attachments near a single gadget. The balls of radius $9$ around one attachment from each group are disjoint, and since $S$ is connected and has points outside each of them, it runs from the centre of each ball to its boundary. Hence
\[
    \operatorname{len}(S)\ge9g\ge\frac43(4g-1)\ge\frac43(r-1),
\]
which gives (iii).

Otherwise, all points $\bar x_j$ lie in the ball $B_s$ of one gadget $s$. We project $S$ onto the square or fallback segment of $s$ to obtain a network $S'$ of no greater length (\cref{fig:gadget-projection}). The projection sends $\bar x_j$ to its port $p_j$ and does not increase distances, so the projected attachment $x_j'\in S'$ lies within distance $\rho$ of $p_j$. We then add the segments $(x_j',p_j)$ to obtain a network connecting $r$ ports of $s$, of length at most $\operatorname{len}(S)+r\rho$.

If $S$ is good, we have connected all ports, at a cost of at least $4-\Delta$ for a square or $\frac43$ for a fallback, proving (i) and (ii). If $S$ is bad, the network connects two or three ports of a square, requiring at least $\sqrt2$ or $1+\sqrt3$, respectively (\cref{fig:local-gadgets}). In either case, this is at least $\frac43(r-1)$, so
\[
    \operatorname{len}(S)\ge\frac43(r-1)-r\rho\ge\Bigl(\frac43-2\rho\Bigr)(r-1),
\]
where the last inequality uses $r\le2(r-1)$, proving (iii).
\end{proof}

We prove that for every short tree $T$ on $P_\rho$, there is a clean network of length at most $\operatorname{len}(T)+O(\rho L)$ by showing that the triples of the good square connectors form a matching whose clean network satisfies this bound.

\begin{lemma}
\label{lem:cleaning}
For every tree $T$ on $P_\rho$ with $\operatorname{len}(T)\le\frac73L$, there is a clean network of length at most $\operatorname{len}(T)+C_2\rho L$, where $C_2$ is an absolute constant.
\end{lemma}

\begin{proof}
Let $q$ be the number of good square connectors. If two of their triples shared an element $u$, both connectors would attach to $T_u$ and $T_0$, creating a cycle in the incidence tree. Their triples therefore form a matching, whose clean network has length $L+\frac43|U|-\Delta q$, which we will now compare with the length of $T$.

By \cref{lem:displaced-attachments}, each good square connector has $r_S-1=3$ and length at least
\[
    4-\Delta-4\rho\ge\Bigl(\frac43-2\rho\Bigr)(r_S-1)-\Delta,
\]
while every other connector has length at least $(\frac43-2\rho)(r_S-1)$. Summing over the connectors and using~\eqref{eq:connector-count} gives
\[
    \sum_S\operatorname{len}(S)\ge\Bigl(\frac43-2\rho\Bigr)|U|-\Delta q.
\]
The tree $\widehat T$ consists precisely of these connectors and the forest $F$, which has length at least $L$. Since we obtained $\widehat T$ from $T$ by adding length $O(\rho L)$ and then removing edges,
\[
    \operatorname{len}(T)+O(\rho L)\ge\operatorname{len}(\widehat T)\ge L+\frac43|U|-\Delta q-2\rho|U|.
\]
Thus the clean network of this matching has length at most $\operatorname{len}(T)+O(\rho L)+2\rho|U|$. Since $|U|<L$, this is at most $\operatorname{len}(T)+C_2\rho L$ for an absolute constant $C_2$.
\end{proof}

\begin{proof}[Proof of the lower bound in \cref{thm:cable-reduction}]
Let $T$ be an optimal tree on $P_\rho$. The clean network using only fallbacks gives $\operatorname{len}(T)\le L+\frac43|U|\le\frac73L$, since $|U|<L$. We can therefore apply \cref{lem:cleaning} to obtain a clean network of length at most $\OPT(P_\rho)+C_2\rho L$. Its $q$ squares form a matching, so $q\le\nu(H)$ and its length is at least $L+\frac43|U|-\Delta\nu(H)$. Rearranging gives the claimed lower bound.
\end{proof}

\FloatBarrier

\section{No PTAS in high dimensions}
\label{sec:growing}

We first route the cables in linear dimension, which by \cref{thm:cable-reduction} gives a constant approximation gap, and then reduce the dimension to $O(\log n)$.

\subsection{Orthogonal routing in linear dimension}
\label{sec:orthogonal-routing}

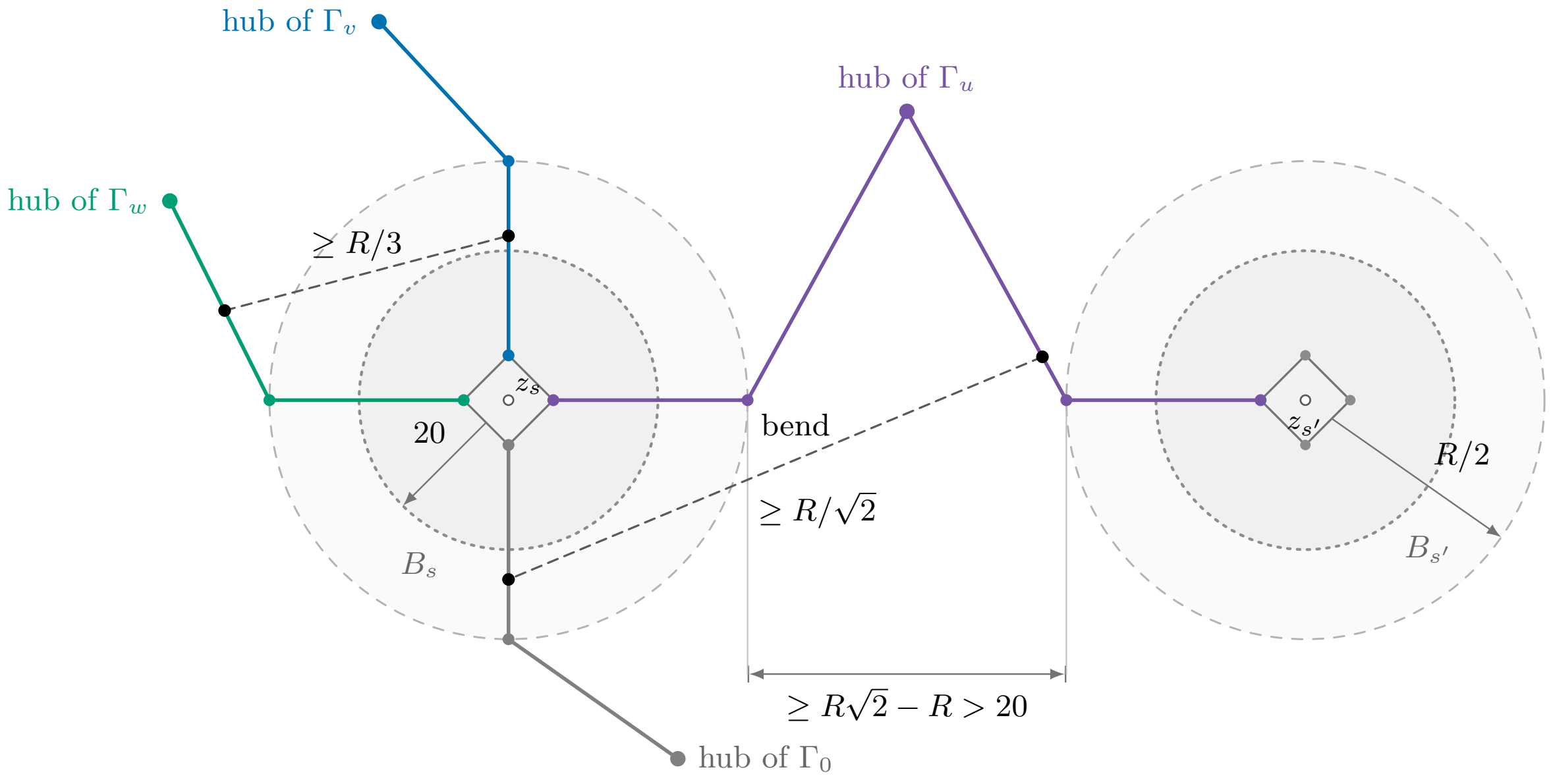
\begin{figure}[!b]
\centering
\definecolor{rsCableW}{HTML}{009E73}
\begin{tikzpicture}[line cap=round,line join=round,
  x=1cm,y=1cm,font=\small,>=latex,
  gseg/.style={draw=black,line width=1.05pt,line cap=round,line join=round},
  hseg/.style={draw=black,line width=1.05pt,line cap=round,line join=round},
  disk/.style={fill=black!2,draw=black!30,dashed,line width=.55pt},
  ball/.style={fill=black!6,draw=black!45,dotted,line width=.8pt},
  dist/.style={draw=black!65,densely dashed,line width=.6pt},
  dim/.style={draw=black!55,line width=.45pt,|<->|,>=latex},
  note/.style={font=\footnotesize},
  pt/.style={circle,fill=black,inner sep=0pt,minimum size=3.6pt}
]
\def\rb{1.5}   %
\def\rg{2.4}   %
\filldraw[disk] (0,0) circle[radius=\rg];
\filldraw[disk] (8,0) circle[radius=\rg];
\filldraw[ball] (0,0) circle[radius=\rb];
\filldraw[ball] (8,0) circle[radius=\rb];
\node[note,text=black!60,anchor=north east] at (-112:\rb) {$B_s$};
\node[note,text=black!60,anchor=north west] at ($(8,0)+(-55:\rb)$) {$B_{s'}$};
\draw[draw=black!55,line width=.45pt,<->] (0,0)--node[note,above left=-1pt] {$20$} (-135:\rb);
\draw[draw=black!55,line width=.45pt,<->] (8,0)--node[note,above right=-1pt,pos=.6] {$R/2$} ($(8,0)+(-35:\rg)$);
\draw[dim] (\rg,-2.75)--node[note,below] {$\ge R\sqrt2-R>20$} (8-\rg,-2.75);
\draw[draw=black!25,line width=.4pt] (\rg,0)--(\rg,-2.75) (8-\rg,0)--(8-\rg,-2.75);
\filldraw[fill=black!5,draw=black!55,line width=.6pt] (.45,0)--(0,.45)--(-.45,0)--(0,-.45)--cycle;
\draw[gseg,draw=cableU] (.45,0)--(\rg,0);       \draw[hseg,draw=cableU] (\rg,0)--(4,2.9);
\draw[gseg,draw=cableV] (0,.45)--(0,\rg);       \draw[hseg,draw=cableV] (0,\rg)--(-1.3,3.8);
\draw[gseg,draw=rsCableW] (-.45,0)--(-\rg,0);     \draw[hseg,draw=rsCableW] (-\rg,0)--(-3.4,2.0);
\draw[gseg,draw=black!50] (0,-.45)--(0,-\rg);   \draw[hseg,draw=black!50] (0,-\rg)--(1.7,-3.6);
\foreach \p/\c in {{(.45,0)}/cableU,{(0,.45)}/cableV,{(-.45,0)}/rsCableW,{(0,-.45)}/black!50,
                   {(\rg,0)}/cableU,{(0,\rg)}/cableV,{(-\rg,0)}/rsCableW,{(0,-\rg)}/black!50}
  \fill[\c] \p circle[radius=1.7pt];
\filldraw[fill=white,draw=black!65,line width=.5pt] (0,0) circle[radius=1.4pt];
\node[font=\scriptsize,anchor=south west,inner sep=1pt] at (.02,.02) {$z_s$};
\fill[cableU] (4,2.9) circle[radius=2.2pt];
\node[note,text=cableU,above=2pt] at (4,2.9) {hub of $\Gamma_u$};
\fill[cableV] (-1.3,3.8) circle[radius=2.2pt];
\node[note,text=cableV,left=2pt] at (-1.3,3.8) {hub of $\Gamma_v$};
\fill[rsCableW] (-3.4,2.0) circle[radius=2.2pt];
\node[note,text=rsCableW,left=2pt] at (-3.4,2.0) {hub of $\Gamma_w$};
\fill[black!50] (1.7,-3.6) circle[radius=2.2pt];
\node[note,text=black!60,right=2pt] at (1.7,-3.6) {hub of $\Gamma_0$};
\node[note,below right=0pt] at (\rg,0) {bend};
\coordinate (X3) at ($(-\rg,0)!.45!(-3.4,2.0)$);
\coordinate (Y3) at (0,1.65);
\draw[dist] (X3)--(Y3);
\node[pt] at (X3) {}; \node[pt] at (Y3) {};
\node[note,anchor=south,inner sep=1pt] at ($(X3)!.5!(Y3)+(-.1,.08)$) {$\ge R/3$};
\begin{scope}[shift={(8,0)}]
  \filldraw[fill=black!5,draw=black!55,line width=.6pt] (.45,0)--(0,.45)--(-.45,0)--(0,-.45)--cycle;
  \foreach \p in {(0,.45),(.45,0),(0,-.45)} \fill[black!45] \p circle[radius=1.5pt];
  \filldraw[fill=white,draw=black!65,line width=.5pt] (0,0) circle[radius=1.4pt];
  \node[font=\scriptsize,below=1pt] at (0,0) {$z_{s'}$};
\end{scope}
\draw[gseg,draw=cableU] (7.55,0)--(5.6,0);
\draw[hseg,draw=cableU] (5.6,0)--(4,2.9);
\fill[cableU] (7.55,0) circle[radius=1.7pt];
\fill[cableU] (5.6,0) circle[radius=1.7pt];
\coordinate (X1) at ($(5.6,0)!.15!(4,2.9)$);
\coordinate (Y1) at (0,-1.8);
\draw[dist] (X1)--(Y1);
\node[pt] at (X1) {}; \node[pt] at (Y1) {};
\node[note,anchor=north west,inner sep=1pt] at ($(X1)!.55!(Y1)+(.05,-.08)$) {$\ge R/\sqrt2$};
\end{tikzpicture}
\caption{Orthogonal routing near gadgets $s$ and $s'$. (a)~Gadget segments lie within distance $R/2$ of their centre (dashed), keeping segments of distinct gadgets more than $20$ apart. (b)~Hub segments are at distance at least $R/3>20$ from every other cable. The cables inside $B_s$ (dotted) therefore consist exactly of the gadget segments of $s$.}
\label{fig:routing-separation}
\end{figure}

By \cref{thm:cable-reduction}, it suffices to construct a separated cable routing of total length $O(m)$. We reserve a distinct coordinate direction for each gadget and each cable, placing the gadgets far apart on their axes and joining the ports of each cable to a common \emph{hub} on its axis. Different cables then come close only near a gadget they share.

Concretely, besides the directions $b_1,b_2$ in which the gadgets are drawn, reserve a unit coordinate vector $e_s$ for each gadget $s$ and $f_i$ for each label $i\in U\cup\{0\}$, all mutually orthogonal; this uses $D=m+|U|+3$ dimensions. Set $R=64$. We place each gadget $s$ at $z_s=Re_s$, and let the hub of $\Gamma_i$ be $Rf_i$. Each port labelled $i$ has the form $p_{s,i}=Re_s+\alpha v$, where $v\in\{\pm b_1,\pm b_2\}$ and $\alpha=1$ for a square or $\alpha=2/3$ for a fallback. The cable has a \emph{gadget segment} from this port in direction $v$ up to the \emph{bend} $Re_s+\frac R2v$, followed by a \emph{hub segment} from the bend straight to the hub, which together form the \emph{branch}
\[
    Re_s+\alpha v\;\longrightarrow\;Re_s+\frac R2v\;\longrightarrow\;Rf_i.
\]
The cable $\Gamma_i$ is the union of these branches over all ports labelled $i$; we call this the \emph{orthogonal routing} (\cref{fig:high-dimensional-routing}).

\begin{lemma}
\label{lem:orthogonal-routing}
The cables of the orthogonal routing are separated and have total length $L=O(m)$.
\end{lemma}

\begin{proof}
We first establish two distance bounds, illustrated in \cref{fig:routing-separation}.
\begin{enumerate}[(a)]
\item The gadget segments of $s$ extend radially away from $z_s$ in the directions $\pm b_1,\pm b_2$ and end at distance $R/2$ from it. Since gadget centres are $R\sqrt2$ apart, segments from distinct gadgets are at least $R\sqrt2-R>20$ apart.
\item Every hub segment is at distance at least $R/3>21$ from every other cable. To see this, write a point on the hub segment from gadget $s$ to $Rf_i$ as
\[
    x=(1-t)\Bigl(Re_s+\frac R2v\Bigr)+tRf_i,
    \qquad 0\le t\le1.
\]
Its $e_s$-, $f_i$- and $v$-coordinates are $(1-t)R$, $tR$ and $(1-t)R/2$. A point $y$ on another cable has zero $f_i$-coordinate. If it does not lie on a branch from $s$, its $e_s$-coordinate is also zero, so
\[
    \|x-y\|\ge\|((1-t)R,tR)\|\ge R/\sqrt2.
\]
Otherwise, the branches containing $x$ and $y$ share the same gadget $s$. Since they belong to different cables, they use different ports of $s$, so the branch containing $y$ leaves in a direction orthogonal or opposite to $v$ and has nonpositive $v$-coordinate. The $f_i$- and $v$-coordinates then give
\[
    \|x-y\|\ge\max\Bigl\{tR,(1-t)\frac R2\Bigr\}\ge R/3.
\]
\end{enumerate}

(S1) Since the centres are $R\sqrt2>60$ apart, the balls $B_s$ are more than $20$ apart. No hub segment enters any of these balls: the coordinates in~(b) give distance at least $R/3>20$ from the centre of its gadget and at least $R$ from every other centre. By~(a), gadget segments from other gadgets also stay outside $B_s$. Thus inside $B_s$, the cables are exactly the straight segments from the ports of $s$ to the boundary. By~(a) and~(b), two points on different cables less than $20$ apart must lie on gadget segments of the same gadget. These segments run in orthogonal or opposite directions from its centre, so both points lie within distance $20$ of it.

(S2) Gadget segments have length $R/2-\alpha\ge31$, and hub segments have length $96$. The branches of each cable meet only at its hub, so the cable is a tree. A gadget segment from $s$ has $e_s$-coordinate $R$, whereas all other branches of its cable have $e_s$-coordinate zero, giving distance at least $R$ between them. Every pair of non-incident segments contains such a gadget segment, so their distance is at least $R$. The angles $\varphi$ at bends and $\varphi'$ at hubs satisfy
\[
    \cos\varphi=\frac{32}{96},
    \qquad
    \cos\varphi'=\frac{32^2\langle v,v'\rangle+64^2}{96^2}\le\frac59,
\]
so both are at least $45^\circ$.

Finally, each of the $4|E|+2|U|$ branches has length less than $128$, giving total cable length $L=O(m)$.
\end{proof}

\begin{lemma}
\label{lem:linear-dimensional-gap}
There are constants $\beta>0$ and $\rho>0$ such that the following holds. Given an instance $H$ of 3DM-3, one can construct in polynomial time $n=\Theta(m)$ terminals $P_\rho\subseteq\BR^{O(n)}$ and a threshold $\tau$ with
\[
    \nu(H)=r\;\Longrightarrow\;\OPT(P_\rho)\le\tau,
    \qquad
    \nu(H)\le(1-\delta)r\;\Longrightarrow\;\OPT(P_\rho)\ge(1+\beta)\tau.
\]
Consequently, Euclidean Steiner Tree is APX-hard already in $O(n)$ dimensions.
\end{lemma}

\begin{proof}
Construct the cables of the orthogonal routing. By \cref{lem:orthogonal-routing}, they are separated and $L\le C_1m$ for a constant $C_1$. Fix a constant
\[
    \rho\le\min\Bigl\{\frac1{100},\frac{\Delta\delta}{12C_1C_2}\Bigr\},
\]
where $C_2$ is the constant in \cref{thm:cable-reduction}, and set $\tau=L+\frac43|U|-\Delta r$.

The first implication is the upper bound in \cref{thm:cable-reduction}. For the second, let $\beta=\frac{\Delta\delta}{2(6C_1+4)}$. Then
\begin{align*}
    \OPT(P_\rho)
    &\ge L+\tfrac43|U|-\Delta\nu(H)-C_2\rho L
        &&\text{by the lower bound in \cref{thm:cable-reduction},}\\
    &\ge L+\tfrac43|U|-\Delta r+\Delta\delta r-C_2\rho L
        &&\text{since }\nu(H)\le(1-\delta)r,\\
    &=\tau+\Delta\delta r-C_2\rho L
        &&\text{by the definition of }\tau,\\
    &\ge\tau+\tfrac{\Delta\delta}{2}r
        &&\text{since }L\le C_1m\le6C_1r\text{ and }\rho\le\tfrac{\Delta\delta}{12C_1C_2},\\
    &=\tau+\beta(6C_1+4)r
        &&\text{by the definition of }\beta,\\
    &\ge(1+\beta)\tau
        &&\text{since }\tau\le L+\tfrac43|U|\le(6C_1+4)r.
\end{align*}

Since $\rho$ is constant, $n=\Theta(L/\rho^2)=\Theta(m)$ and $D=m+|U|+3=O(n)$.
\end{proof}

\subsection{Reducing to logarithmic dimension}
\label{sec:main-dimension-reduction}

Writing $\beta$ for the gap constant in \cref{lem:linear-dimensional-gap}, we now reduce the dimension of the terminal set $P_\rho=\{p_1,\ldots,p_n\}$ constructed there, making no further use of the cables.

The Johnson--Lindenstrauss lemma~\cite{JohnsonL84} states that, for every $0<\eta<1$, any $n$ points in a Euclidean space can be mapped into $O(\eta^{-2}\log n)$ dimensions so that, after rescaling, every pairwise distance changes by a factor between $1$ and $1+\eta$. Using the deterministic construction of Engebretsen, Indyk, and O'Donnell~\cite{DBLP:conf/soda/EngebretsenIO02}, we obtain in polynomial time points $q_1,\ldots,q_n\in\BR^k$, $k=O(\beta^{-2}\log n)=O(\log n)$, with
\begin{equation}
\label{eq:jl-distances}
    \|p_i-p_j\|\le\|q_i-q_j\|\le\Bigl(1+\frac\beta2\Bigr)\|p_i-p_j\|
    \qquad\text{for all }i,j.
\end{equation}

We follow Karthik C.~S.'s argument in Woodruff et al.~\cite[Section~4.2]{DBLP:journals/corr/abs-2602-03837}, using Kirszbraun's theorem to compare optimal Steiner tree lengths. Since these bounds apply only to distances between terminals, we use Kirszbraun's extension theorem~\cite{Kirszbraun34} to extend the map to Steiner points. For $A\subseteq\BR^D$, a map $f\colon A\to\BR^{D'}$ is $\lambda$-Lipschitz if $\|f(x)-f(y)\|\le\lambda\|x-y\|$ for all $x,y\in A$. By Kirszbraun's theorem, $f$ extends to a $\lambda$-Lipschitz map $F\colon\BR^D\to\BR^{D'}$, that is,
\[
    F(a)=f(a)\quad\text{for all }a\in A,
    \qquad
    \|F(x)-F(y)\|\le\lambda\|x-y\|\quad\text{for all }x,y\in\BR^D.
\]
In particular, if $f$ maps a set $A$ of terminals to a lower-dimensional space and stretches distances between terminals by a factor of at most $\lambda$, then the extension $F$ maps also every Steiner point to the lower-dimensional space and stretches distances between a terminal and a Steiner point, or between two Steiner points, by at most the same factor. Applying $F$ to the vertices of a tree and joining the mapped vertices by straight segments therefore yields a connected network whose length is at most $\lambda$ times that of the original tree.

\begin{lemma}
\label{lem:terminal-comparison}
Let $P=\{p_1,\ldots,p_n\}$ and $Q=\{q_1,\ldots,q_n\}$ be finite point sets in Euclidean spaces of possibly different dimensions. If, for some $\lambda\ge1$, their pairwise distances satisfy $\|p_i-p_j\|\le\|q_i-q_j\|\le\lambda\|p_i-p_j\|$ for all $i,j$, then
\[
    \OPT(P)\le\OPT(Q)\le\lambda\OPT(P).
\]
\end{lemma}

\begin{proof}
We prove the two inequalities by applying Kirszbraun's theorem in opposite directions. First, the map $q_i\mapsto p_i$ is $1$-Lipschitz by the left-hand inequality. Extending this map to the Steiner points of an optimal tree on $Q$ gives a network on $P$ of length at most $\OPT(Q)$, so $\OPT(P)\le\OPT(Q)$. Conversely, the map $p_i\mapsto q_i$ is $\lambda$-Lipschitz by the right-hand inequality. Extending this map to the Steiner points of an optimal tree on $P$ gives a network on $Q$ of length at most $\lambda\OPT(P)$, so $\OPT(Q)\le\lambda\OPT(P)$.
\end{proof}

\begin{proof}[Proof of \cref{thm:growing-dimension-hardness}]
Apply \cref{lem:linear-dimensional-gap} to obtain $P_\rho$ and $\tau$, and construct $Q$ as above in $O(\log n)$ dimensions. Combining \eqref{eq:jl-distances} with \cref{lem:terminal-comparison} for $\lambda=1+\beta/2$ gives
\[
    \text{YES:}\quad\OPT(Q)\le\Bigl(1+\frac\beta2\Bigr)\tau,
    \qquad
    \text{NO:}\quad\OPT(Q)\ge(1+\beta)\tau.
\]
We may assume $0<\beta\le1$. A $(1+\beta/4)$-approximation distinguishes these cases: in a YES instance, it returns a network of length at most
\[
    \Bigl(1+\frac\beta4\Bigr)\Bigl(1+\frac\beta2\Bigr)\tau
    <(1+\beta)\tau,
\]
whereas in a NO instance every network has length at least $(1+\beta)\tau$. Thus the reduction preserves a constant approximation gap in $O(\log n)$ dimensions, proving the theorem.
\end{proof}

\subsection{Padding to doubly logarithmic dimension}
\label{sec:main-padding}

\begin{proof}[Proof of \cref{thm:loglog-hardness}]
Let $Q=\{q_1,\ldots,q_n\}$ be a terminal set obtained by the reduction in the proof of \cref{thm:growing-dimension-hardness}. As shown there, a $(1+\beta/4)$-approximation distinguishes the source matching instances. By \cref{lem:gap-source}(iii) and the polynomial size of the reduction, under $\mathrm{NP}\not\subseteq\mathrm{SUBEXP}$ there is a constant $c>0$ such that no such approximation runs in time $2^{n^c}$.

To pad $Q$, set $N=2^{\lceil n^{c/2}\rceil}$, which is at least $n$ for sufficiently large $n$, and add $N-n$ terminals at the position of one original terminal. This leaves the optimum and the approximation gap unchanged. The padded instance has $N$ terminals in $O(\log n)=O(\log\log N)$ dimensions, since $\log N\ge n^{c/2}$ implies $\log n\le\frac2c\log\log N$.

Suppose Euclidean Steiner Tree admits a PTAS in $O(\log\log N)$ dimensions for $N$ terminals. Running it on the padded instance with the same accuracy would distinguish the source cases in time $N^{O(1)}=2^{O(n^{c/2})}$, contradicting the $2^{n^c}$ lower bound.
\end{proof}

\section{Lower bounds in fixed dimension \texorpdfstring{$d\ge3$}{d >= 3}}
\label{sec:fixed}

We prove \cref{thm:fixed-dimension-hardness} for fixed $d\ge3$. We use the gadgets and cable reduction (\cref{thm:cable-reduction}) from \cref{sec:cable}, changing only the routing. In $\BR^d$, the cables cannot all be short, so we route them along a grid drawing of total length $L=O_d(m^{d/(d-1)})$. Choosing $\rho=\Theta(m/L)$ keeps the sampling error $O(\rho L)$ in \cref{thm:cable-reduction} below half the matching gap, so the YES and NO optima remain $\Omega(m)$ apart. Since both are $\Theta(L)$, the relative gap is $\Omega(m/L)$, and hence $\Omega_d(m^{1/(1-d)})$.

\subsection{Cable routing by cube wiring}
\label{sec:cube-wiring}

We use the following consequence of the Cube Wiring Theorem of de Berg, Bodlaender, Kisfaludi-Bak, Marx, and van der Zanden~\cite{deBergBBMZ20}.

\begin{theorem}[{\cite[Theorem~31]{deBergBBMZ18arxiv}}]
\label{thm:cube-wiring}
There exists a constant $c$ such that for all $d\ge3$, any graph with $n$ vertices and maximum degree $2d$ is the topological minor of the $d$-dimensional grid hypercube of side length $cn^{1/(d-1)}$.
\end{theorem}

A graph $G$ is a topological minor of the grid precisely when the grid contains a subdivision of $G$. The theorem therefore gives a \emph{grid drawing} in $\{0,\ldots,S\}^d$, where $S=\lceil cn^{1/(d-1)}\rceil$: vertices lie at distinct grid points, and edges follow internally disjoint grid paths. The construction in~\cite{deBergBBMZ18arxiv} finds such a drawing in polynomial time.

We construct a graph $G$ with a vertex for every gadget. We then compute a grid drawing of $G$ using \cref{thm:cube-wiring} and modify it to create the actual gadgets and cables connecting their appropriate ports.

\paragraph{The graph.}
Let $k_i\ge2$ be the number of gadgets containing a port labelled $i$. For an element $u$, these are its fallback and the squares for triples containing $u$; for the backbone label $0$, they are all gadgets. Construct a tree $T_i$ whose leaves are exactly these gadgets and whose internal vertices have degree three. Take the union of these trees, which share only gadget vertices, and subdivide every edge twice to obtain $G$, ensuring that each original edge of a tree $T_i$ follows a path of at least three unit grid edges in the grid drawing. We modify the routing only along the first grid edge of each path leaving a gadget, leaving an unchanged grid edge between any two modified parts of a cable; we use this to prove (S2). In $G$, square gadgets have degree four, fallback gadgets degree two, and all other vertices degree at most three. Since $\sum_i k_i=4|E|+2|U|$, the total number of vertices is $n=O(m)$.

\begin{figure}[!tb]
\centering
\definecolor{cwCableW}{HTML}{009E73}
\begin{tikzpicture}[line cap=round,line join=round,
  x={(2.35cm,0cm)},y={(.72cm,.85cm)},z={(0cm,2.4cm)},>=latex,font=\small,
  grid/.style={draw=black!25,line width=.45pt},
  gridoff/.style={draw=black!14,line width=.4pt},
  replaced/.style={draw=black!35,densely dashed,line width=.5pt},
  gadget/.style={fill=black!8,draw=black!70,line width=.7pt},
  vtx/.style={draw=black!75,fill=white,line width=.8pt,inner sep=0pt,minimum size=6.5pt},
  gpt/.style={draw=black!55,fill=white,line width=.5pt,inner sep=0pt,minimum size=4pt},
  port/.style={circle,fill=black!80,inner sep=0pt,minimum size=3.2pt},
  cable/.style={draw=black,line width=1.05pt,line cap=round,line join=round},
  dirn/.style={->,line width=.65pt},
  ball/.style={draw=black!50,dotted,line width=.75pt},
  note/.style={font=\footnotesize},
  lead/.style={draw=black!45,line width=.4pt},
  dim/.style={draw=black!60,line width=.45pt,|<->|,>=latex}
]
\def\e{1.2}     %
\def\h{.17}     %
\def\r{.36}     %
\newcommand{\localgrid}{%
  \foreach \t in {-1,0,1} {
    \draw[grid] (-\e,\t,0)--(\e,\t,0);
    \draw[grid] (\t,-\e,0)--(\t,\e,0);
  }
  \draw[gridoff] (0,0,-\e)--(0,0,\e);
  \foreach \p in {(1,0,0),(-1,0,0),(0,1,0),(0,-1,0),(0,0,1),(0,0,-1)} \node[gpt] at \p {};}
\begin{scope}
\localgrid
\draw[cable,draw=black!55] (0,0,0)--(0,-\e,0);
\draw[cable,draw=cableV] (0,0,0)--(0,0,-\e);
\draw[cable,draw=cableU] (0,0,0)--(-\e,0,0);
\draw[cable,draw=cwCableW] (0,0,0)--(0,\e,0);
\node[vtx] at (0,0,0) {};
\node[note,anchor=north west,inner sep=2pt] at (.05,-.05,0) {$z_s$};
\draw[dirn,draw=cableU] (-.3,.22,0)--(-.8,.22,0) node[note,above,pos=.5] {$\delta_u$};
\draw[dirn,draw=cwCableW] (.22,.32,0)--(.22,.82,0) node[note,right,pos=.55] {$\delta_w$};
\draw[dirn,draw=black!60] (-.24,-.32,0)--(-.24,-.82,0) node[note,left,pos=.55] {$\delta_0$};
\draw[dirn,draw=cableV] (.12,0,-.25)--(.12,0,-.75) node[note,right,pos=.5] {$\delta_v$};
\draw[dim] (-.1,0,0)--node[note,left=-1pt] {$100$} (-.1,0,1);
\begin{scope}[shift={(-1.15,-.9,-1.0)},->,line width=.5pt,font=\footnotesize]
  \draw (0,0,0)--(.3,0,0) node[right=-1pt] {$b_1$};
  \draw (0,0,0)--(0,.45,0) node[above right=-2pt] {$b_2$};
  \draw (0,0,0)--(0,0,.28) node[above=-1pt] {$e_3$};
\end{scope}
\node at (0,-.7,-1.45) {(a) Grid drawing};
\end{scope}
\draw[->,draw=black!55,line width=.9pt] (1.45,0,0)--(1.95,0,0);
\begin{scope}[shift={(3.45,0,0)}]
\localgrid
\draw[replaced] (0,0,0)--(0,0,-.5);
\draw[cable,draw=cableV] (\h,0,0)--(.5,0,0)--(.5,0,-.5)--(0,0,-.5)--(0,0,-\e);
\draw[cable,draw=black!55] (0,-\h,0)--(0,-\e,0);
\draw[cable,draw=cableU] (-\h,0,0)--(-\e,0,0);
\draw[cable,draw=cwCableW] (0,\h,0)--(0,\e,0);
\filldraw[gadget] (-\h,0,0)--(0,-\h,0)--(\h,0,0)--(0,\h,0)--cycle;
\foreach \p in {(-\h,0,0),(0,-\h,0),(\h,0,0),(0,\h,0)} \node[port] at \p {};
\draw[ball] (0,0,0) circle[radius=\r*2.4cm];
\draw[ball] plot[domain=0:360,samples=72,smooth] ({\r*cos(\x)},{\r*sin(\x)},0);
\node[note,text=black!60,anchor=south east,inner sep=1pt] at ($(0,0,0)+(125:\r*2.4cm)$) {$B_s$};
\foreach \p in {(.5,0,0),(.5,0,-.5),(0,0,-.5)} \fill[cableV] \p circle[radius=1.4pt];
\node[note,anchor=north west,inner sep=1pt] at (.53,0,-.02) {$z_s+25b_1$};
\node[note,anchor=west,inner sep=1pt] at (.54,0,-.5) {$z_s+25b_1+50\delta_v$};
\node[note,text=cableU,anchor=south] at (-1.0,0,0) {$\Gamma_u$};
\node[note,text=cwCableW,anchor=south east] at (-.02,1.15,0) {$\Gamma_w$};
\node[note,text=black!60,anchor=west] at (.08,-1.1,0) {$\Gamma_0$};
\node[note,text=cableV,anchor=east] at (-.05,0,-1.15) {$\Gamma_v$};
\draw[lead] (\h,0,0)--(.78,0,.2) node[note,anchor=west,inner sep=1pt] {$p_{s,v}=z_s+b_1$};
\node[note,anchor=west] at (.06,0,-1) {$z_s+100\delta_v$};
\node[note,anchor=east] at (-.06,0,-.5) {$z_s+50\delta_v$};
\node at (0,-.7,-1.45) {(b) Cable routing};
\end{scope}
\end{tikzpicture}
\caption{Wiring a square gadget. (a) Scaled grid drawing around $z_s$. (b) $\Gamma_u,\Gamma_w,\Gamma_0$ continue straight from their ports; $\Gamma_v$ leaves along $b_1$ and rejoins its grid segment at $z_s+50\delta_v$. Inside the dotted ball $B_s$, the cables follow the gadget axes, as required by (S1).}
\label{fig:cube-wiring}
\end{figure}
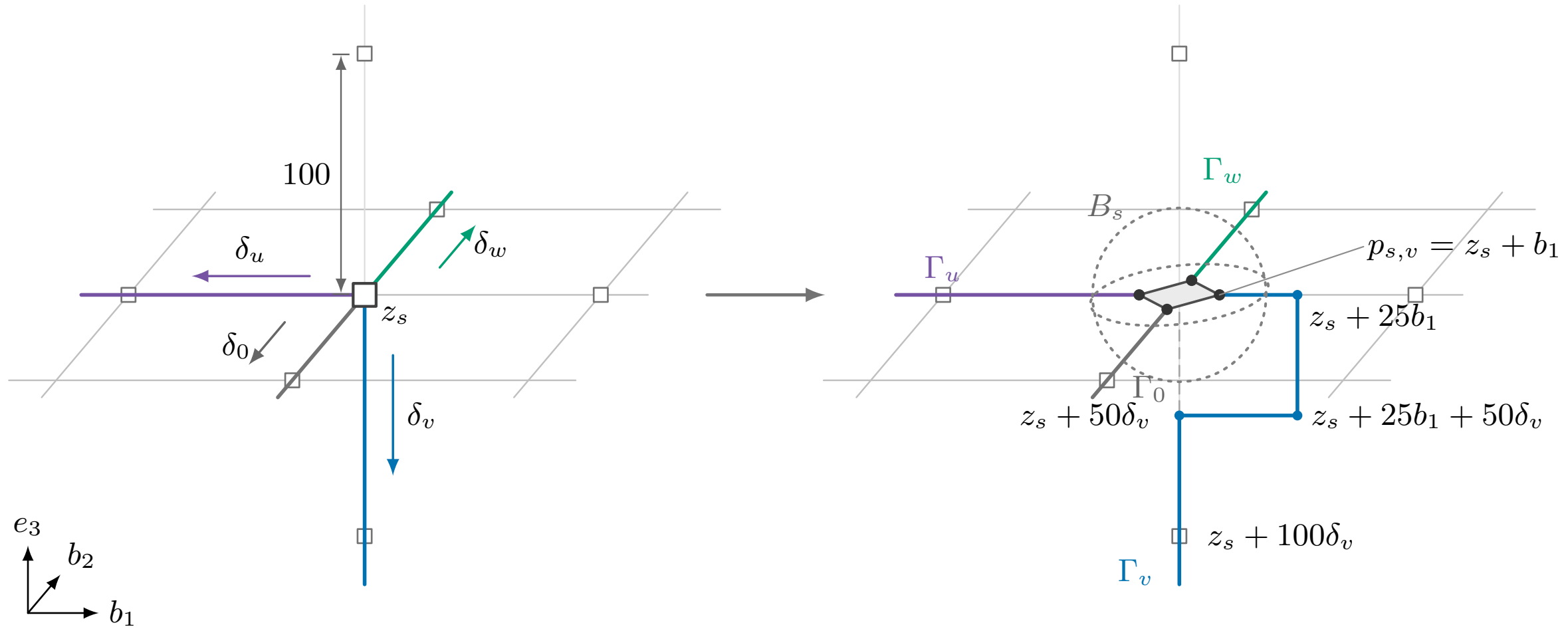

\paragraph{The cables.}
We apply \cref{thm:cube-wiring} to compute a grid drawing of $G$ and scale it by $100$. This assigns each gadget $s$ a grid point $z_s$. Each scaled unit edge is a \emph{grid segment} of length $100$, and grid segments without a common endpoint are at least $100$ apart.

We centre each gadget $s$ at $z_s$ and modify each drawn tree $T_i$ near the gadgets to attach it to the appropriate ports. The resulting cables $\Gamma_i$ form the \emph{cable routing} (\cref{fig:cube-wiring}).

For each label $i$ of a gadget $s$, the drawn tree $T_i$ leaves $z_s$ along the grid segment $(z_s,z_s+100\delta_i)$, where the $\delta_i$ are distinct signed coordinate vectors. We draw the gadgets in planes parallel to the $b_1b_2$-plane, with ports in directions $\pm b_1,\pm b_2$ for a square and $\pm b_1$ for a fallback. We assign each label $i$ a port $p_{s,i}$ and write $\hat c_i\in\{\pm b_1,\pm b_2\}$ for its direction from $z_s$; the assignment and local modifications are as follows (\cref{fig:cube-wiring}(b)).
\begin{enumerate}[(i)]
\item If $\delta_i$ is an available port direction, assign label $i$ the port in that direction, so $\hat c_i=\delta_i$. Replace the grid segment by the straight segment from $p_{s,i}$ to $z_s+100\delta_i$; for example, $\Gamma_u$ has $\delta_u=\hat c_u=-b_1$ in the figure.
\item Otherwise, assign label $i$ one of the remaining ports, so $\hat c_i\perp\delta_i$; for example, $\Gamma_v$ has $\delta_v=-e_3$ and $\hat c_v=b_1$ in the figure. The cable follows $\hat c_i$ from its port to $z_s+25\hat c_i$, continues parallel to the grid segment, and joins it at its midpoint:
\[
    p_{s,i}\;\longrightarrow\;z_s+25\hat c_i\;\longrightarrow\;z_s+25\hat c_i+50\delta_i\;\longrightarrow\;z_s+50\delta_i\;\longrightarrow\;z_s+100\delta_i.
\]
\end{enumerate}
At each port, the cable begins with a segment in direction $\hat c_i$ of length at least $24>20$.

Each local modification adds at most $50$ to a grid segment of length $100$. Since the grid paths are edge-disjoint and the cube contains at most $d(S+1)^d$ unit grid edges, the total cable length satisfies
\[
    L\le150d(S+1)^d=O_d(m^{d/(d-1)}).
\]
The cables can be constructed in polynomial time, and the following lemma shows that they are separated, with its proof deferred to \cref{app:wired-cables}.

\begin{lemma}
\label{lem:wired-cables}
For every fixed $d\ge3$, the cable routing in $\BR^d$ consists of separated cables.
\end{lemma}

For Euclidean Cable Steiner Tree, the minimum connection cost is $\frac43|U|-\Delta\nu(H)$, independently of $L$ (\cref{sec:cable-recovery-proof}). Thus the cable routing gives a constant approximation gap in fixed dimension.

\begin{corollary}
\label{cor:cable-apx}
For every fixed $d\ge3$, Euclidean Cable Steiner Tree is APX-hard in $\BR^d$.
\end{corollary}

\begin{proof}
Given an instance $H$ of 3DM-3, construct the cables above. They are separated by \cref{lem:wired-cables} and have minimum connection cost $\frac43|U|-\Delta\nu(H)$. Since $|U|=3r$, this cost equals $(4-\Delta)r$ if $\nu(H)=r$ and is at least $(4-\Delta+\Delta\delta)r$ if $\nu(H)\le(1-\delta)r$. By \cref{lem:gap-source}, approximating it within the constant factor $1+\Delta\delta/(4-\Delta)$ is therefore NP-hard.
\end{proof}

This gives a negative answer for $d\ge3$ to a question of Bartlmae, J\"unger, and Langetepe~\cite{DBLP:journals/corr/abs-2412-07046}. They gave a planar PTAS for a line and point terminals, and asked whether it extends to segment terminals. It remains open whether Euclidean Cable Steiner Tree is APX-hard in the plane or admits a PTAS.

\subsection{The gap in fixed dimension}
\label{sec:fixed-gap}

\begin{lemma}
\label{lem:fixed-gap}
There are constants $a>0$ and $c_d>0$ such that, given an instance $H$ of 3DM-3, one can compute in polynomial time a set $P\subseteq\BR^d$ of $O_d(m^{(d+2)/(d-1)})$ terminals and a threshold $\tau\le c_dm^{d/(d-1)}$ satisfying
\[
    \nu(H)=r\;\Longrightarrow\;\OPT(P)\le\tau,
    \qquad
    \nu(H)\le(1-\delta)r\;\Longrightarrow\;\OPT(P)\ge\tau+am.
\]
\end{lemma}

\begin{proof}
Use the cable routing from \cref{sec:cube-wiring}, which is separated by \cref{lem:wired-cables} and has total cable length $L\le C_dm^{d/(d-1)}$. Let $C_2$ be the constant in \cref{thm:cable-reduction} and choose
\[
    \rho=\min\Bigl\{\frac1{100},\frac{\Delta\delta m}{12C_2L}\Bigr\},
\]
so that $C_2\rho L\le\frac{\Delta\delta}{12}m$, and set $P=P_\rho$ and $\tau=L+\frac43|U|-\Delta r$.

The YES bound follows from the upper bound in \cref{thm:cable-reduction}. For the NO bound, set $a=\frac{\Delta\delta}{12}$. Then
\begin{align*}
    \OPT(P)
    &\ge L+\tfrac43|U|-\Delta\nu(H)-C_2\rho L
        &&\text{by the lower bound in \cref{thm:cable-reduction},}\\
    &\ge L+\tfrac43|U|-\Delta r+\Delta\delta r-C_2\rho L
        &&\text{since }\nu(H)\le(1-\delta)r,\\
    &\ge\tau+\tfrac{\Delta\delta}6m-C_2\rho L
        &&\text{since }m\le6r,\\
    &\ge\tau+am
        &&\text{since }C_2\rho L\le\tfrac{\Delta\delta}{12}m.
\end{align*}

The threshold satisfies $\tau\le L+\frac43m\le c_dm^{d/(d-1)}$. Since $1/\rho=O_d(L/m+1)=O_d(m^{1/(d-1)})$, sampling gives $|P|=O(L/\rho^2)=O_d(m^{(d+2)/(d-1)})$.
\end{proof}

\begin{proof}[Proof of \cref{thm:fixed-dimension-hardness} for $d\ge3$]
Choose $\varepsilon=(a/(4c_d))m^{1/(1-d)}$ and apply \cref{lem:fixed-gap} to the matching instance. The same gap comparison applies, with $\varepsilon^{-(d-1)}=\Theta_d(m)=\Theta_d(r)$ and polynomially many terminals. Under weak ETH, running time $2^{O(\varepsilon^{-\gamma(d-1)})}\poly(n)$ for a fixed $0<\gamma<1$ would solve the matching gap in time $2^{O(r^\gamma)}$, contradicting \cref{lem:gap-source}.

Under Gap-ETH, the same argument would have ruled out time $2^{o(\varepsilon^{-(d-1)})}\poly(n)$, since this would yield a $2^{o(r)}$-time algorithm for the matching gap, contrary to \cref{lem:gap-source}(i). Under $\mathrm{NP}\not\subseteq\mathrm{SUBEXP}$, take $c>0$ from \cref{lem:gap-source}(iii). Running time $2^{o(\varepsilon^{-c(d-1)})}\poly(n)$ would give a $2^{o(r^c)}$-time algorithm for the matching gap, contradicting this bound. These arguments establish both bounds stated after \cref{thm:fixed-dimension-hardness}, with $c$ independent of $d$ because it depends only on the source matching problem.
\end{proof}

\section{Lower bounds in the plane}
\label{sec:plane-cables}

We now prove \cref{thm:fixed-dimension-hardness} for $d=2$, using the planar construction outlined in \cref{sec:overview-plane}. The main step is the following analogue of \cref{lem:fixed-gap}.

\begin{restatable}{lemma}{PlaneCablesGap}
\label{lem:plane-cables-gap}
There are constants $a>0$ and $c$ such that, given an instance $H$ of 3DM-3, one can compute in polynomial time a set $P\subseteq\BR^2$ of $O(m^4)$ terminals and a threshold $\tau\le cm^2$ satisfying
\[
    \nu(H)=r\;\Longrightarrow\;\OPT(P)\le\tau,
    \qquad
    \nu(H)\le(1-\delta)r\;\Longrightarrow\;\OPT(P)\ge\tau+am.
\]
\end{restatable}

\subsection{The construction}
\label{sec:plane-cables-construction}
\label{sec:plane-cables-idea}

We recall the gadgets and cable connections of \cref{sec:overview-plane}, illustrated in \cref{fig:overview-plane-layout}, before giving their geometric placement in the next subsection.

\begin{figure}[H]
\centering
\begin{tikzpicture}[line cap=round,line join=round,x=1.2cm,y=1.2cm,font=\normalsize,
  row/.style={draw=black!38,line width=1.05pt,line cap=round,line join=round},
  connection/.style={draw=connectionGreen,line width=.95pt,line cap=round,line join=round},
  guide/.style={draw=black!22,densely dashed,line width=0.4pt},
  active/.style={circle,fill=black!30,inner sep=0pt,minimum size=3.4pt},
  steiner/.style={draw=none,fill=none,inner sep=0pt,minimum size=0pt},
  caption/.style={align=center}]
\foreach \sx/\heading in {0/{(a)},2.5/{(b)}} {
  \begin{scope}[xshift=\sx*1.2cm]
    \node at (0,1.3) {\heading};
    \draw[guide] (0,0.5774)--(-0.5,-0.2887)--(0.5,-0.2887)--cycle;
    \draw[row] (0,0.5774)--++(0,0.3);
    \draw[row] (-0.5,-0.2887)--++(-0.26,-0.15);
    \draw[row] (0.5,-0.2887)--++(0.26,-0.15);
    \foreach \p in {(0,0.5774),(-0.5,-0.2887),(0.5,-0.2887)} \node[active] at \p {};
  \end{scope}
}
\draw[connection] (0,0.5774)--(-0.5,-0.2887);
\foreach \p in {(0,0.5774),(-0.5,-0.2887)} \fill \p circle[radius=2pt];
\begin{scope}[xshift=2.5*1.2cm]
  \draw[connection] (0,0)--(0,0.5774) (0,0)--(-0.5,-0.2887) (0,0)--(0.5,-0.2887);
  \node[steiner] at (0,0) {};
  \foreach \p in {(0,0.5774),(-0.5,-0.2887),(0.5,-0.2887)} \fill \p circle[radius=2pt];
\end{scope}
\node[caption] at (0,-1.3) {$\ell$};
\node[caption] at (2.5,-1.3) {$2\cdot\frac{\sqrt3}{2}\ell$};
\foreach \sx/\heading in {5/{(c)},7.5/{(d)},10/{(e)}} {
  \begin{scope}[xshift=\sx*1.2cm]
    \node at (0,1.3) {\heading};
    \draw[guide] (0,0.7071)--(0.7071,0)--(0,-0.7071)--(-0.7071,0)--cycle;
    \foreach \x/\y in {0/1,1/0,0/-1,-1/0}
      \draw[row] (0.7071*\x,0.7071*\y)--(0.95*\x,0.95*\y);
    \foreach \x/\y in {0/0.7071,0.7071/0,0/-0.7071,-0.7071/0} \node[active] at (\x,\y) {};
  \end{scope}
}
\begin{scope}[xshift=5*1.2cm]
  \draw[connection] (-0.7071,0)--(0,0.7071);
  \foreach \p in {(-0.7071,0),(0,0.7071)} \fill \p circle[radius=2pt];
\end{scope}
\begin{scope}[xshift=7.5*1.2cm]
  \draw[connection] (0,0.7071)--(0,0.4082)--(-0.7071,0) (0,0.4082)--(0.7071,0);
  \node[steiner] at (0,0.4082) {};
  \foreach \p in {(-0.7071,0),(0,0.7071),(0.7071,0)} \fill \p circle[radius=2pt];
\end{scope}
\begin{scope}[xshift=10*1.2cm]
  \draw[connection] (0,0.7071)--(0.1494,0.1494)--(0.7071,0)
    (0.1494,0.1494)--(-0.1494,-0.1494)
    (-0.7071,0)--(-0.1494,-0.1494)--(0,-0.7071);
  \node[steiner] at (0.1494,0.1494) {};
  \node[steiner] at (-0.1494,-0.1494) {};
  \foreach \p in {(-0.7071,0),(0,0.7071),(0.7071,0),(0,-0.7071)} \fill \p circle[radius=2pt];
\end{scope}
\node[caption] at (5,-1.3) {$1$};
\node[caption] at (7.5,-1.3) {$2\cdot\frac{1+\sqrt3}{2\sqrt2}$};
\node[caption] at (10,-1.3) {$3\cdot\frac{1+\sqrt3}{3}$};
\end{tikzpicture}
\caption{Shortest networks on $k\ge2$ ports of a triangle of side $\ell$ (a, b) and of a unit square (c--e), with cables in grey and connections in red. Lengths are shown as multiples of $k-1$.}
\label{fig:plane-cables-costs}
\end{figure}
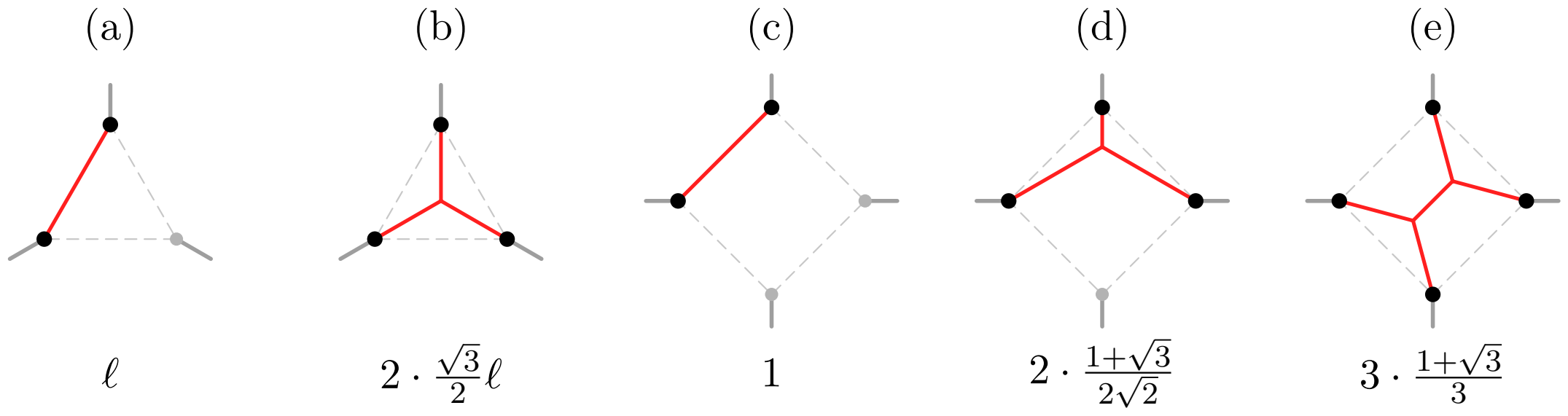

\paragraph{Gadgets.}
Fix $\omega=1/100$. As in \cref{sec:cable}, cables are joined at small \emph{gadgets}, ending at their corners, called \emph{ports}. We use two types of gadgets: equilateral \emph{triangles} of side $1$ or $1-\omega$, and unit \emph{squares}, whose connection costs are shown in \cref{fig:plane-cables-costs}. A \emph{crossover} has an upper and a lower triangle, as in \cref{fig:overview-plane-crossover}(c,~d). We call it an \emph{$\omega$-crossover} when its upper triangle has side $1-\omega$; all remaining triangles have side $1$.

{
\paragraph{Layout.}
We label the elements of $H$ by $1,\ldots,3r$ and write $E=\{e_1,\ldots,e_{|E|}\}$. As in \cref{fig:overview-plane-layout}, the horizontal levels $0,\ldots,3r$ are numbered from bottom to top, with the backbone on level $0$. Each triple $e$ has a square $Q_e$ below level $0$ and three columns, one for each $j\in e$, while its fourth corner connects to the backbone. The column for $j$ runs from $Q_e$ to level $j$, with a crossover on each level $0,\ldots,j-1$. Its top crossover is an $\omega$-crossover, so choosing its upper triangle saves $\sqrt3\,\omega$. Each column has at most $3r$ crossovers, giving a total of $C\le9r|E|$ crossovers.}

\paragraph{Cables.}
At a crossover $x$, let $U_x,D_x$ denote the cables above and below, and $L_x,R_x$ those to the left and right (\cref{fig:overview-plane-crossover}(c,~d)), so $U_x=D_y$ for the crossover $y$ directly above $x$. Within each column, cables connect the square to the lowest crossover and consecutive crossovers to each other, while along each level they connect adjacent crossovers. A column of element $j$ ends on level $j$, and the cable above its top crossover continues along that level; the backbone is part of the leftmost cable of level~$0$. There are $C$ cables below crossovers and $C+3r+1$ along the levels, for a total of $2C+3r+1$ cables. The ports of each gadget lie on distinct cables.

{
\begingroup
\expandafter\patchcmd\csname thmt@original@definition\endcsname{{theorem}}{{definition}}{}{}
\PlaneCleanTree*
\endgroup

Choosing the upper triangle makes a crossover \emph{up}, and choosing the lower makes it \emph{down} (\cref{fig:overview-plane-crossover}(c,~d)). The possible column patterns, including a \emph{switch} from up to down, are shown in \cref{fig:overview-plane-switch}.

First consider a clean tree with no broken crossovers, and let $q$ be the number of used squares and $C_\omega$ the number of up $\omega$-crossovers. It uses one triangle at each crossover, at a cost of $\sqrt3(1-\omega)$ for an up $\omega$-crossover and $\sqrt3$ otherwise, while each used square costs $1+\sqrt3$. Its total length is therefore
\[
    L+\sqrt3\,C+(1+\sqrt3)\,q-\sqrt3\,\omega C_\omega.
\]

The argument in \cref{sec:overview-plane} gives $q\ge r$, since connecting all levels requires at least $r$ squares: each used square has three columns, each joining one pair of consecutive levels. Moreover, at most one $\omega$-crossover per element can be up, since two would create a cycle, so $C_\omega\le3r$. Together, these bounds give length at least
\begin{equation}
\label{eq:plane-cables-tau}
    \tau=L+\sqrt3\,C+(1+\sqrt3)\,r-3\sqrt3\,\omega r.
\end{equation}
Equality requires $q=r$ and $C_\omega=3r$. The up $\omega$-crossovers then already connect all levels, so a switch would create a cycle. The used squares therefore have all their columns up and cover each element exactly once, so their triples form a perfect matching. Conversely, choosing the squares of a perfect matching, making their columns up and all others down, gives a clean tree of length $\tau$ (\cref{fig:overview-plane-layout}).

Since there are no fallbacks, the cleaning argument of \cref{sec:cable} no longer applies. A tree may connect only some cables at a gadget or join cables at different gadgets, and it may leave crossovers broken or miss discounts. After placing and sampling the cables, we count these deviations and show that each costs at least $\sqrt3\,\omega$, while a tree with few deviations still yields a large matching.}

\subsection{Placing and sampling the cables}
\label{sec:plane-cables-placement}

We place the gadgets far apart on a grid and route the cables along its lines, so that each cable leaves its port radially away from the gadget's centre. For a gadget $s$ centred at $z_s$, let $B_s$ be the closed ball of radius $20$ around $z_s$. The cables satisfy the following separation properties; the coordinates and proof are given in \cref{app:plane-placement}.

\begin{lemma}
\label{lem:plane-cables-separated}
The cables have the following properties.
\begin{enumerate}[(i)]
\item The balls $B_s$ are pairwise more than $20$ apart. Inside each $B_s$, the cables consist exactly of the radial segments from the ports of $s$ to the boundary of $B_s$. Any two points on different cables at distance less than $20$ lie in a common ball $B_s$.
\item Every cable segment has length at least $1$; non-incident segments of the same cable are at least $1$ apart, and incident segments meet at an angle of at least $45^\circ$.
\item The cables can be constructed in polynomial time and have total length $L=O(r^2)$.
\end{enumerate}
\end{lemma}

\label{sec:plane-cables-connectors}

For $0<\rho\le1/100$, sample terminals along the cables as in \cref{sec:cable-routing} to obtain $P_\rho$. Cable recovery (\cref{lem:cable-recovery}) applies to every tree of length $O(L)$ on $P_\rho$, since its proof needs only (S2) and separation greater than $2\rho$ between different cables. Here (S2) follows from \cref{lem:plane-cables-separated}(ii), and part~(i) gives a separation of at least $1-\omega>2\rho$.

Let $T$ be an optimal tree on $P_\rho$. Since the cables together with Steiner trees at the gadgets form a connected network of length $O(L)$, we have $\operatorname{len}(T)=O(L)$. As in \cref{sec:cable-reduction-proof}, recovery yields a tree $\widehat T$ of length at most $\operatorname{len}(T)+O(\rho L)$. It consists of disjoint trees $T_i$, each containing all sampled terminals on $\Gamma_i$, with total length at least $L$, and connectors joining these trees (\cref{fig:connectors}). Each connector $S$ has $r_S$ attachments on $r_S$ distinct cables, so with $2C+3r+1$ cables in total, the count~\eqref{eq:connector-count} becomes
\begin{equation}
\label{eq:plane-cables-count}
    \sum_S(r_S-1)=(2C+3r+1)-1=2C+3r.
\end{equation}

Following \cref{sec:cable-reduction-proof}, we distinguish good and bad connectors.

\begin{definition}
\label{def:plane-good-connector}
A connector is
\begin{enumerate}[(i)]
\item \emph{good} at a triangle $s$ with ports $p_{s,1},p_{s,2},p_{s,3}$ if it has exactly three attachments, one within distance $\rho$ of the cable segment inside $B_s$ leaving each port;
\item \emph{good} at a square $s$ with ports $p_{s,1},\ldots,p_{s,4}$ if it has exactly four attachments, one within distance $\rho$ of the cable segment inside $B_s$ leaving each port;
\item \emph{bad} otherwise.
\end{enumerate}
\end{definition}

A bad connector may join only some of the cables at a gadget, for example those at two corners of a triangle, or connect cables at different gadgets. The following analogue of \cref{lem:displaced-attachments} shows that, up to $O(\rho)$ errors, a good connector has length at least the Steiner optimum on its gadget's ports, while a bad connector costs more per joined pair of cables.

\begin{lemma}
\label{lem:plane-cables-attachments}
Let $S$ be a connector with $r$ attachments.
\begin{enumerate}[(i)]
\item If $S$ is good at a triangle of side $\ell$, then $\operatorname{len}(S)\ge\sqrt3\,\ell-3\rho$.
\item If $S$ is good at a square, then $\operatorname{len}(S)\ge1+\sqrt3-4\rho$.
\item If $S$ is bad, then $\operatorname{len}(S)\ge\bigl(\frac{1+\sqrt3}{2\sqrt2}-2\rho\bigr)(r-1)$.
\end{enumerate}
\end{lemma}

\begin{figure}[!t]
\centering
\begin{tikzpicture}[line cap=round,line join=round,x=1.3cm,y=1.3cm,font=\normalsize,>=latex,
  network/.style={draw=connectionGreen,line width=1.05pt,line cap=round,line join=round},
  added/.style={draw=connectionGreen,line width=1.4pt,dash pattern=on 0pt off 2.6pt,line cap=round,line join=round},
  cone/.style={fill=black!9},
  cab/.style={line width=1.05pt,line cap=round,line join=round}]
\begin{scope}[xshift=0cm]
  \fill[cone] (0.000,1.100)--++(30:.85) arc[start angle=30,end angle=150,radius=.85]--cycle;
  \fill[cone] (-0.953,-0.550)--++(150:.85) arc[start angle=150,end angle=270,radius=.85]--cycle;
  \fill[cone] (0.953,-0.550)--++(270:.85) arc[start angle=270,end angle=390,radius=.85]--cycle;
  \filldraw[fill=black!3,draw=black!35] (0.000,1.100)--(-0.953,-0.550)--(0.953,-0.550)--cycle;
  \draw[cableU,cab] (0.000,1.100)--(0.000,1.950);
  \fill[black!55] (0.000,1.100) circle[radius=1.5pt];
  \draw[cableV,cab] (-0.953,-0.550)--(-1.689,-0.975);
  \fill[black!55] (-0.953,-0.550) circle[radius=1.5pt];
  \draw[cableW,cab] (0.953,-0.550)--(1.689,-0.975);
  \fill[black!55] (0.953,-0.550) circle[radius=1.5pt];
  \node[cableU,above] at (0.000,1.950) {$\Gamma_{i_1}$};
  \node[cableV,below left] at (-1.689,-0.975) {$\Gamma_{i_2}$};
  \node[cableW,below right] at (1.689,-0.975) {$\Gamma_{i_3}$};
\end{scope}
\begin{scope}[xshift=6cm]
  \fill[cone] (0.000,1.100)--++(30:.85) arc[start angle=30,end angle=150,radius=.85]--cycle;
  \fill[cone] (-0.953,-0.550)--++(150:.85) arc[start angle=150,end angle=270,radius=.85]--cycle;
  \fill[cone] (0.953,-0.550)--++(270:.85) arc[start angle=270,end angle=390,radius=.85]--cycle;
  \filldraw[fill=black!3,draw=black!35] (0.000,1.100)--(-0.953,-0.550)--(0.953,-0.550)--cycle;
  \draw[cableU,cab] (0.000,1.100)--(0.000,1.950);
  \fill[black!55] (0.000,1.100) circle[radius=1.5pt];
  \draw[cableV,cab] (-0.953,-0.550)--(-1.689,-0.975);
  \fill[black!55] (-0.953,-0.550) circle[radius=1.5pt];
  \draw[cableW,cab] (0.953,-0.550)--(1.689,-0.975);
  \fill[black!55] (0.953,-0.550) circle[radius=1.5pt];
  \node[cableU,above] at (0.000,1.950) {$\Gamma_{i_1}$};
  \node[cableV,below left] at (-1.689,-0.975) {$\Gamma_{i_2}$};
  \node[cableW,below right] at (1.689,-0.975) {$\Gamma_{i_3}$};
\end{scope}
\draw[network] (0.349,0.816)--(0.050,0.050)--(-0.881,-0.106) (0.050,0.050)--(0.533,-0.710);
\node at (-.2,.3) {$S$};
\foreach \PA/\PB/\PC in {{(0,1.2)}/{(0.349,0.816)}/cableU,{(-1.04,-0.6)}/{(-0.881,-0.106)}/cableV,{(1.04,-0.6)}/{(0.533,-0.710)}/cableW} {
  \draw[black!70,line width=1.1pt,dash pattern=on 0pt off 2.2pt,line cap=round] \PA--\PB;
  \fill[\PC] \PA circle[radius=2pt];
  \fill \PB circle[radius=2pt];}
\node[right=3pt] at (0.349,0.816) {$x_1$};
\node[above left=1pt] at (-0.881,-0.106) {$x_2$};
\node[below=3pt] at (0.533,-0.710) {$x_3$};
\node[above right=0pt,font=\small] at (0.17,1.0) {$\le\rho$};
\node[left=3pt] at (0,1.2) {$\bar x_1$};
\node[below=3pt] at (-1.04,-0.6) {$\bar x_2$};
\node[right=3pt] at (1.04,-0.6) {$\bar x_3$};
\node at (0,-1.75) {(a) Before projection};
\begin{scope}[xshift=6cm]
  \draw[network] (0.050,0.050)--(0.274,0.625)--(0.210,0.736);
  \draw[network] (0.050,0.050)--(-0.677,-0.072)--(-0.743,-0.186);
  \draw[network] (0.050,0.050)--(0.431,-0.550)--(0.533,-0.550);
  \node at (-.2,.3) {$S'$};
  \foreach \PA/\PB in {{(0,1.1)}/{(0.210,0.736)},{(-0.953,-0.55)}/{(-0.743,-0.186)},{(0.953,-0.55)}/{(0.533,-0.550)}} {
    \draw[added] \PA--\PB;
    \fill \PB circle[radius=2pt];}
  \node[right=4pt] at (0.210,0.736) {$x'_1$};
  \node[left=3pt] at (-0.743,-0.186) {$x'_2$};
  \node[below=3pt] at (0.533,-0.550) {$x'_3$};
  \node[above right=0pt,font=\small] at (0.17,1.0) {$\le\rho$};
  \node[left=4pt] at (0.000,1.100) {$p_1$};
  \node[below=4pt] at (-0.953,-0.550) {$p_2$};
  \node[above right=1pt] at (0.953,-0.550) {$p_3$};
  \node at (0,-1.75) {(b) Project and join to ports};
\end{scope}
\end{tikzpicture}
\caption{Projection onto a triangle gadget. (a) Each attachment $x_j$ of $S$ lies within distance $\rho$ of a cable point $\bar x_j$. (b) The shaded regions project to the ports, and projection does not increase the length of $S$. The dotted segments, each of length at most $\rho$, connect the projected network $S'$ to the ports.}
\label{fig:plane-cables-projection}
\end{figure}
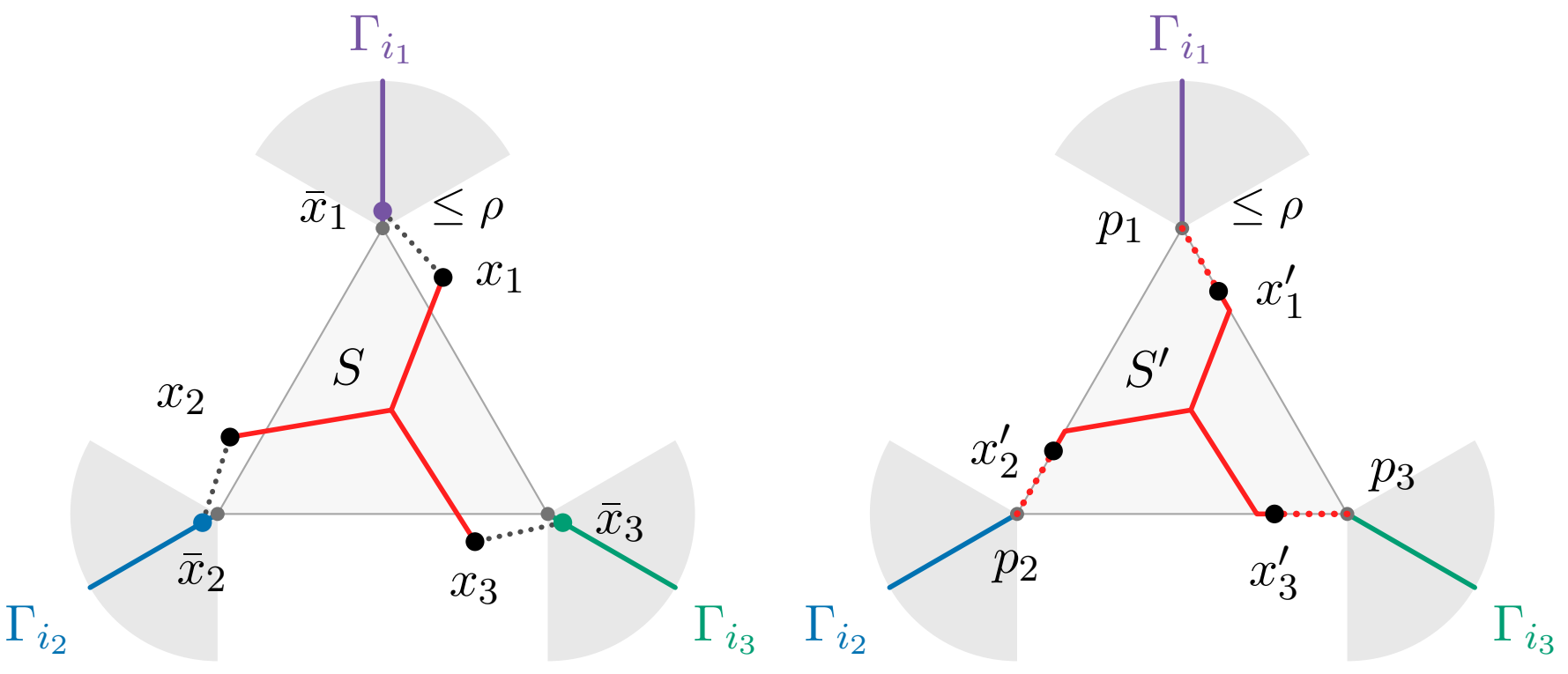

\begin{proof}
Let $x_1,\ldots,x_r$ be the attachments of $S$. Choose the points $\bar x_j$ and group the attachments as in the proof of \cref{lem:displaced-attachments}, using the balls $B_s$ of \cref{lem:plane-cables-separated}. If there are $g\ge2$ groups, then $S$ is bad, and the same disjoint-ball argument gives
\[
    \operatorname{len}(S)\ge9g\ge2(4g-1)\ge2(r-1),
\]
proving (iii).

If there is only one group, all points $\bar x_j$ lie in a single ball $B_s$. Projecting $S$ onto the triangle or square and joining the projected attachments to their ports gives a network on $r$ ports of $s$ of length at most $\operatorname{len}(S)+r\rho$ (\cref{fig:plane-cables-projection}). When $S$ is good, this network joins all ports, at a cost of at least $\sqrt3\,\ell$ for a triangle of side $\ell$ or $1+\sqrt3$ for a square (\cref{fig:plane-cables-costs}(b,~e)), proving (i) and (ii).

Otherwise, the network joins two ports of a triangle, at cost $\ell\ge1-\omega>\frac{1+\sqrt3}{2\sqrt2}$ (\cref{fig:plane-cables-costs}(a)), or two ports of a square, at cost $1>\frac{1+\sqrt3}{2\sqrt2}$ (\cref{fig:plane-cables-costs}(c)) or $\sqrt2>\frac{1+\sqrt3}{2\sqrt2}$. Joining three ports of a square costs $\frac{1+\sqrt3}{\sqrt2}=2\cdot\frac{1+\sqrt3}{2\sqrt2}$ (\cref{fig:plane-cables-costs}(d)). Thus each cost is at least $\frac{1+\sqrt3}{2\sqrt2}(r-1)$. Subtracting $r\rho$ and using $r\le2(r-1)$ gives (iii).
\end{proof}

Extending the terminology for clean trees, we call a crossover $x$ \emph{up} if it has a good upper triangle connector, \emph{down} if it has a good lower triangle connector, and \emph{broken} if it has neither. A crossover cannot be both up and down, since the two good connectors would each join $L_x$ to $R_x$, creating a cycle.

\subsection{The upper bound}
\label{sec:plane-cables-completeness}

Recall the threshold from~\eqref{eq:plane-cables-tau},
\[
    \tau=L+\sqrt3\,C+(1+\sqrt3)\,r-3\sqrt3\,\omega r.
\]

\begin{lemma}
\label{lem:plane-cables-completeness}
If $\nu(H)=r$, then $\OPT(P_\rho)\le\tau$.
\end{lemma}

\begin{proof}
Let $M$ be a perfect matching, and construct the clean tree using the squares of its triples, with all crossovers in their columns up and all remaining crossovers down (\cref{fig:overview-plane-layout}). It contains $P_\rho$ because it includes the edges between consecutive terminals on every cable, and it has length $\tau$ because $q=r$ and $C_\omega=3r$. It is connected because each level is connected and the up $\omega$-crossover for each element $j$ joins levels $j-1$ and $j$.
\end{proof}

\subsection{The lower bound}
\label{sec:plane-cables-soundness}

If $\widehat T$ had the connectors of a clean tree and no broken crossovers, the argument of \cref{sec:plane-cables-idea} would apply. In general, we count the following deviations:
\begin{itemize}
\item $D_{\mathrm{br}}$, the number of broken crossovers;
\item $D_{\mathrm{bad}}=\sum_S(r_S-1)$, summed over all bad connectors $S$, the number of connections made by bad connectors;
\item $D_\omega=\max\{3r-C_\omega,0\}$, the number of missing discounts, where $C_\omega$ is the number of up $\omega$-crossovers.
\end{itemize}
Let $D=D_{\mathrm{br}}+D_{\mathrm{bad}}+D_\omega$ be the total deviation. If $D=0$, then $\widehat T$ has the connectors of a clean tree with all $3r$ discounts, so its used squares give a perfect matching by \cref{sec:plane-cables-idea}. We will show that $\operatorname{len}(T)\ge\tau+\sqrt3\,\omega D-O(\rho L)$, so each deviation adds length, and that $\nu(H)\ge r-5D$, so few deviations still give a large matching.

For each element $j$, let $C_{\omega,j}$ be the number of up $\omega$-crossovers in the columns of $j$, so $C_\omega=\sum_jC_{\omega,j}$. In a clean tree with no broken crossovers, $C_{\omega,j}\le1$ by (P3) {in \cref{sec:overview-plane}}, but broken crossovers can split a level and allow more than one discount for the same element. The next lemma bounds the number of extra discounts in terms of $D_{\mathrm{br}}$.

\begin{lemma}
\label{lem:plane-cables-discounts}
$\sum_{j=1}^{3r}\max\{C_{\omega,j}-1,0\}\le2D_{\mathrm{br}}$. In particular, $C_\omega\le3r+2D_{\mathrm{br}}$.
\end{lemma}

\begin{figure}[!tb]
\centering
\definecolor{subA}{HTML}{E07B00}
\definecolor{subB}{HTML}{1F6FD1}
\definecolor{subC}{HTML}{D6202F}
\definecolor{subD}{HTML}{8A3FC4}
\definecolor{subE}{HTML}{109A4A}
\begin{tikzpicture}[line cap=round,line join=round,x=1cm,y=1cm,font=\small,
  lev/.style={line width=1.3pt,line cap=round},
  col/.style={draw=black!30,line width=1.05pt,line cap=round,line join=round},
  stub/.style={line width=1.3pt,line cap=round},
  tri/.style={draw=black!70,densely dotted,line width=.5pt,line join=round},
  conn/.style={draw=black,line width=.95pt,line cap=round,line join=round},
  link/.style={draw=black,line width=1.1pt,line cap=round},
  warp/.style={draw=black,line width=.45pt},
  note/.style={font=\footnotesize,text=black!70},
  lab/.style={text=black!65,font=\footnotesize}]
\newcommand{\xov}[4]{%
  \begin{scope}[shift={(#1,#2)},x=1.5mm,y=1.5mm]
    \fill[white] (-2.6,-2.8) rectangle (2.6,2.8);
    \draw[stub,draw=#3] (-2.6,0)--(-1.2,.75) (-2.6,0)--(-1.2,-.75);
    \draw[stub,draw=#4] (2.6,0)--(1.2,.75) (2.6,0)--(1.2,-.75);
    \draw[tri] (0,2.8)--(-1.2,.75)--(1.2,.75)--cycle;
    \draw[tri] (0,-2.8)--(-1.2,-.75)--(1.2,-.75)--cycle;
  \end{scope}}
\newcommand{\upst}[2]{\begin{scope}[shift={(#1,#2)},x=1.5mm,y=1.5mm]
  \draw[conn] (0,1.43)--(0,2.8) (0,1.43)--(-1.2,.75) (0,1.43)--(1.2,.75);\end{scope}}
\newcommand{\downst}[2]{\begin{scope}[shift={(#1,#2)},x=1.5mm,y=1.5mm]
  \draw[conn] (0,-1.43)--(0,-2.8) (0,-1.43)--(-1.2,-.75) (0,-1.43)--(1.2,-.75);\end{scope}}
\newcommand{\broken}[2]{%
  \draw[draw=red!75!black,line width=1pt] (#1-0.18,#2+0.18)--(#1+0.18,#2-0.18) (#1-0.18,#2-0.18)--(#1+0.18,#2+0.18);}
\node[lab,anchor=east] at (0.05,1.8) {level $i$};
\node[lab,anchor=east] at (0.05,0) {level $i-1$};
\draw[col] (2.0,-0.8)--(2.0,2.6) (3.2,-0.8)--(3.2,2.6) (5.6,-0.8)--(5.6,2.6) (8.0,-0.8)--(8.0,2.6);
\draw[col] (0.8,-0.8)--(0.8,-0.42) (4.4,-0.8)--(4.4,-0.42) (6.8,-0.8)--(6.8,-0.42) (9.2,-0.8)--(9.2,-0.42);
\draw[lev,draw=subA] (0.2,0)--(2.0,0);
\draw[lev,draw=subB] (2.0,0)--(8.0,0);
\draw[lev,draw=subC] (8.0,0)--(9.8,0);
\draw[lev,draw=subD] (0.2,1.8)--(5.6,1.8);
\draw[lev,draw=subE] (5.6,1.8)--(9.8,1.8);
\draw[link] (0.8,0.42)--(0.8,1.8) (4.4,0.42)--(4.4,1.8) (6.8,0.42)--(6.8,1.8) (9.2,0.42)--(9.2,1.8);
\xov{0.8}{0}{subA}{subA}\upst{0.8}{0}
\xov{2.0}{0}{subA}{subB}
\xov{3.2}{0}{subB}{subB}\downst{3.2}{0}
\xov{4.4}{0}{subB}{subB}\upst{4.4}{0}
\xov{5.6}{0}{subB}{subB}\downst{5.6}{0}
\xov{6.8}{0}{subB}{subB}\upst{6.8}{0}
\xov{8.0}{0}{subB}{subC}
\xov{9.2}{0}{subC}{subC}\upst{9.2}{0}
\foreach \px in {0.8,4.4,6.8,9.2} \draw[warp] (\px,0) circle[radius=5mm];
\xov{2.0}{1.8}{subD}{subD}\downst{2.0}{1.8}
\xov{3.2}{1.8}{subD}{subD}\downst{3.2}{1.8}
\xov{5.6}{1.8}{subD}{subE}
\xov{8.0}{1.8}{subE}{subE}\downst{8.0}{1.8}
\node[note,text=black,anchor=north] at (2.0,-0.82) {broken};
\node[note,text=black,anchor=north] at (8.0,-0.82) {broken};
\node[note,text=black,anchor=south] at (5.6,2.62) {broken};
\end{tikzpicture}
\caption{\Cref{lem:plane-cables-discounts}: four up $\omega$-crossovers of element $i$ (circled) link level $i-1$ to the cables of level $i$ above them. The good connectors of the crossovers that are not broken chain the cables along each level into subtrees (one colour per subtree). Here $D_{\mathrm{br},i-1}=2$ and $D_{\mathrm{br},i}=1$, and the four links join the five subtrees into a tree, so a fifth link would close a cycle.}
\label{fig:plane-cables-fragments}
\end{figure}
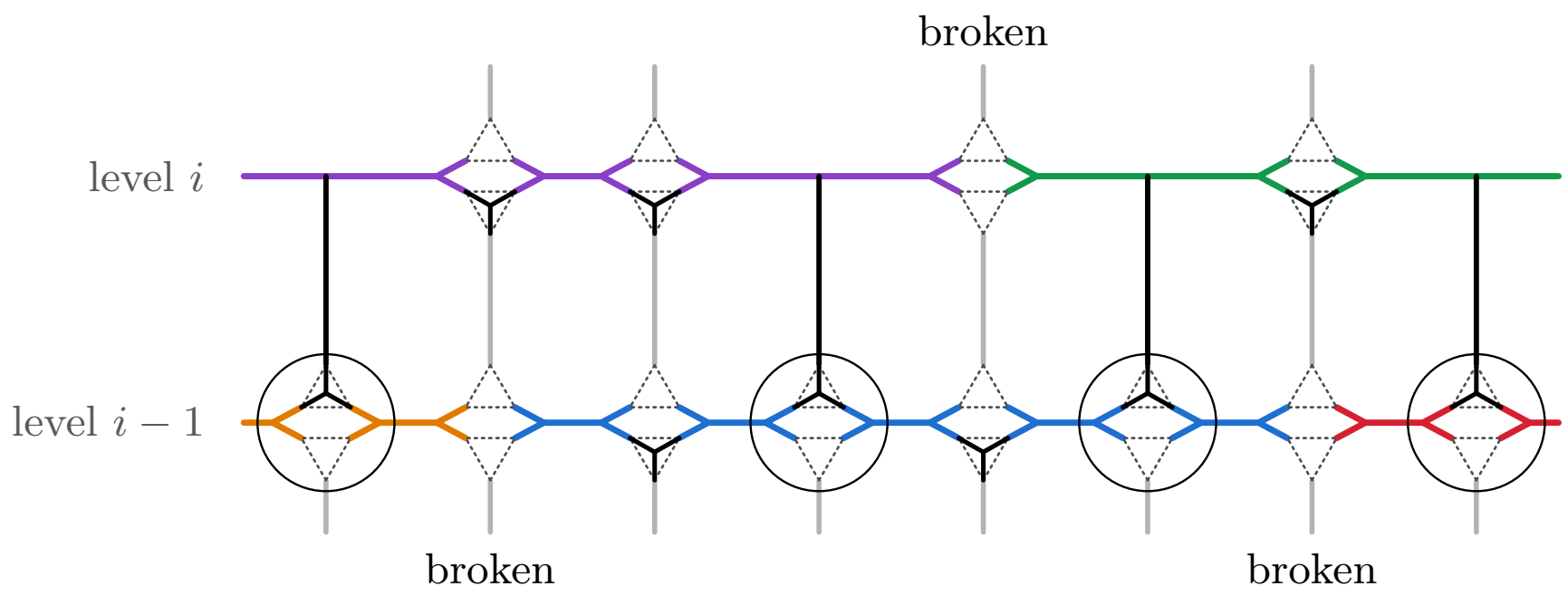

\begin{proof}
Let $D_{\mathrm{br},i}$ be the number of broken crossovers on level $i$. Along each level, the good connectors join its cables into subtrees, with a break at each broken crossover. Thus there are $1+D_{\mathrm{br},i-1}$ subtrees on level $i-1$ and $1+D_{\mathrm{br},i}$ on level $i$. Each of the $C_{\omega,i}$ up $\omega$-crossovers of element $i$ lies on level $i-1$ and links a subtree there to one on level $i$ (\cref{fig:plane-cables-fragments}). These links cannot form a cycle, so their number is at most one less than the total number of subtrees, giving $C_{\omega,i}\le1+D_{\mathrm{br},i-1}+D_{\mathrm{br},i}$.

Hence $\max\{C_{\omega,i}-1,0\}\le D_{\mathrm{br},i-1}+D_{\mathrm{br},i}$. Summing over $i$ counts each $D_{\mathrm{br},i}$ at most twice and proves the first statement. The second follows from $C_\omega\le3r+\sum_i\max\{C_{\omega,i}-1,0\}$.
\end{proof}

\begin{lemma}
\label{lem:plane-cables-cost}
There is an absolute constant $C_3$ such that $\operatorname{len}(T)\ge\tau+\sqrt3\,\omega D-C_3\rho L$.
\end{lemma}

\begin{proof}
Let $q$ be the number of good square connectors. In~\eqref{eq:plane-cables-count}, each good triangle connector contributes $2$, each good square connector contributes $3$, and the bad connectors contribute $D_{\mathrm{bad}}$ in total. There are $C-D_{\mathrm{br}}$ good triangle connectors, so $2(C-D_{\mathrm{br}})+3q+D_{\mathrm{bad}}=2C+3r$, giving
\begin{equation}
\label{eq:plane-cables-squares}
    3q+D_{\mathrm{bad}}=3r+2D_{\mathrm{br}}.
\end{equation}
The cable trees $T_i$ have total length at least $L$. Summing the bounds of \cref{lem:plane-cables-attachments} over the connectors and using $\operatorname{len}(\widehat T)\le\operatorname{len}(T)+O(\rho L)$ gives
\begin{align*}
    \operatorname{len}(T)+O(\rho L)\ge\operatorname{len}(\widehat T)
    &\ge L+(1-\omega)\sqrt3\,C_\omega+\sqrt3\,(C-D_{\mathrm{br}}-C_\omega)\\
    &\qquad+(1+\sqrt3)\,q+\tfrac{1+\sqrt3}{2\sqrt2}\,D_{\mathrm{bad}}-2\rho L.
\end{align*}
Each connector $S$ contributes an error of at most $2\rho(r_S-1)$, so the total error is at most $2\rho\sum_S(r_S-1)=2\rho(2C+3r)\le2\rho L$, since every cable has length at least $1$.

Recall that $\tau=L+\sqrt3\,C+(1+\sqrt3)\,r-3\sqrt3\,\omega r$. By~\eqref{eq:plane-cables-squares}, $q=r+\frac{2D_{\mathrm{br}}-D_{\mathrm{bad}}}{3}$. Also, \cref{lem:plane-cables-discounts} gives $3r-C_\omega\ge D_\omega-2D_{\mathrm{br}}$: this is immediate if $C_\omega\le3r$, and otherwise follows from $C_\omega\le3r+2D_{\mathrm{br}}$. We obtain
\begin{align*}
    \operatorname{len}(T)+O(\rho L)
    &\ge L+\sqrt3\,C-\sqrt3\,\omega C_\omega-\sqrt3\,D_{\mathrm{br}}\\
    &\qquad+(1+\sqrt3)\Bigl(r+\tfrac{2D_{\mathrm{br}}-D_{\mathrm{bad}}}{3}\Bigr)
        +\tfrac{1+\sqrt3}{2\sqrt2}\,D_{\mathrm{bad}}-2\rho L
        &&\text{substituting }q,\\
    &=\tau+\sqrt3\,\omega\,(3r-C_\omega)+\tfrac{2-\sqrt3}{3}\,D_{\mathrm{br}}\\
    &\qquad+\Bigl(\tfrac{1+\sqrt3}{2\sqrt2}-\tfrac{1+\sqrt3}{3}\Bigr)D_{\mathrm{bad}}-2\rho L
        &&\text{by definition of }\tau,\\
    &\ge\tau+\sqrt3\,\omega\,(3r-C_\omega)+0.089\,D_{\mathrm{br}}+0.055\,D_{\mathrm{bad}}-2\rho L
        &&\text{rounding down,}\\
    &\ge\tau+\sqrt3\,\omega\,D_\omega+(0.089-2\sqrt3\,\omega)\,D_{\mathrm{br}}+0.055\,D_{\mathrm{bad}}-2\rho L
        &&\text{by \cref{lem:plane-cables-discounts},}\\
    &\ge\tau+\sqrt3\,\omega\,D_\omega+\sqrt3\,\omega\,D_{\mathrm{br}}+\sqrt3\,\omega\,D_{\mathrm{bad}}-2\rho L
        &&\text{since }\omega=1/100,\\
    &=\tau+\sqrt3\,\omega\,D-2\rho L
        &&\text{by definition of }D.
\end{align*}
Absorbing the error terms into $C_3\rho L$ proves the claim.
\end{proof}

To obtain a matching from $\widehat T$, call a column \emph{good} if all of its crossovers are up and its square has a good connector. In a clean tree with no broken crossovers and all $3r$ discounts, the columns with an up $\omega$-crossover are exactly the good ones, and the triples of their squares form a perfect matching, as in \cref{sec:plane-cables-idea}. In general, each deviation affects only a few columns, so few deviations still leave a nearly perfect matching.

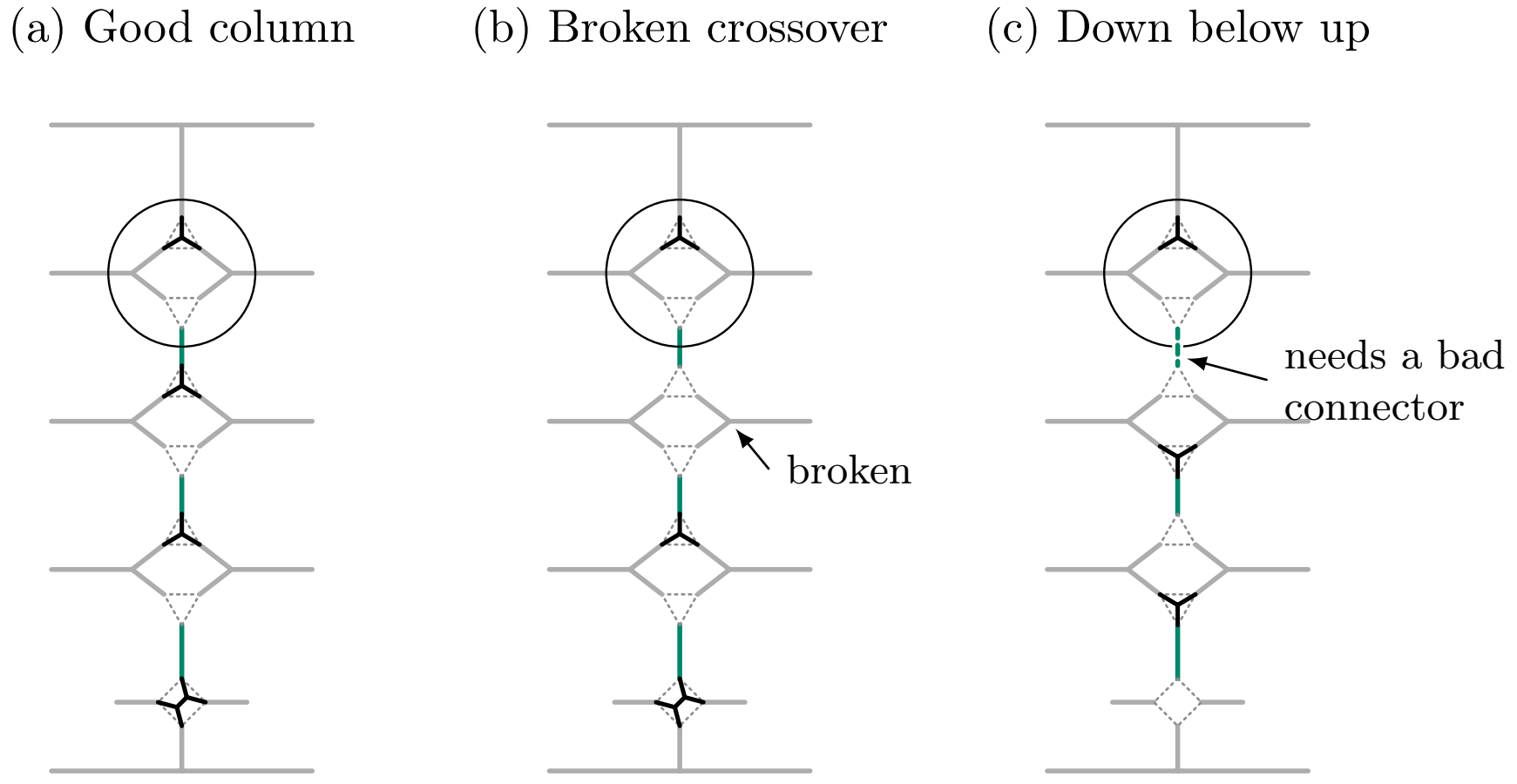
\begin{figure}[H]
\centering
\begin{tikzpicture}[line cap=round,line join=round,
  x=0.9cm,y=0.9cm,font=\small,
  level/.style={draw=black!32,line width=1.05pt,line cap=round,line join=round},
  segment/.style={draw=planeCableW,line width=1.05pt,line cap=round,line join=round},
  loose/.style={draw=planeCableW,line width=1.05pt,line cap=round,dash pattern=on 2pt off 1.5pt},
  pointer/.style={draw=black,line width=.5pt,->,>=latex,shorten >=2pt},
  guide/.style={draw=black!45,densely dotted,line width=0.5pt},
  connection/.style={draw=black,line width=.95pt,line cap=round,line join=round},
  note/.style={font=\footnotesize,align=left}
]
\newcommand{\tri}[1]{%
  \draw[guide] (0,#1+0.47)--(-0.15,#1+0.21)--(0.15,#1+0.21)--cycle;
  \draw[guide] (0,#1-0.47)--(-0.15,#1-0.21)--(0.15,#1-0.21)--cycle;
  \draw[draw=black!32,line width=1.05pt] (-0.15,#1+0.21)--(-0.42,#1)--(-0.15,#1-0.21)
    (0.15,#1+0.21)--(0.42,#1)--(0.15,#1-0.21);}
\newcommand{\fullup}[1]{%
  \draw[connection] (0,#1+0.3)--(0,#1+0.47) (0,#1+0.3)--(-0.15,#1+0.21) (0,#1+0.3)--(0.15,#1+0.21);}
\newcommand{\fulldown}[1]{%
  \draw[connection] (0,#1-0.3)--(0,#1-0.47) (0,#1-0.3)--(-0.15,#1-0.21) (0,#1-0.3)--(0.15,#1-0.21);}
\newcommand{\sqgad}{%
  \draw[guide] (0,-0.92)--(0.2,-1.12)--(0,-1.32)--(-0.2,-1.12)--cycle;}
\newcommand{\sqfull}{%
  \draw[connection] (0,-0.92)--(0.042,-1.078)--(0.2,-1.12) (0.042,-1.078)--(-0.042,-1.162)
    (-0.2,-1.12)--(-0.042,-1.162)--(0,-1.32);}
\foreach \sx/\heading in {0/{(a) Good column},4.2/{(b) Broken crossover},8.4/{(c) Down below up}} {
  \begin{scope}[xshift=\sx*0.9cm]
    \node at (0,4.55) {\heading};
    \draw[level] (-1.1,3.75)--(1.1,3.75);
    \foreach \y in {0,1.25,2.5} \draw[level] (-1.1,\y)--(-0.42,\y) (0.42,\y)--(1.1,\y);
    \draw[level] (0,2.97)--(0,3.75);
    \draw[segment] (0,2.03)--(0,1.72) (0,0.78)--(0,0.47) (0,-0.47)--(0,-0.92);
    \draw[level] (-0.2,-1.12)--(-0.55,-1.12) (0.2,-1.12)--(0.55,-1.12) (0,-1.32)--(0,-1.7) (-1.1,-1.7)--(1.1,-1.7);
    \foreach \y in {0,1.25,2.5} {\tri{\y}}
    \sqgad
    \draw[draw=black,line width=.45pt] (0,2.5) circle[radius=0.62];
  \end{scope}
}
\begin{scope}
  \fullup{2.5} \fullup{1.25} \fullup{0}
  \sqfull
\end{scope}
\begin{scope}[xshift=4.2*0.9cm]
  \fullup{2.5} \fullup{0}
  \sqfull
  \node[note,anchor=west] (nb) at (0.75,0.85) {broken};
  \draw[pointer] (nb.west)--(0.42,1.25);
\end{scope}
\begin{scope}[xshift=8.4*0.9cm]
  \draw[draw=white,line width=2.4pt] (0,2.03)--(0,1.72);
  \draw[loose] (0,2.03)--(0,1.72);
  \fullup{2.5} \fulldown{1.25} \fulldown{0}
  \node[note,anchor=west] (nc) at (0.75,1.6) {needs a bad\\connector};
  \draw[pointer] (nc.west)--(0,1.8);
\end{scope}
\end{tikzpicture}
\caption{Columns with an up $\omega$-crossover (circled). (a) A good column. If a column is not good, it has a broken crossover (b) or a cable requiring a bad connector. In (c), a down crossover lies directly below an up one, so a bad connector must attach to the cable between them.}
\label{fig:plane-cables-propagation}
\end{figure}

\begin{lemma}
\label{lem:plane-cables-decoding}
$\nu(H)\ge r-5D$.
\end{lemma}

\begin{proof}
We first show that at most $D_{\mathrm{br}}+2D_{\mathrm{bad}}$ of the $C_\omega$ columns with an up $\omega$-crossover are not good. Consider a column with an up $\omega$-crossover that is not good. If a crossover is not up, the first such crossover is either broken (\cref{fig:plane-cables-propagation}(b)) or down. In the latter case, an up crossover lies directly above a down one, so the cable between them needs a bad connector (\cref{fig:plane-cables-propagation}(c)). If all crossovers are up, the square has no good connector, so the bottom cable needs a bad connector. Broken crossovers occur in at most $D_{\mathrm{br}}$ columns, and each cable below a crossover belongs to just one column. Since a bad connector $S$ attaches to at most $r_S\le2(r_S-1)$ cables, at most $2D_{\mathrm{bad}}$ columns have such an attachment. Hence at most $D_{\mathrm{br}}+2D_{\mathrm{bad}}$ of these columns are not good.

Hence there are at least $C_\omega-D_{\mathrm{br}}-2D_{\mathrm{bad}}$ good columns. Each of them belongs to the square of one of the $q$ good square connectors. If $q_3$ of these squares have three good columns, then the others have at most two, so there are at most $3q_3+2(q-q_3)=q_3+2q$ good columns. Hence
\begin{align*}
    q_3&\ge C_\omega-D_{\mathrm{br}}-2D_{\mathrm{bad}}-2q\\
    &\ge C_\omega-2r-\tfrac73D_{\mathrm{br}}-2D_{\mathrm{bad}}
        &&\text{as }q\le r+\tfrac23D_{\mathrm{br}}\text{ by~\eqref{eq:plane-cables-squares}}.
\end{align*} The triples of these $q_3$ squares almost form a matching. If such a triple contains an element $j$, then its square has a column of $j$, and this column is good, so its $\omega$-crossover is up. Several of these triples may share an element $j$, but at most $C_{\omega,j}$ of them contain $j$. For every element that lies in several of them, we discard all but one, which discards at most $\sum_j\max\{C_{\omega,j}-1,0\}\le2D_{\mathrm{br}}$ triples by \cref{lem:plane-cables-discounts}. The remaining triples are disjoint, and hence
\begin{align*}
    \nu(H)&\ge q_3-2D_{\mathrm{br}}\\
    &\ge C_\omega-2r-\tfrac{13}3D_{\mathrm{br}}-2D_{\mathrm{bad}}
        &&\text{as }q_3\ge C_\omega-2r-\tfrac73D_{\mathrm{br}}-2D_{\mathrm{bad}},\\
    &\ge r-5D
        &&\text{as }3r-C_\omega\le D_\omega.
\end{align*}
\end{proof}

\subsection{The gap}
\label{sec:plane-cables-proof}

\PlaneCablesGap*

\begin{proof}[Proof of \cref{lem:plane-cables-gap}]
Construct the cables of \cref{lem:plane-cables-separated}, with $L\le C_4m^2$. Let $C_3$ be the constant in \cref{lem:plane-cables-cost}, fix
\[
    \rho=\min\Bigl\{\frac1{100},\frac{\omega\sqrt3\,\delta m}{60C_3L}\Bigr\},
\]
so that $C_3\rho L\le\frac{\omega\sqrt3\,\delta}{60}m$, and set $P=P_\rho$ and $\tau=L+\sqrt3\,C+(1+\sqrt3)\,r-3\sqrt3\,\omega r$.

The first implication is \cref{lem:plane-cables-completeness}. For the second, let $T$ be an optimal tree on $P$ and $a=\frac{\omega\sqrt3\,\delta}{60}$. Then
\begin{align*}
    \OPT(P)=\operatorname{len}(T)
    &\ge\tau+\sqrt3\,\omega D-C_3\rho L
        &&\text{by \cref{lem:plane-cables-cost},}\\
    &\ge\tau+\tfrac{\sqrt3\,\omega}{5}\bigl(r-\nu(H)\bigr)-C_3\rho L
        &&\text{by \cref{lem:plane-cables-decoding},}\\
    &\ge\tau+\tfrac{\omega\sqrt3\,\delta}{30}\,m-C_3\rho L
        &&\text{since }\nu(H)\le(1-\delta)r\text{ and }m\le6r,\\
    &\ge\tau+am
        &&\text{since }C_3\rho L\le\tfrac{\omega\sqrt3\,\delta}{60}m.
\end{align*}

Moreover, $\tau\le L+\sqrt3\,C+(1+\sqrt3)\,r=O(m^2)$, and $1/\rho=O(L/m+1)=O(m)$, so $|P|=O(L/\rho^2)=O(m^4)$.
\end{proof}

\begin{proof}[Proof of \cref{thm:fixed-dimension-hardness} for $d=2$]
Replace \cref{lem:fixed-gap} by \cref{lem:plane-cables-gap} and $c_d$ by $c$ in the proof for $d\ge3$.
\end{proof}

\FloatBarrier

\section{Discussion and open problems}
\label{sec:open-problems}

\paragraph{Lower bounds after the claimed ETH refutations.}
Our lower bounds in fixed dimension are nearly tight under weak ETH; under Gap-ETH, they would have matched the known dependence on accuracy. It remains open whether Gap-ETH can be replaced by another hypothesis compatible with the claimed ETH refutations, under which no $(1+\varepsilon)$-approximation algorithm runs in time $2^{o(\varepsilon^{-(d-1)})}\poly(n)$ for any fixed $d\ge2$.

If the claimed ETH refutations~\cite{OpenAI-ETH26,Claude-ETH26} are correct, many optimality questions previously settled under ETH or Gap-ETH become open again. How many of these lower bounds can be recovered, exactly or nearly, under weaker hypotheses? This question also applies to the Gap-ETH bounds for Euclidean TSP and Rectilinear Steiner Tree of Kisfaludi-Bak, Nederlof, and W\k{e}grzycki~\cite{DBLP:journals/jacm/KisfaludiBakNW25}. Whether weak ETH is a reasonable hypothesis is itself open to debate within the TCS community.

Even if Gap-ETH fails, our reduction suggests that improving the running time of the known PTAS would require highly nontrivial techniques: such a speedup would also give a faster gap-3SAT algorithm, and the claimed ETH refutations rely on sophisticated algebraic tensor methods. These ideas might nevertheless lead to faster geometric approximation schemes, which would be surprising, though so were the claimed refutations of ETH themselves.

\paragraph{The dimension threshold for a PTAS.}
Under $\mathrm{NP}\not\subseteq\mathrm{SUBEXP}$, \cref{thm:loglog-hardness} excludes a PTAS in $O(\log\log n)$ dimensions, while Arora's scheme~\cite{DBLP:journals/jacm/Arora98} gives one in up to $\log\log n/\log\log\log n$ dimensions. Can this $O(\log\log\log n)$ gap in the dimension threshold be closed? The same gap remains for Euclidean TSP: padding Trevisan's APX-hardness reduction~\cite{DBLP:journals/siamcomp/Trevisan00} excludes a PTAS in $O(\log\log n)$ dimensions under the same hypothesis, and Arora's scheme gives a PTAS in up to $\log\log n/\log\log\log n$ dimensions.

\section*{AI methodology}
\label{sec:ai-methodology}

We developed the original proof in linear dimension and asked Astra 6 to extend it to logarithmic dimension, initially envisaging a more ad hoc construction. The model worked out proof details and identified the dimension-reduction argument of Karthik C.~S.~\cite{DBLP:journals/corr/abs-2602-03837}. We subsequently suggested adapting the construction to fixed lower dimensions under Gap-ETH, and it found the connection to the cube-wiring theorem of de Berg, Bodlaender, Kisfaludi-Bak, Marx, and van der Zanden~\cite{deBergBBMZ20}. It also generated a first version of the planar argument, which we then fully rewrote for clarity and simplified using the machinery of the rest of the paper.

We used Astra 6 and Opus 5.5 for literature search, for creating the TikZ figures, for assistance with language formulations, and for working out details based on our insights and general proof strategies. The author verified the correctness and originality of all content, including references.

\bibliography{refs}

\appendix

\section{Omitted proofs}
\label{app:separation}

\subsection{Cable separation in fixed dimension \texorpdfstring{$d\ge3$}{d >= 3}}
\label{app:wired-cables}

\begin{proof}[Proof of \cref{lem:wired-cables}]
The construction preserves the tree structure of each $T_i$, so $\Gamma_i$ is a tree containing exactly the ports labelled $i$. Each replacement path stays within distance $25$ of its original grid segment. Since grid segments with no common endpoint are at least $100$ apart, their replacement paths remain at least $50$ apart. Grid segments from different trees can share an endpoint only at a gadget centre $z_s$, as the trees share only gadget vertices.

On the path replacing $(z_s,z_s+100\delta_i)$, every point has the form $y=z_s+a\hat c_i+b\delta_i$ with $a,b\ge0$. By the port assignment, directions used by different labels are orthogonal or opposite, so points $y,y'$ on different such paths satisfy $\langle y-z_s,y'-z_s\rangle\le0$ and hence
\begin{equation}
\label{eq:arms-apart}
    \|y-y'\|\ge\max\{\|y-z_s\|,\|y'-z_s\|\}.
\end{equation}

\emph{(S1).} The balls $B_s$ are clearly more than $20$ apart, and each contains exactly the straight segments from its gadget's ports to its boundary. If two points on different cables are less than $20$ apart, their original grid segments must share a gadget centre $z_s$, and~\eqref{eq:arms-apart} places both points in $B_s$.

\emph{(S2).} All segments have length at least $24$, and incident segments meet at angles of $90^\circ$ or $180^\circ$. Segments of one cable arising from grid segments with no common endpoint are at least $50$ apart. The double subdivisions ensure that modified grid segments of one cable never share an endpoint, so every remaining non-incident pair lies within a detour of~(ii) or between that detour and the next untouched grid segment. In either case, the detour coordinates give a distance of at least $25$.
\end{proof}

\subsection{Cable separation in the plane}
\label{app:plane-placement}

\begin{figure}[H]
\centering
\colorlet{vcable}{black}
\colorlet{hcable}{black!60}
\begin{tikzpicture}[line cap=round,line join=round,x=1cm,y=1cm,scale=.82,font=\small,>=latex,
  cable/.style={draw=black,line width=1.05pt,line cap=round,line join=round},
  disk/.style={fill=black!5,draw=black!45,dotted,line width=.8pt},
  gad/.style={fill=black!6,draw=black!40,line width=.4pt,line join=round},
  junction/.style={circle,draw=black!60,fill=white,inner sep=0pt,minimum size=4pt,line width=.5pt},
  ctr/.style={circle,fill=black!75,inner sep=0pt,minimum size=2.4pt},
  dim/.style={draw=black!60,line width=.45pt,<->},
  note/.style={font=\footnotesize}]
\def\rt{.24}
\def\cy{1.8}
\def\rd{.9}
\pgfmathsetmacro{\cx}{\rt*cos(30)}
\pgfmathsetmacro{\jx}{\cy*sqrt(3)}
\begin{scope}
\node at (0,4.25) {(a) A crossover $x$};
\filldraw[disk] (0,\cy) circle[radius=\rd];
\filldraw[disk] (0,-\cy) circle[radius=\rd];
\draw[cable,draw=vcable] (0,\cy+\rt)--(0,3.8);
\draw[cable,draw=vcable] (0,-\cy-\rt)--(0,-3.8);
\draw[cable,draw=hcable] (-\cx,\cy-\rt/2)--(-\jx,0)--(-\cx,-\cy+\rt/2) (-\jx,0)--(-4.3,0);
\draw[cable,draw=hcable] (\cx,\cy-\rt/2)--(\jx,0)--(\cx,-\cy+\rt/2) (\jx,0)--(4.3,0);
\filldraw[gad] (0,\cy+\rt)--(-\cx,\cy-\rt/2)--(\cx,\cy-\rt/2)--cycle;
\filldraw[gad] (0,-\cy-\rt)--(-\cx,-\cy+\rt/2)--(\cx,-\cy+\rt/2)--cycle;
\node[junction] at (-\jx,0) {};
\node[junction] at (\jx,0) {};
\node[ctr] at (0,0) {};
\node[note,anchor=west] at (.08,0) {$(X,Y)$};
\node[note,anchor=west] at (\rd+.08,\cy) {$(X,Y+40)$};
\node[note,anchor=west] at (\rd+.08,-\cy) {$(X,Y-40)$};
\node[note,anchor=north east] at (-\jx-.02,-.1) {$(X-40\sqrt3,Y)$};
\node[note,anchor=north west] at (\jx+.02,-.1) {$(X+40\sqrt3,Y)$};
\draw[dim] (0,\cy)--($(0,\cy)+(150:\rd)$);
\node[note,anchor=south east] at ($(0,\cy)+(150:\rd)+(.05,-.05)$) {$20$};
\node[note,text=black!60,anchor=north east] at ($(0,-\cy)+(215:\rd)$) {$B_s$};
\node[text=vcable,anchor=west] at (.1,3.55) {$U_x$};
\node[text=vcable,anchor=west] at (.1,-3.55) {$D_x$};
\node[text=hcable,anchor=south] at (-3.9,.05) {$L_x$};
\node[text=hcable,anchor=south] at (4.0,.05) {$R_x$};
\end{scope}
\begin{scope}[shift={(8.5,0)}]
\node at (0,4.25) {(b) A square $Q_e$};
\def\rs{.26}
\filldraw[disk] (0,-.6) circle[radius=\rd];
\draw[cable,draw=vcable] (0,-.6+\rs)--(0,3.4);
\draw[cable,draw=vcable] (-\rs,-.6)--(-2.0,-.6)--(-2.0,3.4);
\draw[cable,draw=vcable] (\rs,-.6)--(2.0,-.6)--(2.0,3.4);
\draw[cable,draw=black!45] (0,-.6-\rs)--(0,-2.6) (-2.8,-2.6)--(2.8,-2.6);
\filldraw[gad] (\rs,-.6)--(0,-.6+\rs)--(-\rs,-.6)--(0,-.6-\rs)--cycle;
\node[ctr] at (0,-.6) {};
\node[note,anchor=north west] at (\rd-.05,-.65) {$Q_e$};
\node[note,text=black!60,anchor=south east] at ($(0,-.6)+(150:\rd)$) {$B_s$};
\node[note,anchor=north,align=center] at (0,3.95) {three columns of $e$};
\node[note,text=black!60,anchor=north] at (0,-2.7) {backbone};
\end{scope}
\end{tikzpicture}
\caption{Cable routing at (a) a crossover $(X,Y)$ and (b) a square $Q_e$. Cables leave triangle ports radially, with the left and right branches ending at the junctions, and square ports along the diagonals. Inside the dotted balls $B_s$, the cables consist exactly of these radial segments.}
\label{fig:plane-cables-gadgets}
\end{figure}
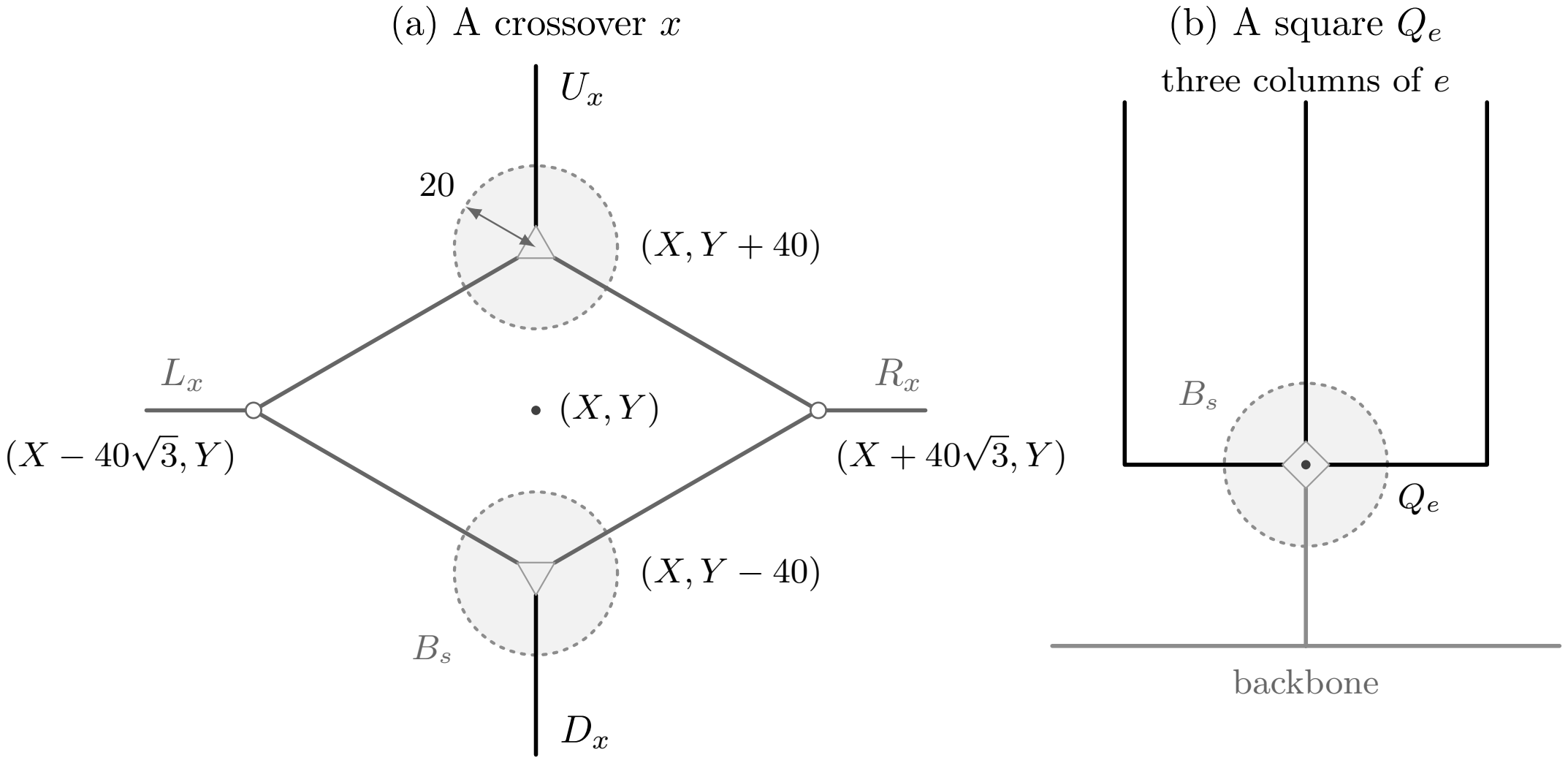

We specify coordinates for the layout in \cref{sec:plane-cables} and verify \cref{lem:plane-cables-separated}. The routing is defined as follows (\cref{fig:plane-cables-gadgets}).
\begin{itemize}
\item Place level $i$ on $y=200i$. For each triple $e_k$, place its three columns on $x=800k$, $x=800k+200$, and $x=800k+400$, and centre $Q_{e_k}$ at $(800k+200,-200)$ with its diagonals parallel to the coordinate axes.
\item At a crossover $(X,Y)$, centre the upper triangle at $(X,Y+40)$ with a corner pointing up, and the lower triangle at $(X,Y-40)$ with a corner pointing down. Join the left corners of both triangles to $(X-40\sqrt3,Y)$, and the right corners to $(X+40\sqrt3,Y)$.
\item Along a column of element $j$ at $x=X$, join consecutive crossovers by their facing ports. Connect the top port of its highest crossover to $(X,200j)$ on level $j$, and the bottom port of its lowest crossover to the nearest port of its square, with a bend at $(X,-200)$ for either outer column.
\item Along level $i$, join successive junctions and endpoints of columns for element $i$ by horizontal segments, omitting the interval between the two junctions of a crossover. Extend level $0$ leftwards to $(0,0)$.
\item Route the backbone from $(0,0)$ to $(0,-400)$ and then along $y=-400$ as far as the last square; connect the bottom port of each square vertically to the backbone.
\end{itemize}

\begin{proof}[Proof of \cref{lem:plane-cables-separated}]
(i) Distinct gadget centres are at least $80$ apart, so the balls $B_s$ are more than $20$ apart. Inside each $B_s$, the construction gives exactly the radial segments from the gadget's ports to the boundary, and any two points on different cables less than $20$ apart lie on radial segments of a common gadget. These segments make angles of at least $90^\circ$, so points at distances $t,t'$ from the centre are at least $\max\{t,t'\}$ apart; a distance of less than $20$ therefore forces both points into $B_s$.

(ii) The grid spacing ensures that every segment has length at least $60$ and that non-incident segments of one cable are at least $60$ apart. At junctions, incident segments meet at $60^\circ$ or $150^\circ$, while all other incident angles are $90^\circ$.

(iii) The total length of the $3r+1$ levels and $3|E|$ columns is $O(r|E|)$, with the remaining segments contributing $O(C+|E|)$. Since $|E|\le3r$ and $C\le9r|E|$, we obtain $L=O(r^2)$, and the explicit coordinates allow the cables to be computed in polynomial time.
\end{proof}

\subsection{The matching gap lemma}
\label{app:gap-source}

\begin{proof}[Proof of \cref{lem:gap-source}]
Petrank~\cite{DBLP:journals/cc/Patrank94}, building on Kann's reduction~\cite{DBLP:journals/ipl/Kann91}, gave a linear-size reduction from bounded-occurrence 3SAT to 3DM-3 that preserves a constant gap. Satisfiable formulas yield perfect matchings, so the PCP theorem gives NP-hardness with perfect completeness.

For (i), Gap-ETH rules out time $2^{o(t)}$ for bounded-occurrence gap-3SAT on instances of size $t$~\cite[Corollary~6.1, full version]{DBLP:journals/jacm/KisfaludiBakNW25}. Since Petrank's reduction gives $r=O(t)$, a $2^{o(r)}$-time algorithm for the matching gap would solve these source instances in time $2^{o(t)}$.

For (ii), Dinur's nearly linear PCP~\cite{Dinur07}, followed by Petrank's reduction, maps a 3SAT formula of size $s$ to a matching gap instance with $r=O(s\,\mathrm{polylog}(s))$. For any fixed $0<\gamma<1$, a $2^{O(r^\gamma)}$-time algorithm would therefore decide 3SAT in time $2^{O(s^\gamma\mathrm{polylog}(s))}\le2^{O(s^{(1+\gamma)/2})}$. Here the polylogarithmic factor is absorbed into $s^{(1-\gamma)/2}$, and $(1+\gamma)/2<1$, contradicting weak ETH.

For (iii), the PCP theorem followed by Petrank's reduction gives $r\le s^k$ for a constant $k$. Under $\mathrm{NP}\not\subseteq\mathrm{SUBEXP}$, 3SAT has no $2^{s^{c'}}$-time algorithm for some $c'>0$. Taking $c=c'/(2k)$, a $2^{r^c}$-time algorithm would decide 3SAT in time $2^{s^{c'/2}}+\poly(s)$, contradicting this bound.
\end{proof}

\end{document}